\documentclass{aa}  
\usepackage{graphicx}
\usepackage[final]{hyperref}
\usepackage{rotating}
\usepackage{float}
\usepackage{txfonts}
\usepackage{float}
\usepackage{siunitx}
\usepackage{tabularx}
\usepackage{color,hyperref}
\usepackage{geometry,fancyhdr}
\usepackage{multicol,multirow}
\usepackage{arydshln}
\usepackage{caption}
\usepackage{subcaption}
\usepackage[flushleft]{threeparttable}
\hypersetup{colorlinks,breaklinks,
            linkcolor=blue,urlcolor=blue,
            anchorcolor=blue,citecolor=blue}
\usepackage{caption}
\usepackage{longtable}
\bibpunct{(}{)}{;}{a}{}{,} 
\usepackage{txfonts}
\begin{document} 

    \title{The Effelsberg survey of FU~Orionis and EX~Lupi objects I.}
    \subtitle{Host environments of FUors/EXors traced by NH$_3$}

   \author{Zs. M. Szabó\inst{1,2,3,4},
          Y. Gong\inst{1},
          K. M. Menten\inst{1},
          W. Yang\inst{1},
          C. J. Cyganowski\inst{2},
          Á. Kóspál\inst{3,4,5,6},
          P. Ábrahám\inst{3,4,5},
          A. Belloche\inst{1},
          \and
          F. Wyrowski\inst{1}
          }
    \institute{Max-Planck-Institut für Radioastronomie, Auf dem Hügel 69, D-53121 Bonn, Germany\\
              \email{zszabo@mpifr-bonn.mpg.de}
    \and
    Scottish Universities Physics Alliance (SUPA), School of Physics and Astronomy, University of St Andrews, North Haugh, St Andrews, KY16 9SS, UK
    \and
    Konkoly Observatory, Research Centre for Astronomy and Earth Sciences, E\"otv\"os Lor\'and Research Network (ELKH), Konkoly-Thege Mikl\'os \'ut 15-17, 1121 Budapest, Hungary
    \and
    CSFK, MTA Centre of Excellence, Budapest, Konkoly Thege Miklós út 15-17., H-1121, Hungary
    \and
    ELTE E\"otv\"os Lor\'and University, Institute of Physics, P\'azm\'any P\'eter s\'et\'any 1/A, H-1117 Budapest, Hungary
    \and
    Max-Planck-Institut für Astronomie, K\"onigstuhl 17, D-69117 Heidelberg, Germany
    }
    

 
  \abstract
   {FU~Orionis (FUor) and EX~Lupi (EXor) type objects represent two small, but rather spectacular groups of low-mass, young eruptive stars. In both cases, outbursts of several magnitudes are observed, which are attributed to enhanced mass accretion from the circumstellar disk onto the central protostar. Although these objects are well studied at optical and near-infrared wavelengths, their host molecular environments are poorly explored because of the scarcity of systematic molecular line observations.}
   {We carried out the first dedicated survey of the molecular environments of a large sample of FUors and EXors, observing a total of 51 sources, including some Gaia~alerts, with the aim of studying the ammonia (NH$_3$) emission in their host environments.}
   {We observed the ammonia $(J,K)$=(1,1), (2,2), and (3,3) inversion transitions at $\sim$23.7\,GHz in position-switching mode using the Effelsberg~100-m radio telescope. For 19 of the 51 sources in our sample we derived H$_2$ column densities and dust temperatures using archival Herschel SPIRE data at 250\,$\mu$m, 300\,$\mu$m, and 500\,$\mu$m.} 
   {We detected the NH$_3$ (1,1) transition toward 28 sources and the (2,2) transition toward 12 sources, while the (3,3) transition was detected toward only two sources in our sample. We find kinetic temperatures between $\sim$12\,K and 21\,K, ammonia column densities from $5.2\times10^{13}\,cm^{-2}$ to $3.2\times10^{15}\,cm^{-2}$, and fractional ammonia abundances with respect to H$_{2}$ from $4.7\times10^{-9}$ to $1.5\times10^{-7}$. These results are comparable to those found in infrared dark clouds (IRDCs). Our kinematic analysis suggests that most of the eruptive stars in our sample reside in rather quiescent (sonic or transonic) host environments.}
   {Our NH$_{3}$ observations and analysis of the SPIRE dust-based H$_{2}$ column density maps confirm the presence of dense material toward 
   7 sources in our sample; additional sources might also harbour dense gas based on their NH$_2$ (2,2) detections, potentially 
   indicating an earlier phase than originally classified.
   Based on our results, we suggest that  observations targeting additional molecular lines would help to refine the evolutionary classification of eruptive stars.}
   \titlerunning{The remnant cores of FUors/EXors traced by NH$_3$}
   \authorrunning{Szab\'o et al.}
   \maketitle
%
\section{Introduction} 
\label{sec:intro}
FU~Orionis (FUor) and EX~Lupi (EXor) type objects are two small, but rather spectacular groups of low-mass pre-main sequence young stellar objects (YSOs). During their formation, low-mass YSOs can undergo violent, episodic outbursts, and the observations of such events as well as the quiescent stages hold crucial information on the early stages of the evolution of Sun-like stars. 
Both FUors and EXors undergo  increases in their optical and near-infrared (NIR) brightnesses, and for both types of objects their eruptions are attributed to enhanced accretion from the circumstellar disk onto the protostar \citep[see, e.g.,][and references therein]{audard2014,fischer2022}. 
FUors may brighten by up to 6 magnitudes at optical wavelengths and stay in a high accretion state for decades, and more likely centuries \citep[][]{paczynski1976,lin1985,kenyon-and-hartmann1988,kenyon-and-hartmann1991,bell1995,turner1997,audard2014,kadam2020,borchert2022}.
EXors, on the other hand, experience outbursts of 1 to 5 magnitudes at optical wavelengths lasting a few months or a few years \citep[see, e.g.,][]{sepic2018}, and their outbursts can be recurring \citep[e.g.,][]{cruz-saenz2022}.
The FUor class was defined by \citet{herbig1977} based on the common properties of FU~Orionis, V1057~Cyg, and V1515~Cyg, often referred to as the classical FUors \citep[see, e.g.,][]{clarke2005,szabo2021,szabo2022}. This class currently includes more than a dozen objects \citep[e.g.,][]{audard2014,szegedi-elek2020}. The EXor class was defined by \citet{herbig1989_eso} based on the properties of EX~Lupi, and this class also consists of about a dozen objects \citep[e.g.,][]{audard2014,park2022}.

The evolution of these peculiar objects starts in their dense cores, i.e.~their birthplaces. However, very little is known in the literature about the host environments of these eruptive objects, despite  dense cores being observable using a variety of molecular lines.  
The paucity of data is partly due to the distances of these objects \citep[see, e.g.,][]{audard2014}.  Another contributing factor is that FUors and EXors have historically been considered too evolved for their dense cores to influence their central YSOs \citep[i.e.\ have been classified as Class\,{\footnotesize II} or T~Tauri stars, see, e.g.,][]{lada1987,herbig1977,kenyon&hartmann1996}.
Indeed, molecular lines were detected toward only $\sim$38\% (13/34) of the target sources in a pioneering ammonia study of T~Tauri stars \citep{lang1979}. 
More recently, however, evidence has emerged that some FUors and EXors are still in more embedded evolutionary phases \citep[e.g.][]{green2013,audard2014}, and that FUors/EXors can also be transition objects (i.e., between  Class\,{\footnotesize 0} and Class\,{\footnotesize I} or between Class\,{\footnotesize I} and Class\,{\footnotesize II}), such as Haro~5a~IRS \citep{kospal2017b}, V1647~Ori \citep{abraham2004b,principe2018}, and HH~354~IRS \citep{bronfman1996,reipurth1997a,reipurth1997b}.
This realization motivates revisiting the temperatures, densities, and dynamic states of the host environments of a large sample of  FUors and EXors, which have not been well-characterized to date due to the scarcity of systematic molecular line observations.  

Ammonia (NH$_3$) was first detected in the interstellar medium by \citet{cheung1968}, and subsequently turned out to be a powerful tool for probing physical conditions in a range of environments, including the environments of low-mass star formation \citep[e.g.,][]{Walmsley1983,Ungerechts1984,tafalla2004}.
The rotational energy levels are described by two principal quantum numbers $(J,K)$, where $J$ is the total angular momentum, and $K$ is its projection along the molecular symmetry axis. 
The spins of the hydrogen atoms can have different orientations, therefore two species of ammonia exist: ortho-NH$_3$, for which all H spins are parallel, and para-NH$_3$, for which they are not. The $(J,K)$ rotational states split into inversion doublets, which split further by hyperfine interactions \citep[for a detailed description see, e.g.,][]{ho1983}. 
The rotational temperature of the gas can be obtained from the intensity ratios of the inversion transitions, and can be used to estimate the kinetic temperature of molecular gas \citep[see][for details]{ho1983}. 
We have chosen ammonia for our study of the host environments of FUors/EXors because it allows for the study of a wide range of properties (i.e.~line widths, temperatures, densities, molecular abundances) and may,  in combination with dust continuum observations, reveal the presence of  dense circumstellar envelopes.
The results of our ammonia survey will help to identify sources with line emission strong enough for follow-up higher angular resolution observations, which in turn will allow us to investigate the relationship between circumstellar envelopes and disk accretion, paving the way toward understanding the eruption mechanisms of FUors and EXors.

This paper is the first part of a survey carried out using the Effelsberg\,100-m radio telescope focusing on FUors and EXors. Here we report observations of the host environments of FUors/EXors in the NH$_3$ (1,1), (2,2), and (3,3) transitions, making this the first dedicated ammonia survey towards these objects.
The second paper in this series will report our search for water (H$_2$O) maser emission towards FUors/EXors.
Our sample consists of 33~FUors, 13~EXors, and a small sample of five Gaia alerts: these are objects with optical variability identified by the Gaia satellite, which are yet to be classified.
Here we note, that although Gaia18dvy is listed with its Gaia alert name (see Table~\ref{tab:non-detection}), this source was classified as an FUor by \citet{szegedi-elek2020} and therefore is not counted within the 5 Gaia alert sources. The criteria for including Gaia alerts in our sample were based on their light curves and luminosity at the time of our proposal submission in 2021 September. In recent years, the Gaia alert system has become an important tool in identifying new outbursting systems \citep[see e.g,][]{hillenbrand2018,hillenbrand2019,szegedi-elek2020,cruz-saenz2022,nagy2022} as well as studying new events in known sources \citep[see, e.g.,][]{nagy2021}.  We specifically chose objects with light curves resembling those of FUors/EXors for inclusion in our sample. 
This paper is organized as follows. 
In Sect.~\ref{sec:observations}, we summarize the observations and describe the data reduction process.
In Sect.~\ref{sec:results_analysis}, we present the analysis and results, including rotational and kinetic temperatures and column densities and abundances for ammonia. 
In Sect.~\ref{dis:ammonia}, we discuss the potential importance of our observational results, and finally our conclusions and summary are given in Sect.~\ref{sec:conclusions}.

\section{Observations and Data Reduction} 
\label{sec:observations}
The three metastable NH$_{3}$ $(J, K) = (1,1), (2,2)$, and $(3,3)$ lines were measured simultaneously with the $J_{K_{\rm a}K_{\rm c}}= 6_{16}-5_{23}$ water maser transition on 2021 November 18, 23, and 2022 January 25 using the Effelsberg~100-m telescope in Germany\footnote{The 100-m telescope at Effelsberg is operated by the Max-Planck- Institut für Radioastronomie (MPIFR) on behalf of the Max-Planck-Gesellschaft (MPG).} (PI: Szabó, project ID: 95-21). 
The observations were performed in position-switching mode with the off-position at an offset of 5\arcmin\, east of each target in azimuth. The 1.3\,cm double beam and dual polarization secondary focus receiver was adopted as the frontend. Fast Fourier Transform Spectrometers (FFTSs) were used as the backend, and each FFTS provides a bandwidth of 300\,MHz and 65536 channels, which gives a channel width of 4.6\,kHz, corresponding to a velocity spacing of 0.06\,km\,s$^{-1}$ at $\sim$23.7\,GHz. The actual spectral resolutions are coarser by a factor of 1.16 \citep{klein2012}. NGC~7027 was used to obtain the pointing and focus corrections at the beginning of each observing session, then W75N was targeted for its well known H$_2$O and NH$_{3}$ transitions in order to verify our spectral setup. We checked the pointing regularly on nearby continuum sources, and the pointing was found to be accurate to about 5$\arcsec$. We also used NGC~7027 as our flux calibrator, which has a continuum flux density of 5.7\,Jy at 23.7\,GHz \citep{ott1994}.
For the spectral calibration, we used the method introduced by \citet{winkel2012}, and the calibration uncertainty is about 15\%. 
The half-power beam width (HPBW) is about 37\arcsec\, and the main beam efficiency is 58.9\% at 24\,GHz. 
The conversion factor from flux density ($S_{\nu}$) to main beam temperature ($T_{\rm mb}$) is $T_{\rm mb}/S_{\nu}$=1.73\,K/Jy. Velocities are presented with respect to the local standard of rest (LSR).

For the data reduction, we used the GILDAS/CLASS software developed by the Institut de Radioastronomie Millim{\' e}trique (IRAM) \footnote{\url{https://www.iram.fr/IRAMFR/GILDAS/}} \citep{gildas_pety2005}. Spectra were averaged for the same target to achieve better sensitivities, and linear baselines were subtracted.

\section{Results and Analysis} 
\label{sec:results_analysis}

Out of our sample of 51 sources, we detected the NH$_3$ (1,1) transition in 28 sources and the (2,2) transition in 12 sources, while the (3,3) transition was only detected in two sources (RNO 1B/1C and V512~Per). This corresponds to detection rates of 54\%, 23\%, and 4\% for NH$_3$ (1,1), (2,2), and (3,3), respectively.  
We did not detect ammonia emission toward the Gaia alert sources. 

\begin{figure*}[htbp]
     \centering
     \begin{subfigure}[b]{0.4\linewidth}
         \centering
         \includegraphics[width=\linewidth]{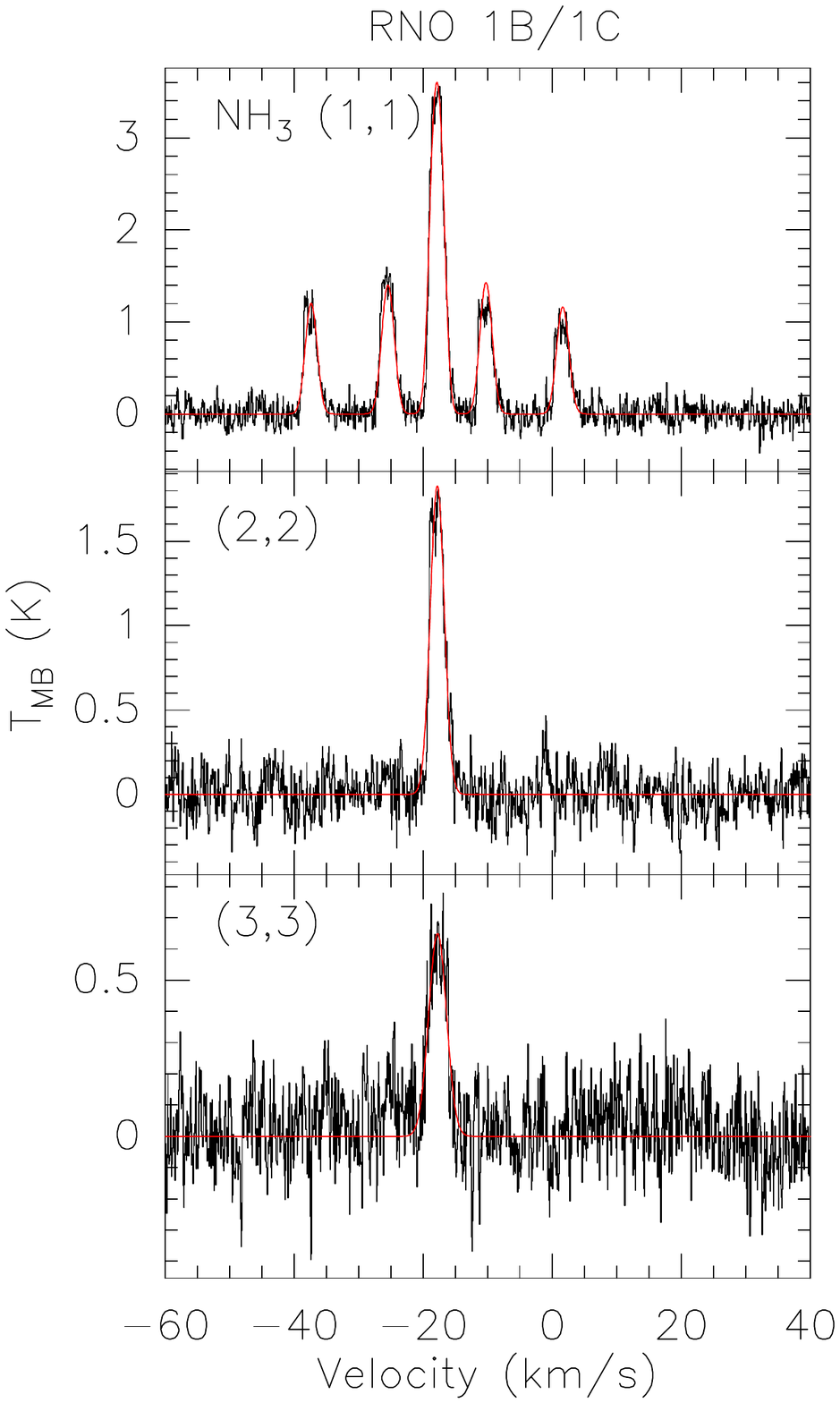}
     \end{subfigure}
     \hspace{-2.5cm}
     \begin{subfigure}[b]{0.4\linewidth}
         \centering
         \includegraphics[width=\linewidth]{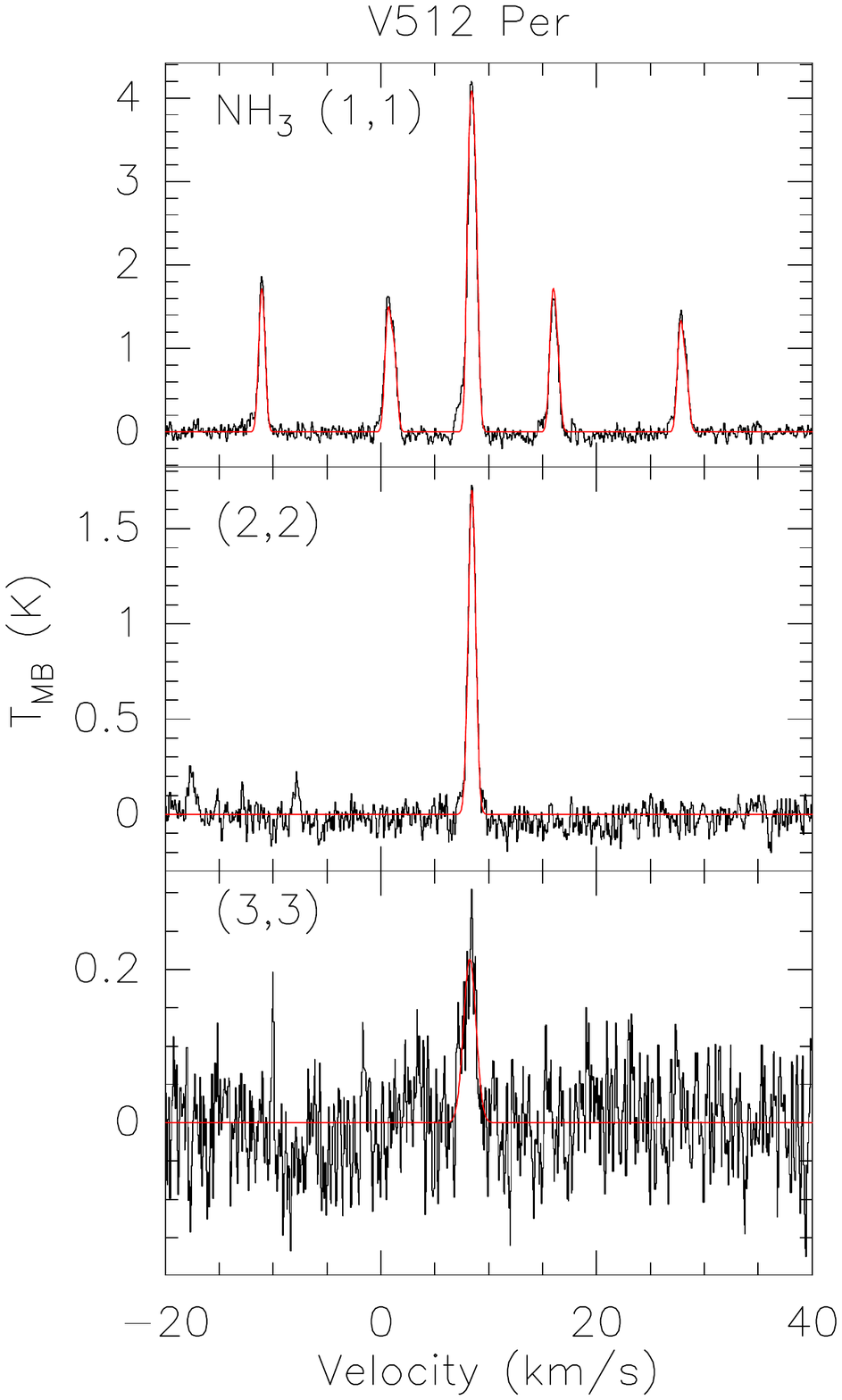}
     \end{subfigure}
     \hspace{-2.5cm}
     \begin{subfigure}[b]{0.4\linewidth}
         \centering
         \includegraphics[width=\linewidth]{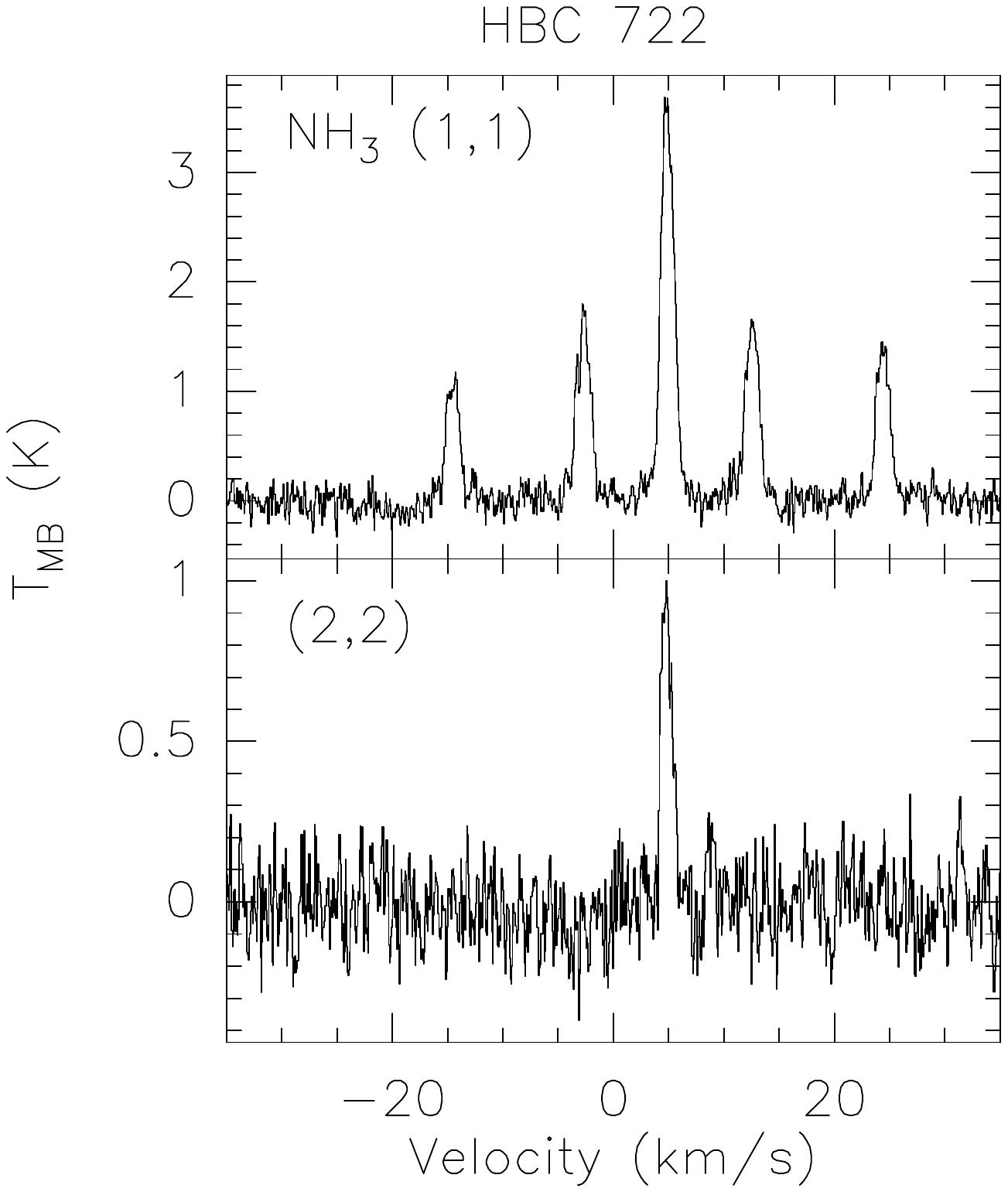}
     \end{subfigure} \\
    \vspace{-2.5cm}
     \begin{subfigure}[b]{0.4\linewidth}
         \centering
         \includegraphics[width=\linewidth]{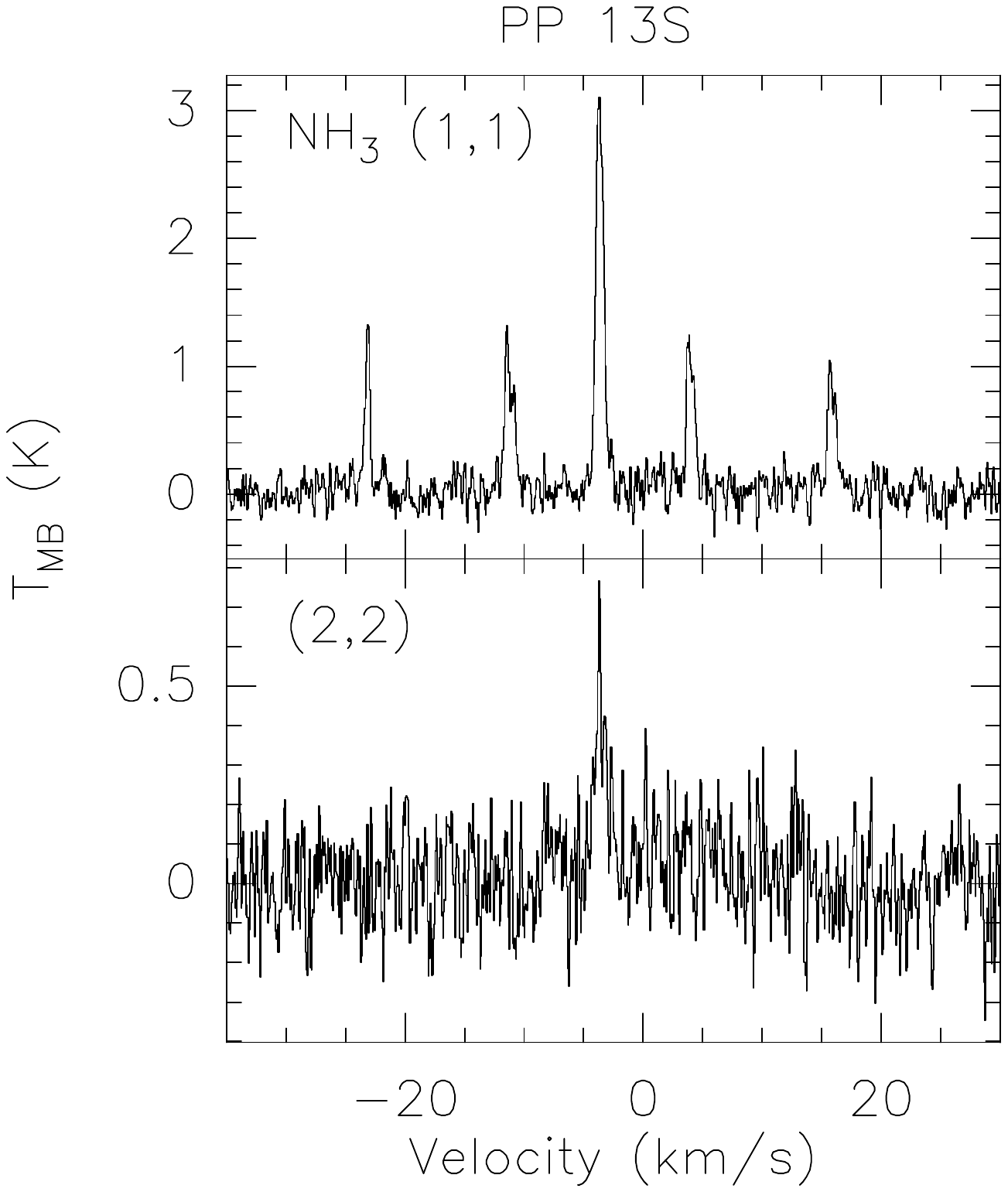}
     \end{subfigure}
     \hspace{-2.5cm}
     \begin{subfigure}[b]{0.4\linewidth}
         \centering
         \includegraphics[width=\linewidth]{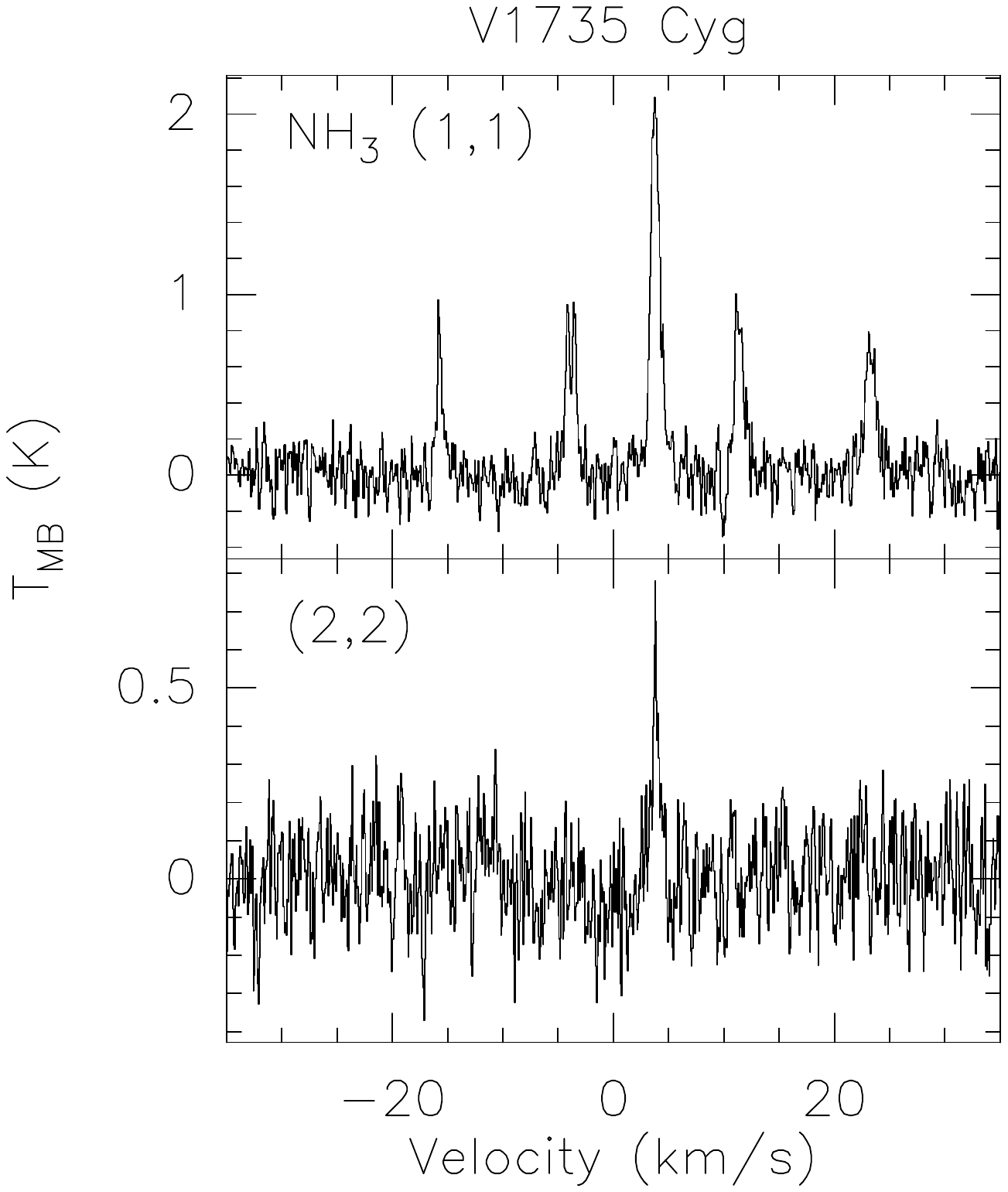}
     \end{subfigure}
     \hspace{-2.5cm}
     \begin{subfigure}[b]{0.4\linewidth}
         \centering
         \includegraphics[width=\linewidth]{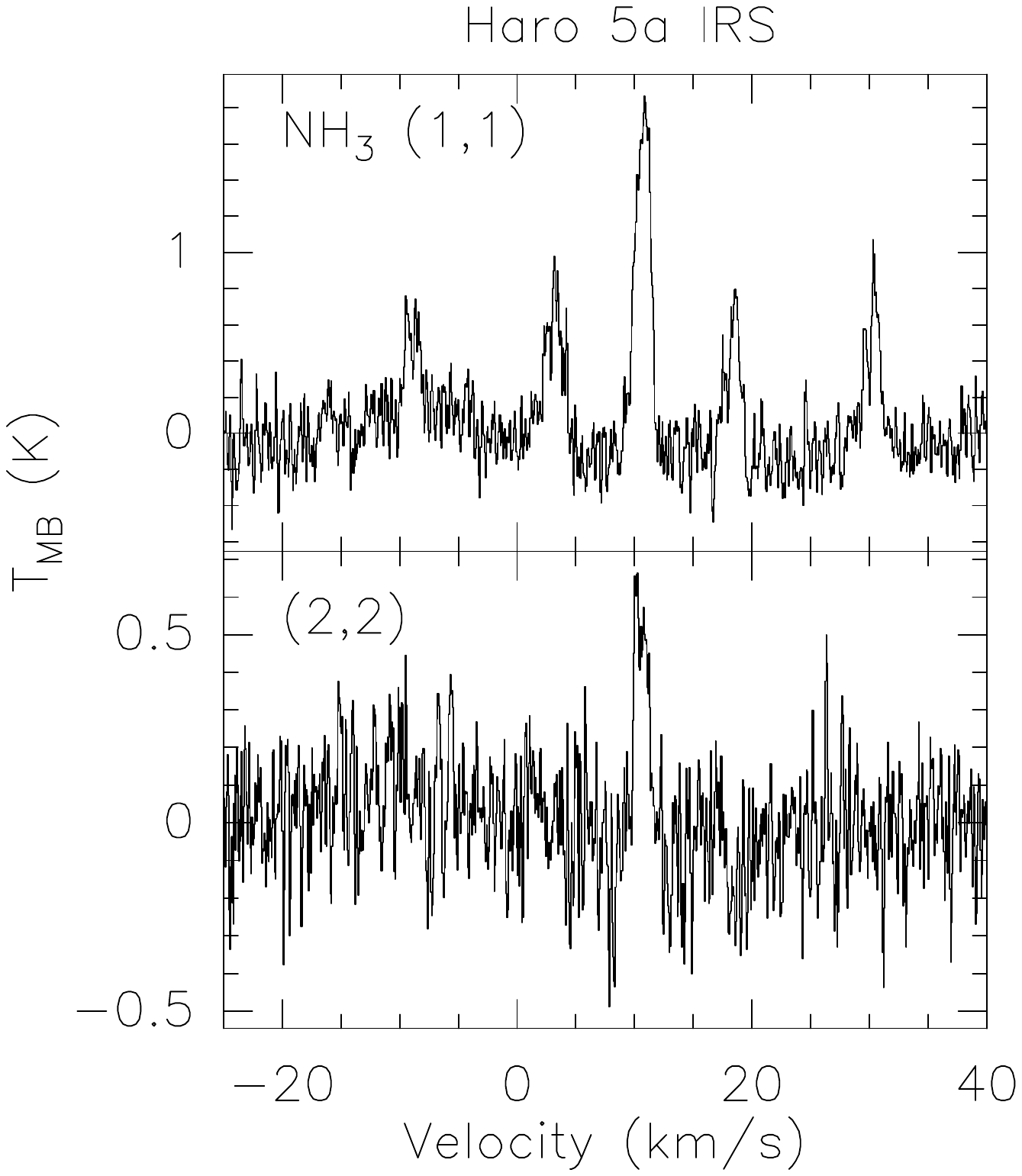}
     \end{subfigure}
     \vspace{-3.5cm}
     \caption{Examples of reduced and calibrated spectra for the NH$_3$ (1,1), (2,2) and (3,3) transitions. The transition is indicated in the upper left corner of each panel. For the first two sources the fits are shown in red. The (3,3) transition was detected only towards RNO~1B/1C and V512~Per (SVS~13A).}
     \label{fig:spectra} 
\end{figure*}

The hyperfine structure (HFS) lines of the NH$_3$ (1,1) transition were fitted using the hyperfine fitting method in CLASS. 
The independent parameters from the fitting are the LSR velocity ($\varv_{\rm LSR}$), line width ($\Delta \varv$) at the full width at half maximum (FWHM) of a Gaussian profile, and the optical depth of the main HFS line ($\tau_{m}$). 
The satellite lines of the (2,2) and (3,3) transitions were too weak to be detected.  Therefore, only the main lines of these transitions were fitted,  assuming a Gaussian function: the derived parameters are the LSR velocity ($\varv_{\rm LSR}$) and FWHM line width ($\Delta \varv$) of the main lines. 
Figure~\ref{fig:spectra} shows examples of the reduced and calibrated spectra.  
The peak main beam brightness temperatures of the NH$_3$ (1,1), (2,2) and (3,3) transitions were derived from the peak intensity of the Gaussian fit to the main line. 

For undetected lines, three times the rms noise (3$\sigma$) was adopted as an upper limit (given in Table~\ref{tab:non-detection} of Appendix~\ref{sec:negative_detections}).
In Tables~\ref{tab:names-nh31,1_parameters_fuors} and \ref{tab:names-nh31,1_parameters_exors}, we list the properties of the detected (1,1) transitions for the FUors and EXors, respectively. In each case, we give the name, position, optical depth ($\tau$), LSR velocity  ($\varv_{\rm LSR}$(1,1)), line width ($\Delta \varv$(1,1)), and the main beam brightness temperature ($T_{\rm MB}$(1,1)). The formal errors of the fits are given in parentheses. 
In Table~\ref{tab:names-nh322-33_parameters_fuors} and Table~\ref{tab:names-nh322-33_parameters_exors}, we list the $\varv_{\rm LSR}$, $\Delta \varv$ and $T_{\rm MB}$ values for the (2,2) and (3,3) transitions, with 3$\sigma$ upper limits given in the case of non-detections.

\begin{table*}[ht!]
\small
    \centering
    \caption{NH$_3$ (1,1) line parameters for the FU~Orionis type objects detected in our survey.}
    \begin{tabular}{cccrrrclll}
    \hline \hline
    \multirow{2}{*}{Name}   & R.A. (J2000)        & Dec. (J2000)       & \multirow{2}{*}{$\tau$(1,1)$^{(a)}$}   & $\varv_{\rm LSR}$(1,1)$^{(b)}$    & $\Delta \varv$(1,1)$^{(c)}$ & $T_{\rm MB}$(1,1)$^{(d)}$ & $\Delta \varv_{\rm th}^{(e)}$ & $\Delta \varv_{\rm nt}^{(f)}$ & \multirow{2}{*}{$Ma^{(g)}$} \\
                            & ($^{\rm h}$ $^{\rm m}$ $^{\rm s}$)   & ($^{\circ}$ $\arcmin$ $\arcsec$)   &                                & (km\,s$^{-1}$)        & (km\,s$^{-1}$) & (K) & (km\,s$^{-1}$) & (km\,s$^{-1}$)  &\\
    \hline
    RNO~1B/1C   & 00 36 46.30 & +63 28 54.0 & $1.16$ $(0.03)$ & $-17.83$ $(0.02)$    &  $2.39$ $(0.04)$ & $3.75$ $(0.32)$ & $0.31$ & $2.36$ & $3.36$ \\
    PP~13S      & 04 10 41.09 & +38 07 54.5 & $0.79$ $(0.05)$ & $-3.62$ $(0.01)$ & $0.75$ $(0.01)$ & $3.09$ $(0.20)$ & $0.24$ & $0.71$ & $1.46$ \\ 
    L1551~IRS~5 & 04 31 34.07 & +18 08 04.9 & $1.93$ $(0.12)$ & $6.35$ $(0.01)$ & $0.87$ $(0.02)$ & $2.78$ $(0.22)$ & $0.23$ & $0.83$ & $1.78$\\ 
    Haro~5a~IRS	& 05 35 26.74 & $-$05 03 55.0 & $1.80$ $(0.16)$ & $10.7$ $(0.01$) & $1.19$ $(0.02)$ & $1.89$ $(0.26)$ & $0.26$ & $1.16$ & $2.21$ \\ 
    V2775~Ori   & 05 42 48.48 &	$-$08 16 34.7 & $1.07$ $(0.24)$ & $3.05$ $(0.01)$ & $0.65$ $(0.03)$ & $1.98$ $(0.21)$ & $0.19^h$ & $0.62^h$ & $1.12^h$ \\ 
    NGC~2071	& 05 47 09.80 & +00 18 00.0	  & $0.10$ $(0.18)$ & $10.4$ $(0.01)$  & $0.52$ $(0.03)$ & $1.06$ $(0.14)$ & $0.19^h$ & $0.48^h$ & $0.92^h$ \\ 
    V899~Mon    & 06 09 19.24 & $-$06 41 55.8 & $0.42$ $(0.19)$ & $9.63$ $(0.01)$ & $0.46$ $(0.02)$ & $1.62$ $(0.15)$ & $0.19^h$ & $0.41^h$ & $0.79^h$ \\ 
    IRAS~06297+1021W & 06 32 28.70 & +10 19 0 & $0.86$ $(0.22)$ & $4.17$ $(0.01)$ & $0.43$ $(0.02)$ & $1.64$ $(0.16)$ & $0.19^h$ & $0.38^h$ & $0.73^h$ \\ 
    AR~6A/6B	    & 06 40 59.30 & +09 35 49.0 & $0.70$ $(0.11)$ & $5.06$ $(0.02)$ & $2.31$ $(0.05)$ & $1.71$ $(0.25)$ & $0.27$ & $2.29$ & $4.10$\\ 
    IRAS~06393+0913	& 06 42 08.13 & +09 10 30.0 & $0.10$ $(0.11)$ & $7.72$ $(0.02)$ & $0.49$ $(0.04)$ & $0.68$ $(0.14)$ & $0.19^h$ & $0.45^h$ & $0.87^h$ \\ 
    V960~Mon    & 06 59 31.58 & $-$04 05 27.7 & $0.27$ $(0.47)$ & $23.8$ $(0.02)$ & $0.70$   $(0.10)$ & $0.93$ $(0.26)$  & $0.19^h$ & $0.67^h$ & $1.29^h$\\ 
    Z~CMa       & 07 03 43.15 & $-$11 33 06.2 & $0.10$ $(0.27)$ & $13.8$ $(0.02)$ & $1.22$   $(0.04)$ & $1.35$ $(0.19)$ & $0.19^h$ & $1.21^h$ & $2.34^h$ \\ 
    iPTF~15AFQ 	& 07 09 21.39 & $-$10 29 34.5 & $1.14$ $(0.28)$ & $13.3$ $(0.01)$ & $0.71$   $(0.04)$ & $1.77$ $(0.30)$ & $0.19^h$ & $0.68^h$ & $1.31^h$ \\
    IRAS~18270-0153W & 18 29 36.90 & $-$01 51 02.0 & $2.95$ $(0.11)$ & $7.61$ $(0.01)$ & $0.55$ $(0.01)$ & $3.80$ $(0.32)$ & $0.23$ & $0.49$ & $1.06$ \\ 
    OO~Ser 	    & 18 29 49.13 & +01 16 20.6 & $1.68$ $(0.08)$ & $8.31$ $(0.01)$ & $0.77$ $(0.01)$ & $3.43$ $(0.28)$ & $0.26$ & $0.73$ & $1.37$ \\ 
    IRAS~18341-0113S & 18 36 45.70 & $-$01 10 29.0 & $2.82$ $(0.17)$ & $9.27$ $(0.01)$ & $0.41$ $(0.01)$ & $2.65$ $(0.16)$ & $0.22$ & $0.34$ & $0.74$ \\ 
    HBC~722		& 20 58 17.00 & +43 53 43.0	& $1.98$ $(0.06)$ & $4.93$ $(0.01)$  & $1.08$ $(0.01)$ & $3.64$ $(0.34)$ & $0.24$ & $1.05$ & $2.09$ \\ 
    V1057~Cyg 	& 20 58 53.73 & +44 15 28.4 & $1.89$ $(0.57)$ & $4.35$ $(0.02)$   & $0.72$ $(0.06)$ & $0.49$ $(0.16)$ & $0.19^h$ & $0.69^h$ & $1.33^h$ \\ 
    V2495~Cyg 	& 21 00 25.24 & +52 30 16.9	& $0.86$ $(0.37)$ & $4.71$ $(0.02)$   & $0.80$ $(0.05)$ & $0.74$ $(0.15)$ & $0.19^h$ & $0.77^h$ & $1.49^h$ \\ 
    RNO~127		& 21 00 31.80 & +52 29 17.0 & $2.37$ $(0.65)$ & $-2.90$  $(0.01)$  & $0.37$ $(0.02)$ & $0.72$ $(0.16)$ & $0.19^h$ & $0.31^h$ & $0.59^h$ \\ 
    CB~230		& 21 17 38.62 & +68 17 34.0	& $2.56$ $(0.15)$ & $2.79$   $(0.01)$ & $0.46$ $(0.01)$ & $2.21$ $(0.19)$ & $0.19^h$ & $0.41^h$ & $0.79^h$ \\ 
    V1735~Cyg	& 21 47 20.66 & +47 32 03.8	& $1.97$ $(0.17)$ & $3.80$   $(0.01)$ & $0.69$ $(0.02)$ & $2.01$ $(0.21)$ & $0.24$ & $0.64$ & $1.29$ \\ 
    HH~354~IRS		& 22 06 50.37 & +59 02 45.9 & $2.35$ $(0.33)$ & $-1.52$  $(0.01)$ & $0.36$ $(0.01)$ & $1.36$ $(0.16)$ & $0.19^h$ & $0.30^h$ & $0.58^h$ \\ 
    V733~Cep	& 22 53 33.25 & +62 32 23.6	& $2.32$ $(0.43)$ & $-8.93$  $(0.01)$ & $0.36$ $(0.02)$ & $0.98$ $(0.15)$ & $0.19^h$ & $0.30^h$ & $0.58^h$ \\ 
    \hline
    \end{tabular}
    \label{tab:names-nh31,1_parameters_fuors}
    \flushleft 
    \tablefoot{\tablefoottext{a}{Optical depth of the (1,1) transition main line.}\tablefoottext{b}{LSR velocity.}\tablefoottext{c}{Line width determined using the HFS fitting method.}\tablefoottext{d}{Main beam brightness temperature.}\tablefoottext{e}{Thermal line width.}\tablefoottext{f}{Non-thermal line width.}\tablefoottext{g}{Mach number.}\tablefoottext{h}{Source with the assumed $T_{\rm kin}$ value of 14.6\,K.}}
\normalsize
\end{table*}

\begin{table*}[htbp]
\small
    \centering
    \caption{NH$_3$ (1,1) line parameters for the EX~Lupi type objects detected in our survey.}
    \begin{tabular}{ccccrcclll}
    \hline \hline
   \multirow{2}{*}{Name}   & R.A. (J2000)        & Dec. (J2000)       & \multirow{2}{*}{$\tau$(1,1)$^{(a)}$}   & $\varv_{\rm LSR}$(1,1)$^{(b)}$    & $\Delta \varv$(1,1)$^{(c)}$ & $T_{\rm MB}$(1,1)$^{(d)}$ & $\Delta \varv_{\rm th}^{(e)}$ & $\Delta \varv_{\rm nt}^{(f)}$ & \multirow{2}{*}{$Ma^{(g)}$} \\
                            & ($^{\rm h}$ $^{\rm m}$ $^{\rm s}$)   & ($^{\circ}$ $\arcmin$ $\arcsec$)   &                                & (km\,s$^{-1}$)        & (km\,s$^{-1}$) & (K) & (km\,s$^{-1}$) & (km\,s$^{-1}$)  &\\
    \hline
    V512~Per (SVS~13)   & 03 29 03.75 & +31 16 03.9 & $1.77$ $(0.03)$ & $8.45$ $(0.01)$ & $0.64$ $(0.01)$ & $4.23$ $(0.35)$ & $0.27$ & $0.58$ & $1.03$ \\
    LDN~1415~IRS     & 04 41 37.50 & +54 19 22.0 & $1.26$ $(0.62)$ & $-5.77$ $(0.02)$ & $0.48$ $(0.04)$ & $0.63$ $(0.16)$ & $0.19^h$ & $0.44^h$ & $0.85^h$ \\
    V371~Ser            & 18 29 51.21 & +01 16 39.4	& $2.20$ $(0.03)$ & $8.34$ $(0.01)$ & $0.78$ $(0.01)$& $3.76$ $(0.34)$ & $0.23$ & $0.74$ & $1.54$ \\
    V2492~Cyg           & 20 51 26.23 & +44 05 23.8 & $0.86$ $(0.37)$ & $4.71$ $(0.02)$  & $0.70$ $(0.05)$ & $0.74$ $(0.15)$ & $0.19^h$ & $0.67^h$ & $1.29^h$ \\
    \hline 
    \end{tabular}
    \label{tab:names-nh31,1_parameters_exors}
    \flushleft
    \tablefoot{\tablefoottext{a}{Optical depth of the (1,1) transition main line.}\tablefoottext{b}{LSR velocity.}\tablefoottext{c}{Line width determined using the HFS fitting method.}\tablefoottext{d}{Main beam brightness temperature.}\tablefoottext{e}{Thermal line width.}\tablefoottext{f}{Non-thermal line width.}\tablefoottext{g}{Mach number.}\tablefoottext{h}{Source with the assumed $T_{\rm kin}$ value of 14.6\,K.}}
\end{table*}

\begin{table*}[h]
\small
    \centering
    \caption{NH$_3$ (2,2) and (3,3) line parameters for the FU~Orionis type objects detected in NH$_3$ (1,1).}
    \begin{tabular}{crccccc}
    \hline \hline
    \multirow{2}{*}{Name}   & $\varv_{\rm LSR}$(2,2)$^{(a)}$    & $\Delta \varv$(2,2)$^{(b)}$ & $T_{\rm MB}$(2,2)$^{(c)}$ & $\varv_{\rm LSR}$(3,3)$^{(d)}$    & $\Delta \varv$(3,3)$^{(e)}$ & $T_{\rm MB}$(3,3)$^{(f)}$ \\
                            & (km\,s$^{-1}$)        & (km\,s$^{-1}$) & (K)               & (km\,s$^{-1}$)        & (km\,s$^{-1}$) & (K)               \\
    \hline
    RNO~1B/1C   & $-17.80$ $(0.01)$ & $2.55$ $(0.03)$ & $1.82$ $(0.11)$ & $-17.75$ $(0.05)$  &  $3.19$ $(0.11)$ & $0.64$ $(0.11)$ \\
    PP~13S      & $-3.57$ $(0.04)$  & $1.32$ $(0.12)$ & $0.74$ $(0.11)$ & $-$                & $-$              & $<0.11$ \\
    L1551~IRS~5 & $6.30$ $(0.03)$   & $0.62$ $(0.08)$ & $0.59$ $(0.11)$ & $-$                & $-$              & $<0.10$ \\
    Haro~5a~IRS & $10.53$ $(0.04)$  & $1.27$ $(0.09)$ & $0.59$ $(0.14)$ & $-$                & $-$              & $<0.16$ \\
    V2775~Ori           & $-$       & $-$             & $<0.12$         & $-$                & $-$              & $<0.13$ \\
    NGC~2071            & $-$       & $-$             & $<0.11$         & $-$                & $-$              & $<0.10$ \\
    V899~Mon            & $-$       & $-$             & $<0.10$         & $-$                & $-$              & $<0.11$ \\
    IRAS~06297$+$1021W   & $-$       & $-$             & $<0.12$         & $-$                & $-$              & $<0.14$ \\
    AR~6A/6B            & $5.28$ $(0.07)$ &  $2.48$ $(0.19)$ & $0.73$ $(0.13)$ & $-$ & $-$ & $<0.16$ \\
    IRAS~06393$+$0913     & $-$ & $-$ & $<0.10$ & $-$ & $-$ & $<0.10$ \\
    V960~Mon            & $-$ & $-$ & $<0.18$ & $-$ & $-$ & $<0.18$ \\
    Z~CMa               & $-$ & $-$ & $<0.12$ & $-$ & $-$ & $<0.13$ \\
    iPTF~15AFQ             & $-$ & $-$ & $<0.19$ & $-$ & $-$ & $<0.21$ \\
    IRAS~18270$-$0153W   & $7.62$ $(0.01)$ & $0.87$ $0.05)$ & $1.14$ $(0.11)$ & $-$ & $-$ & $<0.14$ \\
    OO~Ser              & $8.33$ $(0.02)$ & $1.07$ $0.05)$ & $1.07$ $(0.11)$ & $-$ & $-$ & $<0.11$ \\
    IRAS~18341$-$0113S   & $9.20$ $(0.02)$ & $0.42$ $0.05)$ & $0.59$ $(0.12)$ & $-$ & $-$ & $<0.12$ \\
    HBC~722             & $4.82$ $(0.02)$ & $1.16$ $0.04)$ & $0.98$ $(0.11)$ & $-$ & $-$ & $<0.11$ \\
    V1057~Cyg           & $-$ & $-$ & $<0.10$ & $-$ & $-$ & $<0.10$ \\
    V2495~Cyg           & $-$ & $-$ & $<0.15$ & $-$ & $-$ & $<0.10$ \\
    RNO~127             & $-$ & $-$ & $<0.10$ & $-$ & $-$ & $<0.10$ \\
    CB~230              & $-$ & $-$ & $<0.10$ & $-$ & $-$ & $<0.10$ \\
    V1735~Cyg           & $3.84$ $(0.03)$ & $0.78$ $(0.10)$ & $0.77$ $(0.12)$ & $-$ & $-$ & $<0.33$ \\
    HH~354~IRS          & $-$ & $-$ & $<0.10$ & $-$ & $-$ & $<0.11$ \\
    V733~Cep            & $-$ & $-$ & $<0.10$ & $-$ & $-$ & $<0.10$ \\
    \hline
    \end{tabular}
    \label{tab:names-nh322-33_parameters_fuors}
    \flushleft
    \tablefoot{\tablefoottext{a}{LSR velocity of the (2,2) transition.}\tablefoottext{b}{Line width determined using a single Gaussian.}\tablefoottext{c}{Main beam brightness temperature. The upper limits are 3$\mathbf{\sigma}$.}\tablefoottext{d}{LSR velocity of the (3,3) transition.}\tablefoottext{e}{Line width determined using a single Gaussian.}\tablefoottext{e}{Line width determined using a single Gaussian.}\tablefoottext{f}{Main beam brightness temperature. The upper limits are 3$\mathbf{\sigma}$. The errors are given in parentheses.}}
\end{table*}

\subsection{Molecular excitation}\label{sec.excitation}
The rotational ($T_{\rm rot}$) and kinetic ($T_{\rm kin}$) temperatures, as well as the ammonia column density ($N_{\rm NH_3}$), were determined using the standard method \citep{ho1983,Ungerechts1984}. The results are given in Tables~\ref{tab:trot-tkin-fuors} and \ref{tab:trot-tkin-exors} for the FUors and EXors, respectively.  

The rotational temperature can only be determined for sources in  which both the (1,1) and (2,2) transitions were detected.
To calculate the $T_{\rm rot}$ values, we used the following relation \citep{ho1983}:
\begin{equation}
    T_{\rm rot}=\frac{-41.5}{{\rm ln}\bigg(\frac{-0.282}{\tau_m(1,1)}{\rm ln}\Big(1-\frac{T_{\rm MB}(2,2)}{T_{\rm MB}(1,1)}\Big(1-\exp(-\tau_m(1,1))\Big)\Big)\bigg)}\;,
    \label{eq:t_rot}
\end{equation}
using the optical depth of the (1,1) main line, $\tau_m(1,1)$, and the main beam brightness temperatures, $T_{\rm MB}$, of the (1,1) and (2,2) main lines derived from the Gaussian fitting.

The derived rotational temperatures in our sample range from 11\,K to 18\,K, with an average $T_{\rm rot}$ of 13.2\,K. 

\begin{equation}
    T_{\rm kin}=\frac{T_{\rm rot}(1,2)}{1-\frac{T_{\rm rot}(1,2)}{42\;{\rm K}}{\rm ln}\bigg(1+1.1\exp\Big(\frac{-16\;{\rm K}}{T_{\rm rot}(1,2)}\Big)\bigg)}\;,
    \label{eq:t_kin}
\end{equation}
where $T_{\rm rot}$(1,2) is the rotational temperature determined from the (1,1) and (2,2) inversion transitions, and 42\,K is the energy difference between the (1,1) and (2,2) levels. We find that the host environments are characterised by kinetic temperatures of 12--21\,K with an average kinetic temperature of 14.6\,K, with the highest kinetic temperature found towards RNO~1B/1C. These kinetic temperatures are lower than those founds towards Class\,{\footnotesize II} sources (26--37\,K), but are similar to low-mass and high-mass dense clumps in early evolutionary stages \citep[e.g.,][]{benson1989,pillai2006,zhang2011,wienen2012}.

To calculate the ammonia column density, the rotational temperature derived from Eq.~\ref{eq:t_rot}, $T_{\rm rot}$(1,2), the optical depth, $\tau_m$(1,1), and the linewidth, $\Delta \varv$(1,1), of the (1,1) inversion transition are needed. 

We calculated the $N_{\rm tot}$ values using the column density of the (1,1) level, assuming that the energy levels are populated according to the Boltzmann distribution \citep[see, e.g.,][]{rohlfs2004,wienen2012}.  
For the calculation of the total column density, we used the relation given by \citet{rohlfs2004}, with the assumption that the lowest metastable levels dominate in the population 

\begin{align}
    N_{\rm tot}\approx N(1,1)\Bigg(\frac{1}{3}\exp\Bigg(\frac{23.1}{T_{\rm rot}(1,2)}\Bigg)+1+\frac{5}{3}\exp\Bigg(-\frac{41.2}{T_{\rm rot}(1,2)}\Bigg) \nonumber \\ +\frac{14}{3}\exp\Bigg(-\frac{99.4}{T_{\rm rot}(1,2)}\Bigg)\Bigg)\;,
    \label{eq:N_tot}
\end{align}
where $T_{\rm rot}$(1,2) is the rotational temperature, and $N$(1,1) is the column density of the (1,1) level,

\begin{align}
     N(1,1) & =4.14\times10^3\frac{g_{\rm l}\,\nu^2\,T_{\rm rot}(1,2)}{g_{\rm u}\,A_{\rm ul}}\left(1+{\rm  exp}(\frac{h\nu}{kT_{\rm ex}})\right) \tau_{\rm m}(1,1)\,\Delta \varv\; \notag \\ 
     & \approx 2.72\times 10^{13} T_{\rm rot}(1,2) \tau_{\rm m}(1,1)\,\Delta \varv\;{\rm cm}^{-2},
     \label{eq:N(1,1)}
\end{align}

where $g_{\rm l}$, $g_{\rm u}$ are the statistical weights of the lower/upper levels, $A_{\rm ul}$ is the Einstein coefficient, $\nu$ is the frequency in units of GHz, $T_{\rm ex}$ is the excitation temperature of the (1,1) transition, and $\Delta \varv$ is the linewidth in units of km~s$^{-1}$.  We have used the approximation that $\frac{h\nu}{kT_{\rm ex}}\ll$1 in Eq.~(\ref{eq:N(1,1)}). Assuming  LTE, we used the rotational temperature as the excitation temperature \citep[e.g.,][]{goldsmith1999,wilson2009(tools-of...)}. The values for $g_{\rm l}$, $g_{\rm u}$, $A_{\rm ul}$, and $\nu$ were taken from the Leiden Atomic and Molecular Database \citep[LAMDA,][]{lamda2005}.

The average rotational temperature for all sources (with the (1,1) and the (2,2) emission detected) was found to be 13.2\,K.
For sources detected only in the (1,1) transition, we adopt T$_{\rm rot}=$ 13.2\,K in order to estimate their NH$_3$ column densities and abundances. 
The results are given in Tables~\ref{tab:trot-tkin-fuors} and \ref{tab:trot-tkin-exors}, for the FUors and EXors respectively. Since we find  rotational temperatures between 11\,K and 18\,K, we assume an uncertainty for the average  rotational temperature of $\sim$30\%. 

The ammonia column densities in the sample range from $5.2\times10^{13}\,cm^{-2}$ (IRAS~06393$+$0913) to $3.2\times10^{15}\,cm^{-2}$ (RNO~1B/1C), with an average of $1.18\times10^{15}\,cm^{-2}$ and a median of $1.15\times10^{15}\,cm^{-2}$, respectively.
In infrared dark clouds (IRDCs) ammonia column densities were found to range from $7.6\times10^{14}\,cm^{-2}$ to $6.04\times10^{15}\,cm^{-2}$, with an average value of 
$3\times10^{15}\,cm^{-2}$ \citep{pillai2006}. 
Approximately 2/3 of our sample (18/28 objects $\sim$64\%) falls within this range, with the remaining sources having NH$_3$ column densities lower than observed in IRDCs
(see Table~\ref{tab:trot-tkin-fuors} and \ref{tab:trot-tkin-exors}).

\begin{table*}[htbp]
\small
    \centering
    \caption{NH$_3$ (2,2) and (3,3) line parameters for the EX~Lupi type objects detected in NH$_3$ (1,1). }
    \begin{tabular}{ccccccc}
    \hline \hline
    \multirow{2}{*}{Name}   & $\varv_{\rm LSR}$(2,2)$^{(a)}$    & $\Delta \varv$(2,2)$^{(b)}$ & $T_{\rm MB}$(2,2)$^{(c)}$ & $\varv_{\rm LSR}$(3,3)$^{(d)}$    & $\Delta \varv$(3,3)$^{(e)}$ & $T_{\rm MB}$(3,3)$^{(f)}$ \\
                            & (km\,s$^{-1}$)        & (km\,s$^{-1}$) & (K)               & (km\,s$^{-1}$)        & (km\,s$^{-1}$) & (K)               \\
    \hline
     V512~Per (SVS~13)      & $8.45$ $(0.01)$       & $0.84$ $(0.03)$ & $1.71$ $(0.06)$   & $8.23$ $(0.05)$ & $1.36$ $(0.12)$ & $0.21$ $(0.06)$ \\
     LDN~1415~IRS           & $-$ & $-$ & $<0.11$   & $-$ & $-$ & $<0.12$ \\
     V371~Ser               & $8.43$ $(0.02)$       & $0.97$ $(0.06)$ & $0.80$ $(0.06)$ & $-$ & $-$ & $<0.11$ \\
     V2492~Cyg              & $-$ & $-$ & $<0.10$ & $-$ & $-$ & $<0.10$ \\ 
    \hline
    \end{tabular}
    \label{tab:names-nh322-33_parameters_exors}
    \flushleft
    \tablefoot{\tablefoottext{a}{LSR velocity of the (2,2) transition.}\tablefoottext{b}{Line width determined using a single Gaussian.}\tablefoottext{c}{Main beam brightness temperature. The upper limits are 3$\mathbf{\sigma}$.}\tablefoottext{d}{LSR velocity of the (3,3) transition.}\tablefoottext{e}{Line width determined using a single Gaussian.}\tablefoottext{e}{Line width determined using a single Gaussian.}\tablefoottext{f}{Main beam brightness temperature. The upper limits are 3$\mathbf{\sigma}$. The errors are given in parentheses.}}
\end{table*}

\begin{table*}[h]
\small
    \centering
    \caption{Derived parameters for FU~Orionis type objects detected in NH$_3$ emission.}
    \begin{tabular}{clccccclccc}
    \hline \hline
    \multirow{2}{*}{Name}   & $T_{\rm rot}^{(a)}$    & $T_{\rm kin}^{(b)}$ & $N_{\rm NH_3}^{(c)}$ &  $N_{\rm H_2}^{(d)}$  & T$_{\rm dust}^{(e)}$ &  \multirow{2}{*}{[NH$_3$\,/\,H$_2$]$^{(f)}$} & \multirow{2}{*}{$\rm \eta^{(g)}$}  & Outflow & \multirow{2}{*}{Ref.} &\\
                            & (K)              & (K)           & (cm$^{-2}$)    & (cm$^{-2}$) &  (K)   &  & & (Y/N) & &    \\
    \hline
    RNO~1B/1C               & $17.9$ $(0.9)$ & $21.3$ $(1.1)$ & $(3.2\pm0.2)\times10^{15}$ & $2.1\times10^{23}$ & $14.5$ & $1.5\times10^{-8}$ & $0.16$ & Y & 1 \\
    PP~13S                  & $11.9$ $(0.9)$ & $12.8$ $(1.1)$ & $(6.6\pm1.0)\times10^{14}$ & $2.6\times10^{22}$ & $15.6$ & $2.5\times10^{-8}$ & $0.18$ & Y & 1 \\
    L1551~IRS~5             & $11.3$ $(0.8)$ & $12.1$ $(0.8)$ & $(1.9\pm0.2)\times10^{15}$ & $6.5\times10^{22}$ & $16.4$ & $2.9\times10^{-8}$ & $0.27$ & Y & 1  \\
    Haro~5a~IRS             & $13.6$ $(1.5)$ & $15.1$ $(1.7)$ & $(2.8\pm0.5)\times10^{15}$ & $6.2\times10^{22}$ & $18.1$ & $4.5\times10^{-8}$ & $0.14$ & Y & 1 \\
    V2775~Ori               & $13.2^*$         & $-$ & $(7.5\pm1.9)\times10^{14}$ & $2.6\times10^{22}$ & $14.0$ & $2.8\times10^{-8}$      & $0.15$ & Y/N$^{**}$ & 2, 3 \\
    NGC~2071                & $13.2^*$         & $-$ & $(0.6\pm1.0)\times10^{14}$ & $9.8\times10^{21}$ & $21.8$ & $6.1\times10^{-9}$      & $0.01$ & $-$ &  2, 4\\
    V899~Mon                & $13.2^*$         & $-$ & $(2.2\pm1.0)\times10^{14}$ & $9.1\times10^{21}$ & $13.8$ & $2.4\times10^{-8}$ & $0.05$ & Y & 1\\
    IRAS~06297+1021W         & $13.2^*$         & $-$ & $(4.1\pm1.1)\times10^{14}$ & $1.5\times10^{22}$ & $13.8$ & $2.7\times10^{-8}$ & $0.09$ & Y & 1 \\
    AR~6A/6B                & $15.2$ $(1.9)$ & $17.2$ $(2.2)$ & ($1.8\pm0.4)\times10^{15}$ & $5.1\times10^{22}$ & $15.8$ & $3.5\times10^{-8}$ & $0.06$ & N & 1 \\
    IRAS~06393+0913         & $13.2^*$         & $-$ & $(5.2\pm5.8)\times10^{13}$ & $1.1\times10^{22}$ & $13.6$ & $4.7\times10^{-9}$           & $0.01$ & Y? & 1 \\
    V960~Mon                & $13.2^*$         & $-$ & $(2.1\pm3.5)\times10^{14}$ & $1.3\times10^{22}$ & $14.9$ & $1.6\times10^{-8}$           & $0.02$ & Y & 1 \\
    Z~CMa                   & $13.2^*$         & $-$ & $(1.3\pm3.5)\times10^{14}$ & $-$ & $-$ & $-$ & $0.01$ & Y & 7  \\
    iPTF~15AFQ              & $13.2^*$         & $-$ & $(8.8\pm2.3)\times10^{14}$ & $2.1\times10^{22}$ & $11.8$ & $1.3\times10^{-8}$             & $0.11$  & Y & 1 \\
    IRAS~18270$-$0153~W     & $11.2$ $(0.5)$ & $12.1$ $(0.5)$ & $(1.8\pm0.1)\times10^{15}$ & $5.2\times10^{22}$ & $12.6$ & $3.4\times10^{-8}$ & $0.42$ & $-$ & 1 \\
    OO~Ser                  & $13.8$ $(0.6)$ & $15.3$ $(0.6)$ & $(1.4\pm0.1)\times10^{15}$ & $7.1\times10^{22}$ & $14.7$ & $1.9\times10^{-8}$ & $0.25$ & Y & 2, 5  \\
    IRAS~18341$-$0113~S     & $11.1$ $(0.6)$ & $11.7$ $(0.7)$ & $(1.3\pm0.1)\times10^{15}$ & $2.4\times10^{22}$ & $14.7$ & $5.4\times10^{-8}$ & $0.29$ & $-$ & 2, 5  \\
    HBC~722                 & $12.6$ $(0.6)$ & $13.8$ $(0.7)$ & $(2.3\pm0.2)\times10^{15}$ & $5.5\times10^{22}$ & $13.2$ & $4.1\times10^{-8}$ & $0.31$ & Y & 1\\
    V1057~Cyg               & $13.2^*$         & $-$ & $(1.5\pm0.5)\times10^{15}$ & $1.0\times10^{22}$ & $16.1$ & $1.5\times10^{-7}$ & $0.03$ & Y & 1 \\
    V2495~Cyg               & $13.2^*$         & $-$ & $(1.8\pm0.3)\times10^{15}$ & $3.2\times10^{22}$ & $12.6$ & $5.6\times10^{-8}$ & $0.12$ & Y & 1 \\
    RNO~127                 & $13.2^*$         & $-$ & $(9.5\pm2.9)\times10^{14}$ & $2.1\times10^{22}$ & $11.8$ & $4.5\times10^{-8}$ & $0.06$ & Y & 1 \\
    CB~230                  & $13.2^*$         & $-$ & $(1.3\pm0.1)\times10^{15}$ & $2.1\times10^{22}$ & $15.3$ & $6.2\times10^{-8}$ & $0.19$ & $-$ & 1 \\
    V1735~Cyg               & $12.5$ $(1.5)$ & $13.6$ $(1.6)$ & $(1.5\pm0.3)\times10^{15}$ & $2.3\times10^{22}$ & $15.3$ & $6.5\times10^{-8}$ & $0.17$ & Y & 2, 6 \\
    HH~354~IRS              & $13.2^*$         & $-$ & $(9.3\pm1.7)\times10^{14}$ & $1.8\times10^{22}$ & $16.5$ & $5.1\times10^{-8}$ & $0.11$ & Y & 1  \\
    V733~Cep                & $13.2^*$         & $-$ & $(9.2\pm2.1)\times10^{14}$ & $1.6\times10^{22}$ & $13.8$ & $5.7\times10^{-8}$ & $0.08$ & Y & 1 \\
    \hline
    \end{tabular}
    \label{tab:trot-tkin-fuors}
    \flushleft
    \tablefoot{\tablefoottext{a}{Rotational temperature.}\tablefoottext{b}{Kinetic temperature.}\tablefoottext{c}{NH$_3$ column density.}\tablefoottext{d}{H$_2$ column density.}\tablefoottext{e}{Dust temperature.}\tablefoottext{f}{Ammonia abundance.}\tablefoottext{g}{Beam filling factor.}References are: 1 -- from SED fitting (this work), 2 -- \citet{andre2010}, 3 -- \citet{roy2013}, 4 -- \citet{Schneider2013}, 5 -- \citet{fiorellino2021}, 6 -- \citet{Arzoumanian2011}, 7 -- \citet{evans1994}, and references in Table~\ref{tab:appendix-long}; $^*$Assumed $T_{\rm rot}$ value; $^{**}$The system is seen through a remnant outflow \citep{zurlo2017}}
\end{table*}

\begin{table*}[htbp]
\small
    \centering
    \caption{Derived parameters for EX~Lupi type objects detected in NH$_3$ emission.}
    \begin{tabular}{clcrcccrcc}
    \hline \hline
    \multirow{2}{*}{Name}   & $T_{\rm rot}^{(a)}$    & $T_{\rm kin}^{(b)}$ & $N_{\rm NH_3}^{(c)}$ &  $N_{\rm H_2}^{(d)}$ & T$_{\rm dust}^{(e)}$ & \multirow{2}{*}{[NH$_3$\,/\,H$_2$]$^{(f)}$} &  \multirow{2}{*}{$\rm \eta^{(g)}$} & Outflow & \multirow{2}{*}{Ref.}  \\
                            & (K)              & (K)           & (cm$^{-2}$)    & (cm$^{-2}$)    &    (K)         &  &      & (Y/N) &   \\
    \hline
    V512~Per (SVS~13A)               & $15.1$ $(0.6)$ & $17.1$ $(0.7)$ & $(1.2\pm0.1)\times10^{15}$ &  $8.6\times$10$^{22}$ & $17.5$ & $1.4\times10^{-8}$ & $0.41$ & Y & 2, 3  \\
    LDN~1415~IRS            & $13.2^*$         & $-$ & $(6.6\pm3.4)\times10^{14}$ & $-$ & $-$ & $-$ & $0.04$ & Y & 4 \\
    V371~Ser                & $11.4$ $(0.4)$ & $12.2$ $(0.4)$ & $(1.9\pm0.1)\times10^{15}$ & $6.5\times$10$^{22}$ & $13.4$ & $9.1\times10^{-9}$ & $0.38$ & Y & 2, 5  \\
    V2492~Cyg               & $13.2^*$         &$-$  & $(6.5\pm2.9)\times10^{14}$ & $3\times$10$^{22}$ & $13.4$ & $2.1\times10^{-8}$ & $0.04$ & Y & 1 \\
    \hline
    \end{tabular}
    \label{tab:trot-tkin-exors}
    \flushleft
    \tablefoot{\tablefoottext{a}{Rotational temperature.}\tablefoottext{b}{Kinetic temperature.}\tablefoottext{c}{NH$_3$ column density.}\tablefoottext{d}{H$_2$ column density.}\tablefoottext{e}{Dust temperature.}\tablefoottext{f}{Ammonia abundance.}\tablefoottext{g}{Beam filling factor.}References are: 1 -- from SED fitting (this work), 2 -- \citet{andre2010}, 3 -- \citet{pezzuto2021}, 4 -- \citet{stecklum2007}; 5 -- \citet{fiorellino2021}; $^*$Assumed $T_{\rm rot}$ value}
\end{table*}

\subsection{Beam filling factor} \label{sec:beamfill}
The beam filling factor, $\rm \eta$, gives the fraction of the beam filled by the observed emission, and it can be derived from the radiative transfer equation when the excitation temperature and optical depth of the transition can be determined. For ammonia, optical depths can be derived from the HFS fitting and we assume local thermodynamic equilibrium (LTE) in order to approximate the excitation temperature \citep[see, e.g.,][]{wienen2012,essential-radio-astronomy2016,yan2021}. 
We determined the $\rm \eta$ value for each detected source in our sample using the following equation, with the Rayleigh-Jeans approximation,
\begin{equation}
     \eta = \frac{T_{\rm MB}(1,1)}{\Big(T_{\rm ex}-T_{\rm bg}\Big) \times \Big(1-{\rm exp}(-\rm \tau_m(1,1))\Big)}\;,
    \label{eq:beamfill}
\end{equation}
where T$_{\rm MB}(1,1)$ is the main beam brightness temperature of the (1,1) transition, $\rm \tau(1,1)$ is the optical depth of the (1,1) main line, and T$_{\rm bg}=2.73$\,K. Assuming  LTE, we used the rotational temperature as the excitation temperature \citep[e.g.,][]{goldsmith1999,wilson2009(tools-of...)}. We found that the $\rm \eta$ values in our sample range from $0.06$ to $0.42$ with an average of $0.25$ and a median of $0.26$ (see Tables~\ref{tab:trot-tkin-fuors} and \ref{tab:trot-tkin-exors}).

The low beam filling factors may indicate clumpiness on small scales, which indeed has been revealed by interferometric observations of these ammonia transitions towards dark clouds and high-mass starless cores \citep[see, e.g.,][]{olmi2010,devine2011,ragan2011}. We find no significant differences between the $\rm \eta$ values in this work compared to those in high-mass clumps by \citet{wienen2012}.

\subsection{LSR velocities}
Table~\ref{tab:appendix-long} in the Appendix presents a comparison of the $\varv_{\rm LSR}$ results from our NH$_3$(1,1) observations with published $\varv_{\rm LSR}$ values from the literature.
For most sources, the literature LSR velocities were derived from $^{12}$CO and its rarer isotopologues. 
Where other lines 
or CO observations of nearby clouds were used, this is noted in brackets in Table~\ref{tab:appendix-long}.

Our observations have yielded the first systemic $\varv_{\rm LSR}$ measurements for five sources.  Due to the sensitivity of our observations, we are also able to derive more precise LSR values for a number of sources (see Table~\ref{tab:appendix-long}), which could be helpful for follow-up studies (e.g., comparisons with stellar velocities).
The differences between the NH$_3$ (1,1) velocities and the CO literature values are generally less than 1\,km\,s$^{-1}$, while the velocities derived from the NH$_{3}$ (1,1), (2,2), and (3,3) lines agree within the errors of the fits, generally $<$0.03\,km\,s$^{-1}$.

\subsection{Line widths}
The line widths of the (1,1) inversion transition (where the HFS fitting method was used) range from 0.36\,km\,s$^{-1}$ to 2.39\,km\,s$^{-1}$.  For the (2,2) transition (where Gaussian fitting was used), the line widths are between 0.42\,km\,s$^{-1}$ and 2.55\,km\,s$^{-1}$. We find that the derived line widths for the (2,2) transition are broader than those for the (1,1) transition. The ratios of line width, $\Delta \varv$(2,2)/ $\Delta \varv$(1,1), are between 1.02 and 2.58, with a median of 1.23 and a dispersion of 0.16.
This is consistent with previous results that the line widths obtained from hyperfine structure fitting are smaller than those from Gaussian fitting \citep{wienen2012}. 
We note, that in the following analysis we only use the line width of the (1,1) transition. 

In our sample, RNO~1B/1C and AR~6A/6B have the broadest lines, and both have kinetic temperatures $>$17\,K. 
For RNO 1B/1C, the broad line width may be the result of shock heating and shock-driven turbulence caused by this source's  powerful outflow action \citep[see, e.g.,][]{anglada1994,quanz2007,bae2011}. In contrast, previous observations suggest that AR~6A/6B does not have a CO outflow \citep{moriarty2008}. Based on its environment (see Fig.~\ref{fig:sed1}), we speculate that its increased line width and elevated kinetic temperature are both caused by turbulence.

The observed line width, $\Delta \varv$(1,1), is related to the velocity dispersion, $\sigma_{\rm obs}$, as $\Delta \varv= \sqrt{8\,{\rm ln}\,2} \cdot \sigma_{\rm obs}$. The observed line widths or velocity dispersions of the inversion lines have contributions from both thermal and non-thermal motions \citep[e.g.,][]{hacar2016}, 
\begin{equation}\label{eq:v_int}
    \Delta \varv_{\rm obs} = \sqrt{\Delta \varv_{\rm th}^2 + \Delta \varv_{\rm nt}^2}\;, \hspace{0.3cm}
    \sigma_{\rm obs} = \sqrt{\sigma_{\rm th}^2 + \sigma_{\rm nt}^2}\;,
\end{equation}
where $\Delta \rm \varv_{\rm th}$ and $\sigma_{\rm th}$ are the thermal line width and velocity dispersion, respectively, and $\Delta \rm \varv_{\rm nt}$ and $\sigma_{\rm nt}$ are the non-thermal line width and velocity dispersion, respectively.
The thermal line width and velocity dispersion can be calculated from 
\begin{equation}\label{eq:vthermal}
    \Delta \rm \varv_{\rm th} = \sqrt{\frac{8\,{\rm ln}(2)\,{\rm k}\,T_{\rm kin}}{m}}\;, \hspace{0.3cm} \sigma_{\rm th}=\sqrt{\frac{{\rm k}\,T_{\rm kin}}{m}}
\end{equation}
where $k$ is the Boltzmann constant, $T_{\rm kin}$ is the kinetic temperature, and $m$ is the molecular weight of the given molecule, in this case $m_{\rm NH_{3}} = 17$. According to Eq.~\ref{eq:v_int}, the non-thermal motions can be derived by subtracting the thermal motions from the total.  The kinetic temperatures derived from ammonia are used to derive thermal line widths for sources with both NH$_{3}$ (1,1) and NH$_{3}$ (2,2) detections. For sources with only an NH$_{3}$ (1,1) detection, we assume the average kinetic temperature of 14.6\,K (derived earlier in Sect.~\ref{sec.excitation}) to calculate the contribution of thermal motions. The results are listed in Tables~\ref{tab:names-nh31,1_parameters_fuors} and \ref{tab:names-nh31,1_parameters_exors}.
Based on these results, we conclude that the line widths of most sources are dominated by non-thermal contributions.

Once $\rm \sigma _{\rm nt}$ is determined, the Mach number can be derived in order to distinguish between sonic ($Ma \le 1 $), transonic ($ 1 < Ma \le 2$) and supersonic ($Ma > 2$) environments. 
When calculating the Mach number,  $Ma=f( T_{\rm kin})=\rm \sigma_{\rm nt}/c_{\rm s}$, where $c_{\rm s}$ is the sound speed of  the molecular gas, both terms depend on $T_{\rm kin}$.  Typically, the same $T_{\rm kin}$ is adopted for evaluating $\sigma _{\rm nt}$ and the sound speed \citep[e.g.,][]{hacar2016}.

We calculated the Mach numbers for all sources with detections of both the (1,1) and (2,2) transitions using the values of $T_{\rm kin}$ derived from our observations; the results are listed in Tables~\ref{tab:names-nh31,1_parameters_fuors} and \ref{tab:names-nh31,1_parameters_exors} for FUors and EXors, respectively. 
Using the average kinetic temperature of 14.6\,K, we also derived the Mach numbers for sources with only  (1,1) detections, marked in Tables~\ref{tab:names-nh31,1_parameters_fuors} and \ref{tab:names-nh31,1_parameters_exors}. We find that 10 sources have Mach numbers of $<$1, indicative of sonic motions, while 13 sources show transonic motions. Finally, five FUors in our sample have Mach numbers larger than 2, indicating supersonic environments. Because these 5 FUors are associated with molecular outflows (see Table~\ref{tab:trot-tkin-fuors}), their higher Mach numbers may be attributable to turbulence driven by outflow shocks. 
Interestingly, for the EXors, there is no indication of supersonic environments.
It is also worth noting that there are several examples of sources that have Mach numbers $<$1 despite being known to host outflows (i.e.,~V899~Mon, V2495~Cyg, HH~354~IRS, etc).
Our results show that the host environments of most of the eruptive stars in our sample are dominated by sonic and transonic motions on the scales sampled with the Effelsberg beam ($\sim$37\arcsec), indicating that most eruptive stars reside in rather quiescent host environments.

\subsection{Spectral Energy Distributions (SEDs)}
\label{res:sed}
H$_{2}$ column densities are needed in order to determine ammonia abundances for the sources in our sample. 
For seven sources with NH$_3$ (1,1) detections in our survey,  H$_{2}$ column density and dust temperature maps derived from the \textit{Herschel Gould Belt} survey \citep{andre2010} were available in the literature.  For these sources, we measured column density and dust temperature values for our target sources from the published maps, and these adopted values and the literature references are given in Tables~\ref{tab:trot-tkin-fuors} and \ref{tab:trot-tkin-exors}.  Where H$_{2}$ column density and dust temperature maps were available in the literature for sources undetected in NH$_3$ in our survey, we similarly measured values for our target sources and include these in Table~\ref{tab:non-detection}.  

For the remaining sources with NH$_3$ (1,1) detections, 
we derived H$_{2}$ column density and dust temperature maps using archival \textit{Herschel} SPIRE data at 250\,$\mu$m, 350\,$\mu$m, and 500\,$\mu$m \footnote{The data have been downloaded from \url{http://archives.esac.esa.int/hsa/whsa/}} using the same methods as previous studies \citep[e.g.,][]{andre2010,2017ApJ...840...22L,2017MNRAS.471..100E}.
Prior to pixel-by-pixel spectral energy distribution (SED) fitting, the archival data were convolved to a common angular resolution of 36$\rlap{.}$\arcsec3 (i.e., the beamsize of Herschel/SPIRE at 500\,$\mu$m), which is comparable to the angular resolution of our ammonia observations, and then projected onto the same grid as the 500\,$\mu$m data. 

Assuming that the SED of the dust emission follows the modified-blackbody model, the flux density, $S_{\nu}$, at the frequency $\nu$ can be expressed as
\begin{equation}\label{f.sed}
S_{\nu} = (1-{\rm e}^{-\tau_{\nu}}) B_{\nu}(T_{\rm d}) \Omega \;,
\end{equation}
where $\tau_{\nu}$ is the optical depth at the frequency $\nu$,  $B_{\nu}(T_{\rm d})$ is the Planck function evaluated at the dust temperature $T_{\rm d}$, and $\Omega$ is the solid angle of the beam. The H$_{2}$ column density, $N_{\rm H_{2}}$, is proportional to $\tau_{\nu}$ with the relationship:
\begin{equation}\label{f.h2}
 N_{\rm H_{2}} = \frac{\tau_{\nu}}{\kappa_{\nu}\mu m_{\rm H}} \;,
\end{equation}
where $\mu$ is the molecular weight per hydrogen molecule, taken to be 2.8 \citep{2008A&A...487..993K}, $\kappa_{\nu}$ is the dust opacity per unit (dust+gas) mass, and m$_{\rm H}$ is the mass of hydrogen. 
The dust opacity per unit mass, $\kappa_{\nu}$, is approximated by the power law $\kappa_{\nu} = 0.1 (\nu/1000 {\rm GHz})^{\beta} {\rm cm}^{2}/{\rm g}$ \citep[e.g.,][]{1983QJRAS..24..267H}, where $\beta$ is the dust emissivity index and the canonical gas-to-dust ratio of 100 has been applied. Following previous studies \citep[e.g.][]{andre2010,2017MNRAS.471..100E}, $\beta$ is assumed to be 2 here. 
The SED fitting was performed using the ``LMFIT"\footnote{\url{https://lmfit.github.io/lmfit-py/}} python package \citep{newville_matthew_2014_11813} to fit the two free parameters, $N_{\rm H_{2}}$ and $T_{\rm d}$, for every pixel. The results are shown in Fig.~\ref{fig:sed1}, and the corresponding beam averaged values are given in Tables~\ref{tab:trot-tkin-fuors} and \ref{tab:trot-tkin-exors}. 

For sources in our sample with NH$_3$ (1,1) detections, the dust temperatures range from 11.8\,K to 21.8\,K and the derived H$_{2}$ column densities range from 9.0$\times 10^{21}$\,cm$^{-2}$ to 2.2$\times10^{23}$\,cm$^{-2}$.
We compared the dust temperatures and gas kinetic temperatures for sources in our sample with $T_{\rm kin}$ measurements (i.e.\ those with both NH$_3$ (1,1) and (2,2) detections), and found that in most cases the dust temperature and gas kinetic temperature agree within 3\,K. The only exception is RNO~1B/1C, which has a gas kinetic temperature that is much higher than its dust temperature ($T_{\rm kin}=21.3$\,K, $T_{\rm d}=14.5$\,K). Such a scenario could be caused by inefficient gas-to-dust coupling and gas cooling if the density was $<10^{3.5}$\,cm$^{-3}$ \citep{goldsmith2001}. 
However, RNO~1B/1C is surrounded by deeply embedded objects, forming a small cluster \citep[][]{quanz2007}. Based on the dust temperature map of the source (see Figure~\ref{fig:sed1}) there is significantly warmer dust within $<$1$\arcmin$ of the source position, and this warmer material likely contributes to the higher temperature measured in the Effelsberg beam. 

Interestingly, based on the derived H$_{2}$ column density maps (Fig.~\ref{fig:sed1}), we find that the host environments of the eruptive stars detected in NH$_3$ (1,1) are quite diverse on scales larger than the Effelsberg beam ($\sim$37\arcsec). A few sources are relatively isolated, while others are associated with larger, extended cloud structures. 

\subsection{Ammonia abundance}\label{sec.abundance}
We use the NH$_3$ and H$_{2}$ column densities from Sect.~\ref{sec.excitation} and Sect.~\ref{res:sed} respectively to derive ammonia abundances, $\chi =  N_{\rm NH_3}/N_{\rm H_2}$. The derived abundances, listed in Tables~\ref{tab:trot-tkin-fuors} and \ref{tab:trot-tkin-exors}, range from ${4.7\times10^{-9}}$ to ${1.5\times10^{-7}}$, with an average of ${3.6\times10^{-8}}$ and a median of ${2.8\times10^{-8}}$, respectively.

We compare our results with the NH$_3$ abundances found in other studies, including low-mass, intermediate-mass, and high-mass star forming regions. 
Our derived NH$_3$ abundances are similar to values found in cold, dark clouds \citep[see, e.g.,][]{Ohishi1992}.
With very few exceptions, our derived NH$_3$ abundances fall within the range observed towards IRDCs \citep[e.g., $0.7\times10^{-8}$ to $15.9\times10^{-8}$;][]{pillai2006,zhang2011}. (The exceptions are NGC~2071 and IRAS~06393+0913, which have abundances of $\sim$0.6$\times10^{-8}$ and $\sim$0.5$\times10^{-8}$, respectively; see Table~\ref{tab:trot-tkin-fuors}.) 
Our average abundance of  ${3.6\times10^{-8}}$ is also similar to the average values reported for IRDCs by \citet{pillai2006} ($\sim$${4\times10^{-8}}$) and for high-mass clumps by \citet{dunham2011} (${4.6\times10^{-8}}$).
Interestingly, NH$_3$ abundances observed towards Herbig~Ae/Be stars, which are intermediate-mass pre-main sequence stars, range from $1\times$10$^{-8}$ to $4\times$10$^{-8}$
\citep{fuente1990}, on the lower end of the range observed towards IRDCs and towards our sample.

\section{Discussion}  \label{dis:ammonia}
\subsection{Ammonia in the neighborhood of outbursting systems}
Based on previous studies, ammonia emission can have different origins, corresponding to grain surface and gas-phase chemistry (see \citealt{jorgensen2020} for the most recent review, and references therein).
In the following, we compare our observations with these different scenarios. We first discuss ammonia release from grains, then examine the formation of NH$_3$ via gas-phase chemistry.

During an outburst, the whole disk experiences a temperature increase due to the energy released in the very inner few 0.1\,au of the disk \citep[e.g.,][]{fischer2022}. 
The temperature increase in the outer part of the disk could easily sublimate ammonia molecules off the grains, releasing them back into the gas phase \citep{guesten1988} and enhancing the ammonia abundance.
However, any ammonia set free in this way should not make a significant contribution to our observed ammonia emission: because our FWHM beam size of $\sim$37\arcsec\,(i.e., 5000\,au at 140\,pc) is much larger than typical disk sizes \citep[$\sim$60\,au;][]{2019A&A...621A..76M}, beam dilution effects should result in only a very minor contribution to the ammonia signals we observe. Such enhancements would be better constrained with higher angular resolution observations of ammonia transitions toward the disks around eruptive stars, especially during the bursting phase.

Alternatively, chemical models suggest that ammonia molecules on dust grains can be released into the gas phase through the passage of shocks produced by molecular outflows \citep[e.g.,][]{2017AJ....154...38H}. Such effects have already been confirmed by observations of outflows \citep[e.g.,][]{1995ApJ...443L..37T,1999ApJ...525L.105U,feng2022}, and could explain the relatively high ammonia abundances in some sources, such as RNO~1B/1C. In fact, almost all of the sources in our sample with ammonia (1,1) detections possess CO outflows (see Tables~\ref{tab:trot-tkin-fuors} and \ref{tab:trot-tkin-exors}). 

As suggested by early studies \citep[e.g.,][]{1973ApJ...185..505H,1989A&A...211..413G}, ammonia can also form in cold molecular gas via successive hydrogenation of N$^{+}$ by H$_{2}$ and the subsequent electron recombination of NH$_{4}^{+}$. Hence, it is also possible that the observed ammonia emission is dominated by circumstellar envelopes and/or ambient clouds; as discussed in Sect.~\ref{sec.abundance}, we find that the eruptive stars in our sample generally have NH$_3$ abundances similar to those of IRDCs.
A circumstellar envelope is an important part of any YSO system, being a reservoir of material, replenishing a disk with more matter \citep[e.g.,][]{kenyon&hartmann1996}. 
For example, previous observations of a deeply embedded Class\,{\footnotesize 0} protostar suggest that the ammonia emission is dominated by the circumstellar envelopes on scales of 10$^{4}$\,au, revealed from interferometric observations \citep{tanner2011,jhan2021}. 
Based on Figure~\ref{fig:sed1}, we find that CB~230, HH~354~IRS, L1551~IRS~5, PP~13S, RNO~1B/1C, V960~Mon, and V1057~Cyg coincide with dense dust concentrations, indicative of the presence of dense circumstellar envelopes.
In several cases, our observations are the first NH$_3$ detections of dense circumstellar envelopes identified in other datasets.
The dense circumstellar envelopes of RNO~1B/1C and V1057~Cyg have been confirmed by an interferometric $^{13}$CO and C$^{18}$O survey \citep{feher2017}, with further evidence for an envelope in V1057~Cyg from its SED, using the extensive multi-wavelength data available for this source \citep[][]{szabo2021}.
Similarly, we detect NH$_{3}$ (1,1) and (2,2) toward the EXor type object V371~Ser (also known as EC~53), known to have a dense circumstellar envelope based on millimeter observations and radiative transfer modeling \citep[e.g][and references therein]{baek2020,lee2020-alma}.

Our new ammonia detections, combined with the \textit{Herschel} column density maps, suggest that V2492~Cyg and V2495~Cyg are also associated with dense material. However, we note that it is clear they do not possess the highest column densities/most concentrated peaks.
In the case of AR~6A/6B, our NH$_3$ (1,1) and (2,2) detections are tentative evidence for the presence of dense gas. In the H$_2$ column density map (Figure~\ref{fig:sed1}), however, the dust concentrations appear offset from the target source.
\citet{kospal2017b} found that the CO emission peak at $\varv_{\rm LSR}=5.3\,km\,s^{-1}$ (similar to the 5.06\,km\,s$^{-1}$ derived from the ammonia (1,1) transition) was offset from AR~6A/6B, and suggested, based on the \textit{Herschel}/SPIRE 250\,$\mu$m image, that this source lies in a cavity.
Based on these results and our H$_2$ column density map (Figure~\ref{fig:sed1}), it is most likely that the NH$_3$ emission picked up by the Effelsberg beam originates from material offset from the source. 
For other sources with NH$_3$ detections but without associated dense dust concentration, their ammonia emission might arise from ambient clouds. 

The non-detections of ammonia transitions in our survey could indicate that dense circumstellar envelopes are not present or that the objects are too far away for their envelopes to be detected. 
The distances are known for the majority of sources in our sample \citep[e.g.,][]{audard2014}, allowing us to investigate the second possibility. Interestingly, RNO~1B/1C is the farthest source in our sample, yet NH$_3$ emission was still detected in multiple transitions. 
Similarly, at least NH$_3$ (1,1) was detected towards other sources with large distances, such as Z~CMa, V1735~Cyg and V2495~Cyg, suggesting that distance is unlikely to be a main explanation for NH$_3$ non-detections.
Instead, the non-detections may indicate that dense circumstellar envelopes have already been dispersed. For instance, 
the ammonia non-detection in the case of V1515~Cyg is consistent with a recent multiwavelength SED analysis that found no clear sign of a massive circumstellar envelope \citep{szabo2022}.

\subsection{Is the standard classification scheme reliable for outbursting systems?}
\label{sect:discussion-class}

Based on the standard classification scheme \citep[e.g.,][]{1994ApJ...434..614G,2009ApJS..181..321E}, the sources in our sample have been classified as Class\,{\footnotesize I}, Class\,{\footnotesize II}, or transition objects (i.e.~Class\,{\footnotesize 0/I} or Class\,{\footnotesize I/II}, see Table~\ref{tab:appendix-long}). Because Class\,{\footnotesize II} objects are thought to be beyond the embedded phase \citep[see the recent review by][and references therein]{fischer2022}, their host environments are not expected to harbor as much dense gas as those of younger sources.
In our sample, the Class\,{\footnotesize 0/I} and Class\,{\footnotesize I} sources have the highest ammonia detection rates: 16 sources (3 Class\,{\footnotesize 0/I} and 13 Class\,{\footnotesize I}) are detected in NH$_3$ (1,1), corresponding to detection rates of 100\% for Class\,{\footnotesize 0/I} sources and 81\% for Class\,{\footnotesize I} sources.  Seven of these sources are also detected in NH$_3$ (2,2).
Notably, however, we also detect NH$_3$ (1,1) toward 9 sources classified as older than Class {\footnotesize I}, 4 of which are also detected in (2,2) emission. 
We detect ammonia towards 1 Class\,{\footnotesize I/II} and 8 Class\,{\footnotesize II} objects in our survey, corresponding to detection rates of 33\% and 47\%, respectively.   
We also note that some of the Class\,{\footnotesize II} sources (namely HBC~722, V1057~Cyg, V1735~Cyg, and RNO~1B/1C) have higher NH$_{3}$ and H$_{2}$ column densities than some sources classified as Class\,{\footnotesize I} or Class\,{\footnotesize 0/I} or Class\,{\footnotesize I/II} transition objects.

Our results show that, as expected, the younger systems have significantly higher ammonia detection rates.
However, based on the dust and ammonia evidence, some sources classified as older systems, i.e.~Class\,{\footnotesize II} sources, can still be associated with high concentrations of their dense cores, indicative of a younger evolutionary stage. 
Interestingly, it is puzzling that many younger sources were not detected in our survey (see Table~\ref{tab:appendix-long}). We emphasize the need for future interferometric studies to better understand the effects of the outbursts on the dense cores of young eruptive stars.  Higher-angular resolution NH$_3$ observations could potentially probe whether there is a connection/ongoing accretion from cloud/filament down to core scales \citep[e.g.,][]{redaelli2022}, since NH$_3$ can be used to identify the presence of dense gas. 

As already proposed by \citet{quanz2007b}, the standard classification scheme for low mass YSOs might not be able to adequately  classify FUors, since they might represent an in-between evolutionary stage in the standard classification scheme. Furthermore, FUors might undergo several outburst events \citep{herbig1977,hartmann1985}, just like EXors. After several outbursts, the envelope vanishes in about several hundred thousand years \citep[as discussed above supplying the accretion disk with more material, e.g.,][]{fischer2022}. As a consequence, the objects enter a low accretion state permanently (i.e.~become T~Tauri stars), as discussed by \citet{takami2018,takami2019}. \citet{weintraub1991} and \citet{sandell2001} also suggest that FUors are younger than T~Tauri stars and might be an important link between the more embedded Class\,{\footnotesize I} and the more evolved Class\,{\footnotesize II} stages (the latter closer to or being T~Tauri stars.) Additionally, some FUors have features of both Class\,{\footnotesize I} and Class\,{\footnotesize II} sources: warm continuum consistent with Class\,{\footnotesize II} sources, but rotational line emission typical of Class\,{\footnotesize I}, far higher than Class\,{\footnotesize II} sources with similar mass/luminosity \citep{green2013}.

Compared to the later evolutionary stages, one of the important features of the embedded phase is the presence of dense circumstellar envelopes around YSOs. The appearance of the 10\,$\mu$m silicate feature in absorption has been regarded as a signature for such a circumstellar envelope \citep{quanz2007b}, and dust continuum emission also traces the cold envelopes around YSOs. 
Molecular line tracers like ammonia can provide another tool to investigate the surrounding environments. Because the effective critical densities of the NH$_{3}$ (1,1) and (2,2) lines are 7.9$\times 10^{2}$\,cm$^{-3}$ and 1.6$\times 10^{4}$\,cm$^{-3}$, respectively, the detection of NH$_{3}$ (1,1) and (2,2) would suggest the presence of dense gas at a H$_{2}$ density of $\sim1\times 10^{4}$\,cm$^{-3}$ \citep{2015PASP..127..299S}, which would in turn indicate the embedded phase. The presence of dense gas ($\sim 1\times 10^{4}$\,cm$^{-3}$) indicates that some of the eruptive stars in our sample lie at an earlier phase than previously classified (see Table~\ref{tab:appendix-long}).
For instance, our result from the ammonia observations agrees well with previous studies on V371~Ser (EXor), which was classified as a Class\,{\footnotesize I} object based on its spectral index and bolometric temperature \citep[e.g.,][]{2015ApJS..220...11D}, but ALMA observations revealed that its envelope has a much higher mass than its disk and protostar, suggesting that the source might actually be a Class\,{\footnotesize 0} object \citep{lee2020-alma}. 
We suggest that incorporating more data regarding the presence of dense material surrounding these peculiar objects into the standard classification scheme could better illuminate the evolutionary stages of eruptive FUors and EXors.

By the original definition, the young eruptive star classes of FUors and EXors are Class\,{\footnotesize II} objects, therefore T~Tauri stars \citep[see, e.g.,][]{adams1987,lada1987,kenyon-and-hartmann1991,kenyon&hartmann1996}.
This was further suggested by the only available pre-outburst spectra for two FUors: V1057~Cyg and HBC~722, which both showed properties reminiscent of classical T~Tauri stars (CTTS) prior to their outbursts \citep[][]{herbig1977,miller2011}.
However, nowadays there are many examples of more embedded young eruptive stars,  which were also part of our sample, i.e.~Haro~5a~IRS, HH~354~IRS, L1551~IRS~5 \citep[see, e.g.,][]{audard2014,connelley2018}.

Apart from a single dish study by \citet{lang1979}, there are no dedicated surveys investigating the dense environments specifically focusing only on T~Tauri stars, the closest objects to the ones in our study.  The sample of \citet{lang1979} consisted of 34 T~Tauri stars, located in Taurus-Auriga and the young star cluster NGC~2264, accessible with the Arecibo telescope. Out of the 34 sources they detected at least the (1,1) transition toward 13 sources, a detection rate of 38\%. 
In our case, the sample consisted of 17 Class\,{\footnotesize II} sources (see Table~\ref{tab:appendix-long}), and we have detected at least the (1,1) transition toward 8 of them, which is $\sim$47\%.
\citet{lang1979} found kinetic temperatures from 26\,K to 37\,K, and column densities between 1 and 5.9$\times$10$^{14}$\,cm$^{-2}$. In our sample, the Class\,{\footnotesize II} sources (see Table~\ref{tab:appendix-long}) have kinetic temperatures between 13.63\,K and 21.35\,K and column densities from $1.3\times10^{14}\,cm^{-2}$ to $1.8\times10^{15}\,cm^{-2}$.
When compared to \citet{lang1979}, we found, that for a few of the Class\,{\footnotesize II} sources, namely V899~Mon and V960~Mon, the column densities are within the same order of magnitude, i.e.,~$\sim$10$^{14}\,\rm cm^{-2}$.
However, there are other Class\,{\footnotesize II} sources which have $\sim$1 order of magnitude higher column density values $(i.e., \sim10^{15}\,\rm cm^{-2})$, namely RNO~1B/1C, AR~6A/6B, HBC~722, V1057~Cyg, and V1735~Cyg.
The similar column densities suggest that ammonia does not probe the part of the envelope impacted by the outburst. 

We also compared our results to NH$_3$ observations of Herbig~Ae/Be stars, YSOs that are the intermediate mass counterparts of T~Tauri stars \citep[see, e.g.,][]{waters1998}. These YSOs have similar properties to the objects in our sample, such as P~Cygni profiles indicating mass loss \citep{strom1972}, stellar winds \citep{canto1984}, and are usually illuminating nebulosities (just like the first few FUor examples) \citep{herbig1960}, however outflows are more typical and better understood in low-mass YSOs \citep[e.g.,][]{pezzuto1997,tambovtseva2016,fischer2022}.
\citet{fuente1990} found ammonia column densities ranging between $0.5\times$10$^{14}$ and $2.9\times$10$^{14}$\,cm$^{-2}$, which values are within the same range for 5 sources (NGC~2071, V899~Mon, IRAS~06393$+$0913, V960~Mon and Z~CMa) in our sample (see Table~\ref{tab:trot-tkin-fuors} and \ref{tab:trot-tkin-exors}).
In their study \citet{fuente1990} also obtained maps and found that in HD~200775, a source illuminating an extended reflection nebula in NGC~7023, three different clumps could be traced with the NH$_3$ emission, with varying rotational temperatures and column densities. High-angular resolution observations in the future of a selected sample of eruptive objects could reveal similar clumpiness of the ammonia emission in the host environments of FUors/EXors.

\section{Conclusions} 
\label{sec:conclusions}
In this paper, we presented the results of the first dedicated ammonia survey of low-mass young eruptive stars, in order to investigate their host environments.  Our sample included a total of 51 objects, including FUors, EXors and Gaia alerts, the latter of which are yet to be classified. 
Our observations using the Effelsberg~100-m radio telescope resulted in the detection of NH$_{3}$ (1,1) in 28 sources (24 FUors, 4 EXors), NH$_{3}$ (2,2) in 12 sources (10 FUors, 2 EXors), and NH$_{3}$ (3,3) in 2 sources (the FUor-type object RNO~1B/1C and the EXor-type object V512~Per, the latter more commonly known as SVS~13). Ammonia emission was not detected toward any of the Gaia alert sources. 
Our analysis leads to the following conclusions: 
\begin{itemize}
    \item Based on the results for the 12 sources with both NH$_{3}$ (1,1) and NH$_{3}$ (2,2) detections the kinetic temperatures range from  $\sim$12\,K to $\sim$21\,K, which is slightly lower than the $T_{\rm kin}$ values reported toward classical T~Tauri stars.  The ammonia column densities for sources in our sample detected in NH$_3$ (1,1) range from  
    $5.2\times10^{13}\,cm^{-2}$ to $3.2\times10^{15}\,cm^{-2}$. The average value for our sample, $1.18\times10^{15}\,cm^{-2}$,   
    is higher than the ammonia column densities found toward T~Tauri stars.
    The ammonia abundances with respect to H$_{2}$ for our sample range from $4.7\times10^{-9}$ to $1.5\times10^{-7}$ with an average of $3.6\times10^{-8}$ and a median of $2.8\times10^{-8}$, 
    comparable to IRDCs.
    
    \item  Most of the eruptive stars in our sample reside in rather quiescent (sonic or transonic) host environments, with the exception of 5 FUors (RNO~1B/1C, Haro~5a~IRS, AR~6A/6B, Z~CMa and HBC~722) that exhibit supersonic motions. The supersonic motions  might be caused by associated outflows.
    
    \item We investigate the origin of the observed ammonia emission in the outbursting systems. 
    Comparing with dust-based H$_{2}$ column density maps, we find that circumstellar envelopes are present and likely contribute to the observed ammonia emission in seven sources: CB~230, HH~354~IRS, L1551~IRS~5, PP~13S, RNO~1B/1C, V960~Mon, and V1057~Cyg. Outflow shocks could contribute to the relatively high ammonia abundances in sources like RNO~1B/1C.

    \item Additional eruptive stars potentially harbour dense gas based on their NH$_3$ (2,2) detections, which could indicate  an earlier phase than originally classified. 
    Our results add to the growing evidence that low-mass young eruptive stars occupy a wide range of evolutionary stages \citep[see also e.g.,][]{green2013}.
\end{itemize}

Our Effelsberg ammonia observations have investigated the host environments of eruptive low-mass stars on scales of $\sim$37\arcsec, much larger than the disks surrounding our targets \citep[e.g.,][]{cieza2018,kospal2021,2021ApJ...923..270L}. 
For the majority of these young eruptive stars, their environments are still poorly constrained on small scales and further high angular resolution observations are needed to shed light on the relationship between young eruptive stars, their disks, and their potential circumstellar envelopes. 

Such observations will be important for expanding the standard classification scheme of YSOs, and for studying the effects of the outburst on the host environments of young eruptive stars. 

\begin{acknowledgements}
Based on observations (Project ID: 95-21, PI: Szab\'o) with the 100-m telescope of the MPIfR (Max-Planck-Institut für Radioastronomie) at Effelsberg.
Zs.M.Sz. acknowledges funding from a St Leonards scholarship from the University of St Andrews.
For the purpose of open access, the author has applied a Creative Commons
Attribution (CC BY) licence to any Author Accepted Manuscript version arising.
This research has made use of data from the Herschel Gould Belt survey (HGBS) project (http://gouldbelt-herschel.cea.fr). The HGBS is a Herschel Key Programme jointly carried out by SPIRE Specialist Astronomy Group 3 (SAG 3), scientists of several institutes in the PACS Consortium (CEA Saclay, INAF-IFSI Rome and INAF-Arcetri, KU Leuven, MPIA Heidelberg), and scientists of the Herschel Science Center (HSC).
This project has received funding from the European Research Council (ERC) under the European Union's Horizon 2020 research and innovation programme under grant agreement No 716155 (SACCRED).
\end{acknowledgements}

%
%

\bibliographystyle{aa}
\bibliography{paper}

\begin{thebibliography}{151}
\expandafter\ifx\csname natexlab\endcsname\relax\def\natexlab#1{#1}\fi

\bibitem[{{{\'A}brah{\'a}m} {et~al.}(2004){{\'A}brah{\'a}m}, {K{\'o}sp{\'a}l},
  {Csizmadia}, {Kun}, {Mo{\'o}r}, \& {Prusti}}]{abraham2004b}
{{\'A}brah{\'a}m}, P., {K{\'o}sp{\'a}l}, {\'A}., {Csizmadia}, S., {et~al.}
  2004, \aap, 428, 89

\bibitem[{{{\'A}brah{\'a}m} {et~al.}(2018){{\'A}brah{\'a}m}, {K{\'o}sp{\'a}l},
  {Kun}, {Feh{\'e}r}, {Zsidi}, {Acosta-Pulido}, {Carnerero},
  {Garc{\'\i}a-{\'A}lvarez}, {Mo{\'o}r}, {Cseh}, {Hajdu}, {Hanyecz}, {Kelemen},
  {Kriskovics}, {Marton}, {Mez{\H{o}}}, {Moln{\'a}r}, {Ordasi},
  {Rodr{\'\i}guez-Coira}, {S{\'a}rneczky}, {S{\'o}dor}, {Szak{\'a}ts},
  {Szegedi-Elek}, {Szing}, {Farkas-Tak{\'a}cs}, {Vida}, \&
  {Vink{\'o}}}]{abraham2018}
{{\'A}brah{\'a}m}, P., {K{\'o}sp{\'a}l}, {\'A}., {Kun}, M., {et~al.} 2018,
  \apj, 853, 28

\bibitem[{{Adams} {et~al.}(1987){Adams}, {Lada}, \& {Shu}}]{adams1987}
{Adams}, F.~C., {Lada}, C.~J., \& {Shu}, F.~H. 1987, \apj, 312, 788

\bibitem[{{ALMA Partnership} {et~al.}(2015){ALMA Partnership}, {Brogan},
  {P{\'e}rez}, {Hunter}, {Dent}, {Hales}, {Hills}, {Corder}, {Fomalont},
  {Vlahakis}, {Asaki}, {Barkats}, {Hirota}, {Hodge}, {Impellizzeri}, {Kneissl},
  {Liuzzo}, {Lucas}, {Marcelino}, {Matsushita}, {Nakanishi}, {Phillips},
  {Richards}, {Toledo}, {Aladro}, {Broguiere}, {Cortes}, {Cortes}, {Espada},
  {Galarza}, {Garcia-Appadoo}, {Guzman-Ramirez}, {Humphreys}, {Jung}, {Kameno},
  {Laing}, {Leon}, {Marconi}, {Mignano}, {Nikolic}, {Nyman}, {Radiszcz},
  {Remijan}, {Rod{\'o}n}, {Sawada}, {Takahashi}, {Tilanus}, {Vila Vilaro},
  {Watson}, {Wiklind}, {Akiyama}, {Chapillon}, {de Gregorio-Monsalvo}, {Di
  Francesco}, {Gueth}, {Kawamura}, {Lee}, {Nguyen Luong}, {Mangum}, {Pietu},
  {Sanhueza}, {Saigo}, {Takakuwa}, {Ubach}, {van Kempen}, {Wootten},
  {Castro-Carrizo}, {Francke}, {Gallardo}, {Garcia}, {Gonzalez}, {Hill},
  {Kaminski}, {Kurono}, {Liu}, {Lopez}, {Morales}, {Plarre}, {Schieven},
  {Testi}, {Videla}, {Villard}, {Andreani}, {Hibbard}, \&
  {Tatematsu}}]{alma2015}
{ALMA Partnership}, {Brogan}, C.~L., {P{\'e}rez}, L.~M., {et~al.} 2015, \apjl,
  808, L3

\bibitem[{{Andr{\'e}} {et~al.}(2010){Andr{\'e}}, {Men'shchikov}, {Bontemps},
  {K{\"o}nyves}, {Motte}, {Schneider}, {Didelon}, {Minier}, {Saraceno},
  {Ward-Thompson}, {di Francesco}, {White}, {Molinari}, {Testi}, {Abergel},
  {Griffin}, {Henning}, {Royer}, {Mer{\'\i}n}, {Vavrek}, {Attard},
  {Arzoumanian}, {Wilson}, {Ade}, {Aussel}, {Baluteau}, {Benedettini},
  {Bernard}, {Blommaert}, {Cambr{\'e}sy}, {Cox}, {di Giorgio}, {Hargrave},
  {Hennemann}, {Huang}, {Kirk}, {Krause}, {Launhardt}, {Leeks}, {Le Pennec},
  {Li}, {Martin}, {Maury}, {Olofsson}, {Omont}, {Peretto}, {Pezzuto}, {Prusti},
  {Roussel}, {Russeil}, {Sauvage}, {Sibthorpe}, {Sicilia-Aguilar}, {Spinoglio},
  {Waelkens}, {Woodcraft}, \& {Zavagno}}]{andre2010}
{Andr{\'e}}, P., {Men'shchikov}, A., {Bontemps}, S., {et~al.} 2010, \aap, 518,
  L102

\bibitem[{{Anglada} {et~al.}(1994){Anglada}, {Rodriguez}, {Girart},
  {Estalella}, \& {Torrelles}}]{anglada1994}
{Anglada}, G., {Rodriguez}, L.~F., {Girart}, J.~M., {Estalella}, R., \&
  {Torrelles}, J.~M. 1994, \apjl, 420, L91

\bibitem[{{Arzoumanian} {et~al.}(2011){Arzoumanian}, {Andr{\'e}}, {Didelon},
  {K{\"o}nyves}, {Schneider}, {Men'shchikov}, {Sousbie}, {Zavagno}, {Bontemps},
  {di Francesco}, {Griffin}, {Hennemann}, {Hill}, {Kirk}, {Martin}, {Minier},
  {Molinari}, {Motte}, {Peretto}, {Pezzuto}, {Spinoglio}, {Ward-Thompson},
  {White}, \& {Wilson}}]{Arzoumanian2011}
{Arzoumanian}, D., {Andr{\'e}}, P., {Didelon}, P., {et~al.} 2011, \aap, 529, L6

\bibitem[{{Audard} {et~al.}(2014){Audard}, {{\'A}brah{\'a}m}, {Dunham},
  {Green}, {Grosso}, {Hamaguchi}, {Kastner}, {K{\'o}sp{\'a}l}, {Lodato},
  {Romanova}, {Skinner}, {Vorobyov}, \& {Zhu}}]{audard2014}
{Audard}, M., {{\'A}brah{\'a}m}, P., {Dunham}, M.~M., {et~al.} 2014, in
  Protostars and Planets VI, ed. H.~{Beuther}, R.~S. {Klessen}, C.~P.
  {Dullemond}, \& T.~{Henning}, 387

\bibitem[{{Bae} {et~al.}(2011){Bae}, {Kim}, {Youn}, {Kim}, {Byun}, {Kang}, \&
  {Oh}}]{bae2011}
{Bae}, J.-H., {Kim}, K.-T., {Youn}, S.-Y., {et~al.} 2011, \apjs, 196, 21

\bibitem[{{Baek} {et~al.}(2020){Baek}, {MacFarlane}, {Lee}, {Stamatellos},
  {Herczeg}, {Johnstone}, {Pe{\~n}a}, {Varricatt}, {Hodapp}, {Chen}, \&
  {Kang}}]{baek2020}
{Baek}, G., {MacFarlane}, B.~A., {Lee}, J.-E., {et~al.} 2020, \apj, 895, 27

\bibitem[{{Bailer-Jones} {et~al.}(2021){Bailer-Jones}, {Rybizki}, {Fouesneau},
  {Demleitner}, \& {Andrae}}]{bailerjones-edr3}
{Bailer-Jones}, C.~A.~L., {Rybizki}, J., {Fouesneau}, M., {Demleitner}, M., \&
  {Andrae}, R. 2021, \aj, 161, 147

\bibitem[{{Banzatti} {et~al.}(2014){Banzatti}, {Meyer}, {Manara},
  {Pontoppidan}, \& {Testi}}]{banzatti2014}
{Banzatti}, A., {Meyer}, M.~R., {Manara}, C.~F., {Pontoppidan}, K.~M., \&
  {Testi}, L. 2014, \apj, 780, 26

\bibitem[{{Bell} {et~al.}(1995){Bell}, {Lin}, {Hartmann}, \&
  {Kenyon}}]{bell1995}
{Bell}, K.~R., {Lin}, D.~N.~C., {Hartmann}, L.~W., \& {Kenyon}, S.~J. 1995,
  \apj, 444, 376

\bibitem[{{Benson} \& {Myers}(1989)}]{benson1989}
{Benson}, P.~J. \& {Myers}, P.~C. 1989, \apjs, 71, 89

\bibitem[{{Borchert} {et~al.}(2022){Borchert}, {Price}, {Pinte}, \&
  {Cuello}}]{borchert2022}
{Borchert}, E. M.~A., {Price}, D.~J., {Pinte}, C., \& {Cuello}, N. 2022,
  \mnras, 510, L37

\bibitem[{{Bronfman} {et~al.}(1996){Bronfman}, {Nyman}, \&
  {May}}]{bronfman1996}
{Bronfman}, L., {Nyman}, L.~A., \& {May}, J. 1996, \aaps, 115, 81

\bibitem[{{Canto} {et~al.}(1984){Canto}, {Rodriguez}, {Calvet}, \&
  {Levreault}}]{canto1984}
{Canto}, J., {Rodriguez}, L.~F., {Calvet}, N., \& {Levreault}, R.~M. 1984,
  \apj, 282, 631

\bibitem[{{Cao} {et~al.}(2019){Cao}, {Qiu}, {Zhang}, {Wang}, {Hu}, \&
  {Liu}}]{cao2019}
{Cao}, Y., {Qiu}, K., {Zhang}, Q., {et~al.} 2019, \apjs, 241, 1

\bibitem[{{Chen} {et~al.}(2007){Chen}, {Launhardt}, \& {Henning}}]{chen2007}
{Chen}, X., {Launhardt}, R., \& {Henning}, T. 2007, \apj, 669, 1058

\bibitem[{{Cheung} {et~al.}(1968){Cheung}, {Rank}, {Townes}, {Thornton}, \&
  {Welch}}]{cheung1968}
{Cheung}, A.~C., {Rank}, D.~M., {Townes}, C.~H., {Thornton}, D.~D., \& {Welch},
  W.~J. 1968, \prl, 21, 1701

\bibitem[{{Cieza} {et~al.}(2018){Cieza}, {Ru{\'\i}z-Rodr{\'\i}guez}, {Perez},
  {Casassus}, {Williams}, {Zurlo}, {Principe}, {Hales}, {Prieto}, {Tobin},
  {Zhu}, \& {Marino}}]{cieza2018}
{Cieza}, L.~A., {Ru{\'\i}z-Rodr{\'\i}guez}, D., {Perez}, S., {et~al.} 2018,
  \mnras, 474, 4347

\bibitem[{{Clarke} {et~al.}(2005){Clarke}, {Lodato}, {Melnikov}, \&
  {Ibrahimov}}]{clarke2005}
{Clarke}, C.~J., {Lodato}, G., {Melnikov}, S.~Y., \& {Ibrahimov}, M.~A. 2005,
  \mnras, 361, 942

\bibitem[{{Condon} \& {Ransom}(2016)}]{essential-radio-astronomy2016}
{Condon}, J.~J. \& {Ransom}, S.~M. 2016, {Essential Radio Astronomy}

\bibitem[{{Connelley} \& {Greene}(2010)}]{connelley2010}
{Connelley}, M.~S. \& {Greene}, T.~P. 2010, \aj, 140, 1214

\bibitem[{{Connelley} \& {Reipurth}(2018)}]{connelley2018}
{Connelley}, M.~S. \& {Reipurth}, B. 2018, \apj, 861, 145

\bibitem[{{Cruz-S{\'a}enz de Miera} {et~al.}(2023){Cruz-S{\'a}enz de Miera},
  {K{\'o}sp{\'a}l}, {{\'A}brah{\'a}m}, {Csengeri}, {F{\'e}her}, {G{\"u}sten},
  \& {Henning}}]{cruz-saenz2023}
{Cruz-S{\'a}enz de Miera}, F., {K{\'o}sp{\'a}l}, {\'A}., {{\'A}brah{\'a}m}, P.,
  {et~al.} 2023, arXiv e-prints, arXiv:2301.03387

\bibitem[{{Cruz-S{\'a}enz de Miera} {et~al.}(2022){Cruz-S{\'a}enz de Miera},
  {K{\'o}sp{\'a}l}, {{\'A}brah{\'a}m}, {Park}, {Nagy}, {Siwak}, {Kun},
  {Fiorellino}, {Szab{\'o}}, {Antoniucci}, {Giannini}, {Nisini}, {Szabados},
  {Kriskovics}, {Ordasi}, {Szak{\'a}ts}, {Vida}, {Vink{\'o}}, {Zieli{\'n}ski},
  {Wyrzykowski}, {Garc{\'\i}a-{\'A}lvarez}, {Dr{\'o}{\.z}d{\.z}}, {Og{\l}oza},
  \& {Sonbas}}]{cruz-saenz2022}
{Cruz-S{\'a}enz de Miera}, F., {K{\'o}sp{\'a}l}, {\'A}., {{\'A}brah{\'a}m}, P.,
  {et~al.} 2022, \apj, 927, 125

\bibitem[{{Devine} {et~al.}(2011){Devine}, {Chandler}, {Brogan}, {Churchwell},
  {Indebetouw}, {Shirley}, \& {Borg}}]{devine2011}
{Devine}, K.~E., {Chandler}, C.~J., {Brogan}, C., {et~al.} 2011, \apj, 733, 44

\bibitem[{{Di Francesco} {et~al.}(2020){Di Francesco}, {Keown}, {Fallscheer},
  {Andr{\'e}}, {Ladjelate}, {K{\"o}nyves}, {Men'shchikov}, {Stephens-Whale},
  {Nguyen-Luong}, {Martin}, {Sadavoy}, {Pezzuto}, {Fiorellino}, {Benedettini},
  {Schneider}, {Bontemps}, {Arzoumanian}, {Palmeirim}, {Kirk}, \&
  {Ward-Thompson}}]{di-francesco2020}
{Di Francesco}, J., {Keown}, J., {Fallscheer}, C., {et~al.} 2020, \apj, 904,
  172

\bibitem[{{Diaz-Rodriguez} {et~al.}(2022){Diaz-Rodriguez}, {Anglada},
  {Bl{\'a}zquez-Calero}, {Osorio}, {G{\'o}mez}, {Fuller}, {Estalella},
  {Torrelles}, {Cabrit}, {Rodr{\'\i}guez}, {Lef{\`e}vre}, {Mac{\'\i}as},
  {Carrasco-Gonz{\'a}lez}, {Zapata}, {de Gregorio-Monsalvo}, \&
  {Ho}}]{diaz-rodriguez2022}
{Diaz-Rodriguez}, A.~K., {Anglada}, G., {Bl{\'a}zquez-Calero}, G., {et~al.}
  2022, \apj, 930, 91

\bibitem[{{Dong} {et~al.}(2022){Dong}, {Liu}, {Cuello}, {Pinte},
  {{\'A}brah{\'a}m}, {Vorobyov}, {Hashimoto}, {K{\'o}sp{\'a}l}, {Chiang},
  {Takami}, {Chen}, {Dunham}, {Fukagawa}, {Green}, {Hasegawa}, {Henning},
  {Pavlyuchenkov}, {Pyo}, \& {Tamura}}]{dong2022}
{Dong}, R., {Liu}, H.~B., {Cuello}, N., {et~al.} 2022, Nature Astronomy, 6, 331

\bibitem[{{Dunham} {et~al.}(2011){Dunham}, {Rosolowsky}, {Evans}, {Cyganowski},
  \& {Urquhart}}]{dunham2011}
{Dunham}, M.~K., {Rosolowsky}, E., {Evans}, Neal~J., I., {Cyganowski}, C., \&
  {Urquhart}, J.~S. 2011, \apj, 741, 110

\bibitem[{{Dunham} {et~al.}(2015){Dunham}, {Allen}, {Evans},
  {Broekhoven-Fiene}, {Cieza}, {Di Francesco}, {Gutermuth}, {Harvey},
  {Hatchell}, {Heiderman}, {Huard}, {Johnstone}, {Kirk}, {Matthews}, {Miller},
  {Peterson}, \& {Young}}]{2015ApJS..220...11D}
{Dunham}, M.~M., {Allen}, L.~E., {Evans}, Neal~J., I., {et~al.} 2015, \apjs,
  220, 11

\bibitem[{{Elia} {et~al.}(2017){Elia}, {Molinari}, {Schisano}, {Pestalozzi},
  {Pezzuto}, {Merello}, {Noriega-Crespo}, {Moore}, {Russeil}, {Mottram},
  {Paladini}, {Strafella}, {Benedettini}, {Bernard}, {Di Giorgio}, {Eden},
  {Fukui}, {Plume}, {Bally}, {Martin}, {Ragan}, {Jaffa}, {Motte}, {Olmi},
  {Schneider}, {Testi}, {Wyrowski}, {Zavagno}, {Calzoletti}, {Faustini},
  {Natoli}, {Palmeirim}, {Piacentini}, {Piazzo}, {Pilbratt}, {Polychroni},
  {Baldeschi}, {Beltr{\'a}n}, {Billot}, {Cambr{\'e}sy}, {Cesaroni},
  {Garc{\'\i}a-Lario}, {Hoare}, {Huang}, {Joncas}, {Liu}, {Maiolo}, {Marsh},
  {Maruccia}, {M{\`e}ge}, {Peretto}, {Rygl}, {Schilke}, {Thompson},
  {Traficante}, {Umana}, {Veneziani}, {Ward-Thompson}, {Whitworth}, {Arab},
  {Bandieramonte}, {Becciani}, {Brescia}, {Buemi}, {Bufano}, {Butora},
  {Cavuoti}, {Costa}, {Fiorellino}, {Hajnal}, {Hayakawa}, {Kacsuk}, {Leto}, {Li
  Causi}, {Marchili}, {Martinavarro-Armengol}, {Mercurio}, {Molinaro},
  {Riccio}, {Sano}, {Sciacca}, {Tachihara}, {Torii}, {Trigilio}, {Vitello}, \&
  {Yamamoto}}]{2017MNRAS.471..100E}
{Elia}, D., {Molinari}, S., {Schisano}, E., {et~al.} 2017, \mnras, 471, 100

\bibitem[{{Evans} {et~al.}(1994){Evans}, {Balkum}, {Levreault}, {Hartmann}, \&
  {Kenyon}}]{evans1994}
{Evans}, Neal~J., I., {Balkum}, S., {Levreault}, R.~M., {Hartmann}, L., \&
  {Kenyon}, S. 1994, \apj, 424, 793

\bibitem[{{Evans} {et~al.}(2009){Evans}, {Dunham}, {J{\o}rgensen}, {Enoch},
  {Mer{\'\i}n}, {van Dishoeck}, {Alcal{\'a}}, {Myers}, {Stapelfeldt}, {Huard},
  {Allen}, {Harvey}, {van Kempen}, {Blake}, {Koerner}, {Mundy}, {Padgett}, \&
  {Sargent}}]{2009ApJS..181..321E}
{Evans}, Neal~J., I., {Dunham}, M.~M., {J{\o}rgensen}, J.~K., {et~al.} 2009,
  \apjs, 181, 321

\bibitem[{{Feh{\'e}r} {et~al.}(2017){Feh{\'e}r}, {K{\'o}sp{\'a}l},
  {{\'A}brah{\'a}m}, {Hogerheijde}, \& {Brinch}}]{feher2017}
{Feh{\'e}r}, O., {K{\'o}sp{\'a}l}, {\'A}., {{\'A}brah{\'a}m}, P.,
  {Hogerheijde}, M.~R., \& {Brinch}, C. 2017, \aap, 607, A39

\bibitem[{{Feng} {et~al.}(2022){Feng}, {Liu}, {Caselli}, {Burkhardt}, {Du},
  {Bachiller}, {Codella}, \& {Ceccarelli}}]{feng2022}
{Feng}, S., {Liu}, H.~B., {Caselli}, P., {et~al.} 2022, \apjl, 933, L35

\bibitem[{{Fiorellino} {et~al.}(2021){Fiorellino}, {Elia}, {Andr{\'e}},
  {Men'shchikov}, {Pezzuto}, {Schisano}, {K{\"o}nyves}, {Arzoumanian},
  {Benedettini}, {Ward-Thompson}, {Bracco}, {Di Francesco}, {Bontemps}, {Kirk},
  {Motte}, \& {Molinari}}]{fiorellino2021}
{Fiorellino}, E., {Elia}, D., {Andr{\'e}}, P., {et~al.} 2021, \mnras, 500, 4257

\bibitem[{{Fischer} {et~al.}(2022){Fischer}, {Hillenbrand}, {Herczeg},
  {Johnstone}, {K{\'o}sp{\'a}l}, \& {Dunham}}]{fischer2022}
{Fischer}, W.~J., {Hillenbrand}, L.~A., {Herczeg}, G.~J., {et~al.} 2022, arXiv
  e-prints, arXiv:2203.11257

\bibitem[{{Fuente} {et~al.}(1990){Fuente}, {Martin-Pintado}, {Cernicharo}, \&
  {Bachiller}}]{fuente1990}
{Fuente}, A., {Martin-Pintado}, J., {Cernicharo}, J., \& {Bachiller}, R. 1990,
  \aap, 237, 471

\bibitem[{{Fuller} {et~al.}(1995){Fuller}, {Ladd}, {Padman}, {Myers}, \&
  {Adams}}]{fuller1995}
{Fuller}, G.~A., {Ladd}, E.~F., {Padman}, R., {Myers}, P.~C., \& {Adams}, F.~C.
  1995, \apj, 454, 862

\bibitem[{{Galloway} \& {Herbst}(1989)}]{1989A&A...211..413G}
{Galloway}, E.~T. \& {Herbst}, E. 1989, \aap, 211, 413

\bibitem[{{Giannini} {et~al.}(2016){Giannini}, {Lorenzetti}, {Antoniucci},
  {Arkharov}, {Larionov}, {Di Paola}, {Bisogni}, \& {Marchetti}}]{gianni2016}
{Giannini}, T., {Lorenzetti}, D., {Antoniucci}, S., {et~al.} 2016, \apjl, 819,
  L5

\bibitem[{{Goldsmith}(2001)}]{goldsmith2001}
{Goldsmith}, P.~F. 2001, \apj, 557, 736

\bibitem[{{Goldsmith} \& {Langer}(1999)}]{goldsmith1999}
{Goldsmith}, P.~F. \& {Langer}, W.~D. 1999, \apj, 517, 209

\bibitem[{{Gramajo} {et~al.}(2014){Gramajo}, {Rod{\'o}n}, \&
  {G{\'o}mez}}]{gramajo2014}
{Gramajo}, L.~V., {Rod{\'o}n}, J.~A., \& {G{\'o}mez}, M. 2014, \aj, 147, 140

\bibitem[{{Green} {et~al.}(2013){Green}, {Evans}, {K{\'o}sp{\'a}l}, {Herczeg},
  {Quanz}, {Henning}, {van Kempen}, {Lee}, {Dunham}, {Meeus}, {Bouwman},
  {Chen}, {G{\"u}del}, {Skinner}, {Liebhart}, \& {Merello}}]{green2013}
{Green}, J.~D., {Evans}, Neal~J., I., {K{\'o}sp{\'a}l}, {\'A}., {et~al.} 2013,
  \apj, 772, 117

\bibitem[{{Greene} {et~al.}(1994){Greene}, {Wilking}, {Andre}, {Young}, \&
  {Lada}}]{1994ApJ...434..614G}
{Greene}, T.~P., {Wilking}, B.~A., {Andre}, P., {Young}, E.~T., \& {Lada},
  C.~J. 1994, \apj, 434, 614

\bibitem[{{Guesten} \& {Fiebig}(1988)}]{guesten1988}
{Guesten}, R. \& {Fiebig}, D. 1988, \aap, 204, 253

\bibitem[{{Hacar} {et~al.}(2016){Hacar}, {Alves}, {Burkert}, \&
  {Goldsmith}}]{hacar2016}
{Hacar}, A., {Alves}, J., {Burkert}, A., \& {Goldsmith}, P. 2016, \aap, 591,
  A104

\bibitem[{{Hartmann} \& {Kenyon}(1985)}]{hartmann1985}
{Hartmann}, L. \& {Kenyon}, S.~J. 1985, \apj, 299, 462

\bibitem[{{Hartmann} \& {Kenyon}(1996)}]{kenyon&hartmann1996}
{Hartmann}, L. \& {Kenyon}, S.~J. 1996, \araa, 34, 207

\bibitem[{{Harvey} {et~al.}(2008){Harvey}, {Huard}, {J{\o}rgensen},
  {Gutermuth}, {Mamajek}, {Bourke}, {Mer{\'\i}n}, {Cieza}, {Brooke}, {Chapman},
  {Alcal{\'a}}, {Allen}, {Evans}, {Di Francesco}, \& {Kirk}}]{harvey2008}
{Harvey}, P.~M., {Huard}, T.~L., {J{\o}rgensen}, J.~K., {et~al.} 2008, \apj,
  680, 495

\bibitem[{{Herbig}(1960)}]{herbig1960}
{Herbig}, G.~H. 1960, \apjs, 4, 337

\bibitem[{{Herbig}(1977)}]{herbig1977}
{Herbig}, G.~H. 1977, \apj, 217, 693

\bibitem[{{Herbig}(1989)}]{herbig1989_eso}
{Herbig}, G.~H. 1989, in European Southern Observatory Conference and Workshop
  Proceedings, Vol.~33, European Southern Observatory Conference and Workshop
  Proceedings, 233--246

\bibitem[{{Herbig}(1990)}]{herbig1990}
{Herbig}, G.~H. 1990, \apj, 360, 639

\bibitem[{{Herbst} \& {Klemperer}(1973)}]{1973ApJ...185..505H}
{Herbst}, E. \& {Klemperer}, W. 1973, \apj, 185, 505

\bibitem[{{Hildebrand}(1983)}]{1983QJRAS..24..267H}
{Hildebrand}, R.~H. 1983, \qjras, 24, 267

\bibitem[{{Hillenbrand} {et~al.}(2018){Hillenbrand}, {Contreras Pe{\~n}a},
  {Morrell}, {Naylor}, {Kuhn}, {Cutri}, {Rebull}, {Hodgkin}, {Froebrich}, \&
  {Mainzer}}]{hillenbrand2018}
{Hillenbrand}, L.~A., {Contreras Pe{\~n}a}, C., {Morrell}, S., {et~al.} 2018,
  \apj, 869, 146

\bibitem[{{Hillenbrand} {et~al.}(2013){Hillenbrand}, {Miller}, {Covey},
  {Carpenter}, {Cenko}, {Silverman}, {Muirhead}, {Fischer}, {Crepp}, {Bloom},
  \& {Filippenko}}]{hillenbrand2013}
{Hillenbrand}, L.~A., {Miller}, A.~A., {Covey}, K.~R., {et~al.} 2013, \aj, 145,
  59

\bibitem[{{Hillenbrand} {et~al.}(2019){Hillenbrand}, {Reipurth}, {Connelley},
  {Cutri}, \& {Isaacson}}]{hillenbrand2019}
{Hillenbrand}, L.~A., {Reipurth}, B., {Connelley}, M., {Cutri}, R.~M., \&
  {Isaacson}, H. 2019, \aj, 158, 240

\bibitem[{{Ho} \& {Townes}(1983)}]{ho1983}
{Ho}, P.~T.~P. \& {Townes}, C.~H. 1983, \araa, 21, 239

\bibitem[{{Holdship} {et~al.}(2017){Holdship}, {Viti}, {Jim{\'e}nez-Serra},
  {Makrymallis}, \& {Priestley}}]{2017AJ....154...38H}
{Holdship}, J., {Viti}, S., {Jim{\'e}nez-Serra}, I., {Makrymallis}, A., \&
  {Priestley}, F. 2017, \aj, 154, 38

\bibitem[{{Jhan} \& {Lee}(2021)}]{jhan2021}
{Jhan}, K.-S. \& {Lee}, C.-F. 2021, \apj, 909, 11

\bibitem[{{J{\o}rgensen} {et~al.}(2020){J{\o}rgensen}, {Belloche}, \&
  {Garrod}}]{jorgensen2020}
{J{\o}rgensen}, J.~K., {Belloche}, A., \& {Garrod}, R.~T. 2020, \araa, 58, 727

\bibitem[{{Jurdana-{\v{S}}epi{\'c}} {et~al.}(2018){Jurdana-{\v{S}}epi{\'c}},
  {Munari}, {Antoniucci}, {Giannini}, \& {Lorenzetti}}]{sepic2018}
{Jurdana-{\v{S}}epi{\'c}}, R., {Munari}, U., {Antoniucci}, S., {Giannini}, T.,
  \& {Lorenzetti}, D. 2018, \aap, 614, A9

\bibitem[{{Kadam} {et~al.}(2020){Kadam}, {Vorobyov}, {Reg{\'a}ly},
  {K{\'o}sp{\'a}l}, \& {{\'A}brah{\'a}m}}]{kadam2020}
{Kadam}, K., {Vorobyov}, E., {Reg{\'a}ly}, Z., {K{\'o}sp{\'a}l}, {\'A}., \&
  {{\'A}brah{\'a}m}, P. 2020, \apj, 895, 41

\bibitem[{{Kauffmann} {et~al.}(2008){Kauffmann}, {Bertoldi}, {Bourke}, {Evans},
  \& {Lee}}]{2008A&A...487..993K}
{Kauffmann}, J., {Bertoldi}, F., {Bourke}, T.~L., {Evans}, N.~J., I., \& {Lee},
  C.~W. 2008, \aap, 487, 993

\bibitem[{{Kenyon} {et~al.}(1988){Kenyon}, {Hartmann}, \&
  {Hewett}}]{kenyon-and-hartmann1988}
{Kenyon}, S.~J., {Hartmann}, L., \& {Hewett}, R. 1988, \apj, 325, 231

\bibitem[{{Kenyon} \& {Hartmann}(1991)}]{kenyon-and-hartmann1991}
{Kenyon}, S.~J. \& {Hartmann}, L.~W. 1991, \apj, 383, 664

\bibitem[{{Klein} {et~al.}(2012){Klein}, {Hochg{\"u}rtel}, {Kr{\"a}mer},
  {Bell}, {Meyer}, \& {G{\"u}sten}}]{klein2012}
{Klein}, B., {Hochg{\"u}rtel}, S., {Kr{\"a}mer}, I., {et~al.} 2012, \aap, 542,
  L3

\bibitem[{{K{\"o}nyves} {et~al.}(2020){K{\"o}nyves}, {Andr{\'e}},
  {Arzoumanian}, {Schneider}, {Men'shchikov}, {Bontemps}, {Ladjelate},
  {Didelon}, {Pezzuto}, {Benedettini}, {Bracco}, {Di Francesco}, {Goodwin},
  {Rygl}, {Shimajiri}, {Spinoglio}, {Ward-Thompson}, \& {White}}]{konyves2020}
{K{\"o}nyves}, V., {Andr{\'e}}, P., {Arzoumanian}, D., {et~al.} 2020, \aap,
  635, A34

\bibitem[{{K{\'o}sp{\'a}l} {et~al.}(2011){K{\'o}sp{\'a}l}, {{\'A}brah{\'a}m},
  {Acosta-Pulido}, {Ar{\'e}valo Morales}, {Carnerero}, {Elek}, {Kelemen},
  {Kun}, {P{\'a}l}, {Szak{\'a}ts}, \& {Vida}}]{kospal2011}
{K{\'o}sp{\'a}l}, {\'A}., {{\'A}brah{\'a}m}, P., {Acosta-Pulido}, J.~A.,
  {et~al.} 2011, \aap, 527, A133

\bibitem[{{K{\'o}sp{\'a}l} {et~al.}(2016){K{\'o}sp{\'a}l}, {{\'A}brah{\'a}m},
  {Acosta-Pulido}, {Dunham}, {Garc{\'\i}a-{\'A}lvarez}, {Hogerheijde}, {Kun},
  {Mo{\'o}r}, {Farkas}, {Hajdu}, {Hodos{\'a}n}, {Kov{\'a}cs}, {Kriskovics},
  {Marton}, {Moln{\'a}r}, {P{\'a}l}, {S{\'a}rneczky}, {S{\'o}dor},
  {Szak{\'a}ts}, {Szalai}, {Szegedi-Elek}, {Szing}, {T{\'o}th}, {Vida}, \&
  {Vink{\'o}}}]{kospal2016}
{K{\'o}sp{\'a}l}, {\'A}., {{\'A}brah{\'a}m}, P., {Acosta-Pulido}, J.~A.,
  {et~al.} 2016, \aap, 596, A52

\bibitem[{{K{\'o}sp{\'a}l} {et~al.}(2008){K{\'o}sp{\'a}l}, {{\'A}brah{\'a}m},
  {Apai}, {Ardila}, {Grady}, {Henning}, {Juh{\'a}sz}, {Miller}, \&
  {Mo{\'o}r}}]{kospal2008}
{K{\'o}sp{\'a}l}, {\'A}., {{\'A}brah{\'a}m}, P., {Apai}, D., {et~al.} 2008,
  \mnras, 383, 1015

\bibitem[{{K{\'o}sp{\'a}l} {et~al.}(2017){K{\'o}sp{\'a}l}, {{\'A}brah{\'a}m},
  {Csengeri}, {Henning}, {Mo{\'o}r}, \& {G{\"u}sten}}]{kospal2017b}
{K{\'o}sp{\'a}l}, {\'A}., {{\'A}brah{\'a}m}, P., {Csengeri}, T., {et~al.} 2017,
  \apj, 836, 226

\bibitem[{{K{\'o}sp{\'a}l} {et~al.}(2015){K{\'o}sp{\'a}l}, {{\'A}brah{\'a}m},
  {Mo{\'o}r}, {Haas}, {Chini}, \& {Hackstein}}]{kospal2015}
{K{\'o}sp{\'a}l}, {\'A}., {{\'A}brah{\'a}m}, P., {Mo{\'o}r}, A., {et~al.} 2015,
  \apjl, 801, L5

\bibitem[{{K{\'o}sp{\'a}l} {et~al.}(2006){K{\'o}sp{\'a}l}, {{\'A}brah{\'a}m},
  {Prusti}, {Siebenmorgen}, {Acosta-Pulido}, \& {Mo{\'o}r}}]{kospal2006}
{K{\'o}sp{\'a}l}, {\'A}., {{\'A}brah{\'a}m}, P., {Prusti}, T., {et~al.} 2006,
  in Astronomical Society of the Pacific Conference Series, Vol. 349,
  Astrophysics of Variable Stars, ed. C.~{Aerts} \& C.~{Sterken}, 269

\bibitem[{{K{\'o}sp{\'a}l} {et~al.}(2021){K{\'o}sp{\'a}l}, {Cruz-S{\'a}enz de
  Miera}, {White}, {{\'A}brah{\'a}m}, {Chen}, {Csengeri}, {Dong}, {Dunham},
  {Feh{\'e}r}, {Green}, {Hashimoto}, {Henning}, {Hogerheijde}, {Kudo}, {Liu},
  {Takami}, \& {Vorobyov}}]{kospal2021}
{K{\'o}sp{\'a}l}, {\'A}., {Cruz-S{\'a}enz de Miera}, F., {White}, J.~A.,
  {et~al.} 2021, \apjs, 256, 30

\bibitem[{{Lada}(1987)}]{lada1987}
{Lada}, C.~J. 1987, in Star Forming Regions, ed. M.~{Peimbert} \& J.~{Jugaku},
  Vol. 115, 1

\bibitem[{{Lang} \& {Willson}(1979)}]{lang1979}
{Lang}, K.~R. \& {Willson}, R.~F. 1979, \apj, 227, 163

\bibitem[{{Lee} {et~al.}(2020){Lee}, {Lee}, {Aikawa}, {Herczeg}, \&
  {Johnstone}}]{lee2020-alma}
{Lee}, S., {Lee}, J.-E., {Aikawa}, Y., {Herczeg}, G., \& {Johnstone}, D. 2020,
  \apj, 889, 20

\bibitem[{{Lin} \& {Papaloizou}(1985)}]{lin1985}
{Lin}, D.~N.~C. \& {Papaloizou}, J. 1985, in Protostars and Planets II, ed.
  D.~C. {Black} \& M.~S. {Matthews}, 981--1072

\bibitem[{{Lin} {et~al.}(2017){Lin}, {Liu}, {Dale}, {Li}, {Busquet}, {Zhang},
  {Ginsburg}, {Galv{\'a}n-Madrid}, {Kov{\'a}cs}, {Koch}, {Qian}, {Wang},
  {Longmore}, {Chen}, \& {Walker}}]{2017ApJ...840...22L}
{Lin}, Y., {Liu}, H.~B., {Dale}, J.~E., {et~al.} 2017, \apj, 840, 22

\bibitem[{{Liu} {et~al.}(2018){Liu}, {Dunham}, {Pascucci}, {Bourke}, {Hirano},
  {Longmore}, {Andrews}, {Carrasco-Gonz{\'a}lez}, {Forbrich},
  {Galv{\'a}n-Madrid}, {Girart}, {Green}, {Ju{\'a}rez}, {K{\'o}sp{\'a}l},
  {Manara}, {Palau}, {Takami}, {Testi}, \& {Vorobyov}}]{liu2018}
{Liu}, H.~B., {Dunham}, M.~M., {Pascucci}, I., {et~al.} 2018, \aap, 612, A54

\bibitem[{{Liu} {et~al.}(2021){Liu}, {Tsai}, {Chen}, {Liu}, {Zhang}, {Ma},
  {Elbakyan}, {Green}, {Hales}, {Liu}, {Takami}, {P{\'e}rez}, {Vorobyov}, \&
  {Yang}}]{2021ApJ...923..270L}
{Liu}, H.~B., {Tsai}, A.-L., {Chen}, W.~P., {et~al.} 2021, \apj, 923, 270

\bibitem[{{Mathieu} {et~al.}(1996){Mathieu}, {Martin}, \&
  {Magazzu}}]{mathieau1996}
{Mathieu}, R.~D., {Martin}, E.~L., \& {Magazzu}, A. 1996, in American
  Astronomical Society Meeting Abstracts, Vol. 188, American Astronomical
  Society Meeting Abstracts \#188, 60.05

\bibitem[{{Maury} {et~al.}(2019){Maury}, {Andr{\'e}}, {Testi}, {Maret},
  {Belloche}, {Hennebelle}, {Cabrit}, {Codella}, {Gueth}, {Podio}, {Anderl},
  {Bacmann}, {Bontemps}, {Gaudel}, {Ladjelate}, {Lef{\`e}vre}, {Tabone}, \&
  {Lefloch}}]{2019A&A...621A..76M}
{Maury}, A.~J., {Andr{\'e}}, P., {Testi}, L., {et~al.} 2019, \aap, 621, A76

\bibitem[{{Miller} {et~al.}(2015){Miller}, {Hillenbrand}, {Bilgi}, {Cao},
  {Duggan}, {Arcavi}, {Hosseinzadeh}, {Howell}, \& {McCully}}]{miller2015}
{Miller}, A.~A., {Hillenbrand}, L.~A., {Bilgi}, P., {et~al.} 2015, The
  Astronomer's Telegram, 7428, 1

\bibitem[{{Miller} {et~al.}(2011){Miller}, {Hillenbrand}, {Covey}, {Poznanski},
  {Silverman}, {Kleiser}, {Rojas-Ayala}, {Muirhead}, {Cenko}, {Bloom},
  {Kasliwal}, {Filippenko}, {Law}, {Ofek}, {Dekany}, {Rahmer}, {Hale}, {Smith},
  {Quimby}, {Nugent}, {Jacobsen}, {Zolkower}, {Velur}, {Walters}, {Henning},
  {Bui}, {McKenna}, {Kulkarni}, {Klein}, {Kandrashoff}, \&
  {Morton}}]{miller2011}
{Miller}, A.~A., {Hillenbrand}, L.~A., {Covey}, K.~R., {et~al.} 2011, \apj,
  730, 80

\bibitem[{{Moriarty-Schieven} {et~al.}(2008){Moriarty-Schieven}, {Aspin}, \&
  {Davis}}]{moriarty2008}
{Moriarty-Schieven}, G.~H., {Aspin}, C., \& {Davis}, G.~R. 2008, \aj, 136, 1658

\bibitem[{{Nagy} {et~al.}(2022){Nagy}, {{\'A}brah{\'a}m}, {K{\'o}sp{\'a}l},
  {Park}, {Siwak}, {Cruz-S{\'a}enz de Miera}, {Fiorellino},
  {Garc{\'\i}a-{\'A}lvarez}, {Szab{\'o}}, {Antoniucci}, {Giannini}, {Giunta},
  {Kriskovics}, {Kun}, {Marton}, {Mo{\'o}r}, {Nisini}, {P{\'a}l}, {Szabados},
  {Zieli{\'n}ski}, \& {Wyrzykowski}}]{nagy2022}
{Nagy}, Z., {{\'A}brah{\'a}m}, P., {K{\'o}sp{\'a}l}, {\'A}., {et~al.} 2022,
  \mnras, 515, 1774

\bibitem[{{Nagy} {et~al.}(2021){Nagy}, {Szegedi-Elek}, {{\'A}brah{\'a}m},
  {K{\'o}sp{\'a}l}, {B{\'o}di}, {Bouvier}, {Kun}, {Mo{\'o}r}, {Cseh},
  {Farkas-Tak{\'a}cs}, {Hanyecz}, {Hodgkin}, {Ign{\'a}cz}, {Kiss},
  {K{\"o}nyves-T{\'o}th}, {Kriskovics}, {Marton}, {M{\'e}sz{\'a}ros}, {Ordasi},
  {P{\'a}l}, {Sarkis}, {S{\'a}rneczky}, {S{\'o}dor}, {Szabados}, {Szab{\'o}},
  {Szak{\'a}ts}, {Tarczay-Neh{\'e}z}, {Vida}, \& {Zsidi}}]{nagy2021}
{Nagy}, Z., {Szegedi-Elek}, E., {{\'A}brah{\'a}m}, P., {et~al.} 2021, \mnras,
  504, 185

\bibitem[{Newville {et~al.}(2014)Newville, Stensitzki, Allen, \&
  Ingargiola}]{newville_matthew_2014_11813}
Newville, M., Stensitzki, T., Allen, D.~B., \& Ingargiola, A. 2014, {LMFIT:
  Non-Linear Least-Square Minimization and Curve-Fitting for Python}

\bibitem[{{Ohishi} {et~al.}(1992){Ohishi}, {Irvine}, \& {Kaifu}}]{Ohishi1992}
{Ohishi}, M., {Irvine}, W.~M., \& {Kaifu}, N. 1992, in Astrochemistry of Cosmic
  Phenomena, ed. P.~D. {Singh}, Vol. 150, 171

\bibitem[{{Olmi} {et~al.}(2010){Olmi}, {Araya}, {Chapin}, {Gibb}, {Hofner},
  {Martin}, \& {Poventud}}]{olmi2010}
{Olmi}, L., {Araya}, E.~D., {Chapin}, E.~L., {et~al.} 2010, \apj, 715, 1132

\bibitem[{{Ott} {et~al.}(1994){Ott}, {Witzel}, {Quirrenbach}, {Krichbaum},
  {Standke}, {Schalinski}, \& {Hummel}}]{ott1994}
{Ott}, M., {Witzel}, A., {Quirrenbach}, A., {et~al.} 1994, \aap, 284, 331

\bibitem[{{Paczynski}(1976)}]{paczynski1976}
{Paczynski}, B. 1976, in IAU Symposium, Vol.~73, Structure and Evolution of
  Close Binary Systems, ed. P.~{Eggleton}, S.~{Mitton}, \& J.~{Whelan}, 75

\bibitem[{{Park} {et~al.}(2022){Park}, {K{\'o}sp{\'a}l}, {{\'A}brah{\'a}m},
  {Cruz-S{\'a}enz de Miera}, {Fiorellino}, {Siwak}, {Nagy}, {Giannini},
  {Carini}, {Szab{\'o}}, {Lee}, {Lee}, {Vitali}, {Kun}, {Cseh}, {Krezinger},
  {Kriskovics}, {Ordasi}, {P{\'a}l}, {Szak{\'a}ts}, {Vida}, \&
  {Vink{\'o}}}]{park2022}
{Park}, S., {K{\'o}sp{\'a}l}, {\'A}., {{\'A}brah{\'a}m}, P., {et~al.} 2022,
  \apj, 941, 165

\bibitem[{{Park} {et~al.}(2021){Park}, {K{\'o}sp{\'a}l}, {Cruz-S{\'a}enz de
  Miera}, {Siwak}, {Dr{\'o}{\.z}d{\.z}}, {Ign{\'a}cz}, {Jaffe},
  {K{\"o}nyves-T{\'o}th}, {Kriskovics}, {Lee}, {Lee}, {Mace}, {Og{\l}oza},
  {P{\'a}l}, {Potter}, {Szab{\'o}}, {Sefako}, \& {Worters}}]{park2021}
{Park}, S., {K{\'o}sp{\'a}l}, {\'A}., {Cruz-S{\'a}enz de Miera}, F., {et~al.}
  2021, \apj, 923, 171

\bibitem[{{Parsamian} \& {Mujica}(2004)}]{parsamian2004}
{Parsamian}, E.~S. \& {Mujica}, R. 2004, Astrophysics, 47, 433

\bibitem[{{Pety}(2005)}]{gildas_pety2005}
{Pety}, J. 2005, in SF2A-2005: Semaine de l'Astrophysique Francaise, ed.
  F.~{Casoli}, T.~{Contini}, J.~M. {Hameury}, \& L.~{Pagani}, 721

\bibitem[{{Pezzuto} {et~al.}(2021){Pezzuto}, {Benedettini}, {Di Francesco},
  {Palmeirim}, {Sadavoy}, {Schisano}, {Li Causi}, {Andr{\'e}}, {Arzoumanian},
  {Bernard}, {Bontemps}, {Elia}, {Fiorellino}, {Kirk}, {K{\"o}nyves},
  {Ladjelate}, {Men'shchikov}, {Motte}, {Piccotti}, {Schneider}, {Spinoglio},
  {Ward-Thompson}, \& {Wilson}}]{pezzuto2021}
{Pezzuto}, S., {Benedettini}, M., {Di Francesco}, J., {et~al.} 2021, \aap, 645,
  A55

\bibitem[{{Pezzuto} {et~al.}(1997){Pezzuto}, {Strafella}, \&
  {Lorenzetti}}]{pezzuto1997}
{Pezzuto}, S., {Strafella}, F., \& {Lorenzetti}, D. 1997, \apj, 485, 290

\bibitem[{{Pillai} {et~al.}(2006){Pillai}, {Wyrowski}, {Carey}, \&
  {Menten}}]{pillai2006}
{Pillai}, T., {Wyrowski}, F., {Carey}, S.~J., \& {Menten}, K.~M. 2006, \aap,
  450, 569

\bibitem[{{Principe} {et~al.}(2018){Principe}, {Cieza}, {Hales}, {Zurlo},
  {Williams}, {Ru{\'\i}z-Rodr{\'\i}guez}, {Canovas}, {Casassus},
  {Mu{\v{z}}i{\'c}}, {Perez}, {Tobin}, \& {Zhu}}]{principe2018}
{Principe}, D.~A., {Cieza}, L., {Hales}, A., {et~al.} 2018, \mnras, 473, 879

\bibitem[{{Quanz} {et~al.}(2007{\natexlab{a}}){Quanz}, {Henning}, {Bouwman},
  {Linz}, \& {Lahuis}}]{quanz2007}
{Quanz}, S.~P., {Henning}, T., {Bouwman}, J., {Linz}, H., \& {Lahuis}, F.
  2007{\natexlab{a}}, \apj, 658, 487

\bibitem[{{Quanz} {et~al.}(2007{\natexlab{b}}){Quanz}, {Henning}, {Bouwman},
  {van Boekel}, {Juh{\'a}sz}, {Linz}, {Pontoppidan}, \& {Lahuis}}]{quanz2007b}
{Quanz}, S.~P., {Henning}, T., {Bouwman}, J., {et~al.} 2007{\natexlab{b}},
  \apj, 668, 359

\bibitem[{{Ragan} {et~al.}(2011){Ragan}, {Bergin}, \& {Wilner}}]{ragan2011}
{Ragan}, S.~E., {Bergin}, E.~A., \& {Wilner}, D. 2011, \apj, 736, 163

\bibitem[{{Redaelli} {et~al.}(2022){Redaelli}, {Bovino}, {Sanhueza}, {Morii},
  {Sabatini}, {Caselli}, {Giannetti}, \& {Li}}]{redaelli2022}
{Redaelli}, E., {Bovino}, S., {Sanhueza}, P., {et~al.} 2022, \apj, 936, 169

\bibitem[{{Reipurth} \& {Aspin}(1997)}]{reipurth1997a}
{Reipurth}, B. \& {Aspin}, C. 1997, \aj, 114, 2700

\bibitem[{{Reipurth} {et~al.}(2012){Reipurth}, {Aspin}, \&
  {Herbig}}]{reipurth2012}
{Reipurth}, B., {Aspin}, C., \& {Herbig}, G.~H. 2012, \apjl, 748, L5

\bibitem[{{Reipurth} {et~al.}(1997){Reipurth}, {Bally}, \&
  {Devine}}]{reipurth1997b}
{Reipurth}, B., {Bally}, J., \& {Devine}, D. 1997, \aj, 114, 2708

\bibitem[{{Rohlfs} \& {Wilson}(2004)}]{rohlfs2004}
{Rohlfs}, K. \& {Wilson}, T.~L. 2004, {Tools of radio astronomy}

\bibitem[{{Roy} {et~al.}(2013){Roy}, {Martin}, {Polychroni}, {Bontemps},
  {Abergel}, {Andr{\'e}}, {Arzoumanian}, {Di Francesco}, {Hill}, {Konyves},
  {Nguyen-Luong}, {Pezzuto}, {Schneider}, {Testi}, \& {White}}]{roy2013}
{Roy}, A., {Martin}, P.~G., {Polychroni}, D., {et~al.} 2013, \apj, 763, 55

\bibitem[{{Ru{\'\i}z-Rodr{\'\i}guez} {et~al.}(2017){Ru{\'\i}z-Rodr{\'\i}guez},
  {Cieza}, {Williams}, {Tobin}, {Hales}, {Zhu}, {Mu{\v{z}}i{\'c}}, {Principe},
  {Canovas}, {Zurlo}, {Casassus}, {Perez}, \& {Prieto}}]{ruiz-rodriguez2017}
{Ru{\'\i}z-Rodr{\'\i}guez}, D., {Cieza}, L.~A., {Williams}, J.~P., {et~al.}
  2017, \mnras, 466, 3519

\bibitem[{{Sandell} \& {Aspin}(1998)}]{sandell1998}
{Sandell}, G. \& {Aspin}, C. 1998, \aap, 333, 1016

\bibitem[{{Sandell} \& {Weintraub}(2001)}]{sandell2001}
{Sandell}, G. \& {Weintraub}, D.~A. 2001, \apjs, 134, 115

\bibitem[{{Schneider} {et~al.}(2013){Schneider}, {Andr{\'e}}, {K{\"o}nyves},
  {Bontemps}, {Motte}, {Federrath}, {Ward-Thompson}, {Arzoumanian},
  {Benedettini}, {Bressert}, {Didelon}, {Di Francesco}, {Griffin}, {Hennemann},
  {Hill}, {Palmeirim}, {Pezzuto}, {Peretto}, {Roy}, {Rygl}, {Spinoglio}, \&
  {White}}]{Schneider2013}
{Schneider}, N., {Andr{\'e}}, P., {K{\"o}nyves}, V., {et~al.} 2013, \apjl, 766,
  L17

\bibitem[{{Sch{\"o}ier} {et~al.}(2005){Sch{\"o}ier}, {van der Tak}, {van
  Dishoeck}, \& {Black}}]{lamda2005}
{Sch{\"o}ier}, F.~L., {van der Tak}, F.~F.~S., {van Dishoeck}, E.~F., \&
  {Black}, J.~H. 2005, \aap, 432, 369

\bibitem[{{Shirley}(2015)}]{2015PASP..127..299S}
{Shirley}, Y.~L. 2015, \pasp, 127, 299

\bibitem[{{Stecklum} {et~al.}(2007){Stecklum}, {Melnikov}, \&
  {Meusinger}}]{stecklum2007}
{Stecklum}, B., {Melnikov}, S.~Y., \& {Meusinger}, H. 2007, \aap, 463, 621

\bibitem[{{Stojimirovi{\'c}} {et~al.}(2008){Stojimirovi{\'c}}, {Snell}, \&
  {Narayanan}}]{stojimirovic2008}
{Stojimirovi{\'c}}, I., {Snell}, R.~L., \& {Narayanan}, G. 2008, \apj, 679, 557

\bibitem[{{Strom} {et~al.}(1972){Strom}, {Strom}, {Yost}, {Carrasco}, \&
  {Grasdalen}}]{strom1972}
{Strom}, S.~E., {Strom}, K.~M., {Yost}, J., {Carrasco}, L., \& {Grasdalen}, G.
  1972, \apj, 173, 353

\bibitem[{{Szab{\'o}} {et~al.}(2021){Szab{\'o}}, {K{\'o}sp{\'a}l},
  {{\'A}brah{\'a}m}, {Park}, {Siwak}, {Green}, {Mo{\'o}r}, {P{\'a}l},
  {Acosta-Pulido}, {Lee}, {Cseh}, {Cs{\"o}rnyei}, {Hanyecz},
  {K{\"o}nyves-T{\'o}th}, {Krezinger}, {Kriskovics}, {Ordasi}, {S{\'a}rneczky},
  {Seli}, {Szak{\'a}ts}, {Szing}, \& {Vida}}]{szabo2021}
{Szab{\'o}}, Z.~M., {K{\'o}sp{\'a}l}, {\'A}., {{\'A}brah{\'a}m}, P., {et~al.}
  2021, \apj, 917, 80

\bibitem[{{Szab{\'o}} {et~al.}(2022){Szab{\'o}}, {K{\'o}sp{\'a}l},
  {{\'A}brah{\'a}m}, {Park}, {Siwak}, {Green}, {P{\'a}l}, {Acosta-Pulido},
  {Lee}, {Ibrahimov}, {Grankin}, {Kov{\'a}cs}, {Bora}, {B{\'o}di}, {Cseh},
  {Cs{\"o}rnyei}, {Dr{\'o}{\.z}d{\.z}}, {Hanyecz}, {Ign{\'a}cz}, {Kalup},
  {K{\"o}nyves-T{\'o}th}, {Krezinger}, {Kriskovics}, {Og{\l}oza}, {Ordasi},
  {S{\'a}rneczky}, {Seli}, {Szak{\'a}ts}, {S{\'o}dor}, {Szing}, {Vida}, \&
  {Vink{\'o}}}]{szabo2022}
{Szab{\'o}}, Z.~M., {K{\'o}sp{\'a}l}, {\'A}., {{\'A}brah{\'a}m}, P., {et~al.}
  2022, \apj, 936, 64

\bibitem[{{Szegedi-Elek} {et~al.}(2020){Szegedi-Elek}, {{\'A}brah{\'a}m},
  {Wyrzykowski}, {Kun}, {K{\'o}sp{\'a}l}, {Chen}, {Marton}, {Mo{\'o}r}, {Kiss},
  {P{\'a}l}, {Szabados}, {Varga}, {Varga-Vereb{\'e}lyi}, {Andreas}, {Bachelet},
  {Bischoff}, {B{\'o}di}, {Breedt}, {Burgaz}, {Butterley}, {Carrasco},
  {{\v{C}}epas}, {Damljanovic}, {Gezer}, {Godunova}, {Gromadzki}, {Gurgul},
  {Hardy}, {Hildebrandt}, {Hoffmann}, {Hundertmark}, {Ihanec}, {Janulis},
  {Kalup}, {Kaczmarek}, {K{\"o}nyves-T{\'o}th}, {Krezinger}, {Kruszy{\'n}ska},
  {Littlefair}, {Maskoli{\={u}}nas}, {M{\'e}sz{\'a}ros}, {Miko{\l}ajczyk},
  {Mugrauer}, {Netzel}, {Ordasi}, {Pak{\v{s}}tien{\.{e}}}, {Rybicki},
  {S{\'a}rneczky}, {Seli}, {Simon}, {{\v{S}}i{\v{s}}kauskait{\.{e}}},
  {S{\'o}dor}, {Sokolovsky}, {Stenglein}, {Street}, {Szak{\'a}ts}, {Tomasella},
  {Tsapras}, {Vida}, {Zdanavi{\v{c}}ius}, {Zieli{\'n}ski}, {Zieli{\'n}ski}, \&
  {Zi{\'o}{\l}kowska}}]{szegedi-elek2020}
{Szegedi-Elek}, E., {{\'A}brah{\'a}m}, P., {Wyrzykowski}, {\L}., {et~al.} 2020,
  \apj, 899, 130

\bibitem[{{Tafalla} \& {Bachiller}(1995)}]{1995ApJ...443L..37T}
{Tafalla}, M. \& {Bachiller}, R. 1995, \apjl, 443, L37

\bibitem[{{Tafalla} {et~al.}(2004){Tafalla}, {Myers}, {Caselli}, \&
  {Walmsley}}]{tafalla2004}
{Tafalla}, M., {Myers}, P.~C., {Caselli}, P., \& {Walmsley}, C.~M. 2004, \aap,
  416, 191

\bibitem[{{Takami} {et~al.}(2019){Takami}, {Chen}, {Liu}, {Hirano},
  {K{\'o}sp{\'a}l}, {{\'A}brah{\'a}m}, {Vorobyov}, {Cruz-S{\'a}enz de Miera},
  {Csengeri}, {Green}, {Hogerheijde}, {Hsieh}, {Karr}, {Dong}, {Trejo}, \&
  {Chen}}]{takami2019}
{Takami}, M., {Chen}, T.-S., {Liu}, H.~B., {et~al.} 2019, \apj, 884, 146

\bibitem[{{Takami} {et~al.}(2018){Takami}, {Fu}, {Liu}, {Karr}, {Hashimoto},
  {Kudo}, {Vorobyov}, {K{\'o}sp{\'a}l}, {Scicluna}, {Dong}, {Tamura}, {Pyo},
  {Fukagawa}, {Tsuribe}, {Dunham}, {Henning}, \& {de Leon}}]{takami2018}
{Takami}, M., {Fu}, G., {Liu}, H.~B., {et~al.} 2018, \apj, 864, 20

\bibitem[{{Tambovtseva} \& {Grinin}(2016)}]{tambovtseva2016}
{Tambovtseva}, L. \& {Grinin}, V. 2016, in Accretion Processes in Cosmic
  Sources, 56

\bibitem[{{Tanner} \& {Arce}(2011)}]{tanner2011}
{Tanner}, J.~D. \& {Arce}, H.~G. 2011, \apj, 726, 40

\bibitem[{{Torrelles} {et~al.}(1986){Torrelles}, {Ho}, {Moran}, {Rodriguez}, \&
  {Canto}}]{torrelles1986}
{Torrelles}, J.~M., {Ho}, P.~T.~P., {Moran}, J.~M., {Rodriguez}, L.~F., \&
  {Canto}, J. 1986, \apj, 307, 787

\bibitem[{{Turner} {et~al.}(1997){Turner}, {Bodenheimer}, \&
  {Bell}}]{turner1997}
{Turner}, N.~J.~J., {Bodenheimer}, P., \& {Bell}, K.~R. 1997, \apj, 480, 754

\bibitem[{{Umemoto} {et~al.}(1999){Umemoto}, {Mikami}, {Yamamoto}, \&
  {Hirano}}]{1999ApJ...525L.105U}
{Umemoto}, T., {Mikami}, H., {Yamamoto}, S., \& {Hirano}, N. 1999, \apjl, 525,
  L105

\bibitem[{{Ungerechts} \& {Guesten}(1984)}]{Ungerechts1984}
{Ungerechts}, H. \& {Guesten}, R. 1984, \aap, 131, 177

\bibitem[{{Walmsley} \& {Ungerechts}(1983)}]{Walmsley1983}
{Walmsley}, C.~M. \& {Ungerechts}, H. 1983, \aap, 122, 164

\bibitem[{{Waters} \& {Waelkens}(1998)}]{waters1998}
{Waters}, L.~B.~F.~M. \& {Waelkens}, C. 1998, \araa, 36, 233

\bibitem[{{Weintraub} {et~al.}(1991){Weintraub}, {Sandell}, \&
  {Duncan}}]{weintraub1991}
{Weintraub}, D.~A., {Sandell}, G., \& {Duncan}, W.~D. 1991, \apj, 382, 270

\bibitem[{{White} {et~al.}(2019){White}, {K{\'o}sp{\'a}l}, {Rab},
  {{\'A}brah{\'a}m}, {Cruz-S{\'a}enz de Miera}, {Csengeri}, {Feh{\'e}r},
  {G{\"u}sten}, {Henning}, {Vorobyov}, {Audard}, \& {Postel}}]{white2019}
{White}, J.~A., {K{\'o}sp{\'a}l}, {\'A}., {Rab}, C., {et~al.} 2019, \apj, 877,
  21

\bibitem[{{Wienen} {et~al.}(2012){Wienen}, {Wyrowski}, {Schuller}, {Menten},
  {Walmsley}, {Bronfman}, \& {Motte}}]{wienen2012}
{Wienen}, M., {Wyrowski}, F., {Schuller}, F., {et~al.} 2012, \aap, 544, A146

\bibitem[{{Wilson} {et~al.}(2009){Wilson}, {Rohlfs}, \&
  {H{\"u}ttemeister}}]{wilson2009(tools-of...)}
{Wilson}, T.~L., {Rohlfs}, K., \& {H{\"u}ttemeister}, S. 2009, {Tools of Radio
  Astronomy}

\bibitem[{{Winkel} {et~al.}(2012){Winkel}, {Kraus}, \& {Bach}}]{winkel2012}
{Winkel}, B., {Kraus}, A., \& {Bach}, U. 2012, \aap, 540, A140

\bibitem[{{Wouterloot} \& {Brand}(1989)}]{wouterloot1989}
{Wouterloot}, J.~G.~A. \& {Brand}, J. 1989, \aaps, 80, 149

\bibitem[{{Yan} {et~al.}(2021){Yan}, {Yang}, {Yang}, {Sun}, \&
  {Wang}}]{yan2021}
{Yan}, Q.-Z., {Yang}, J., {Yang}, S., {Sun}, Y., \& {Wang}, C. 2021, \apj, 910,
  109

\bibitem[{{Zapata} {et~al.}(2015){Zapata}, {Galv{\'a}n-Madrid},
  {Carrasco-Gonz{\'a}lez}, {Curiel}, {Palau}, {Rodr{\'\i}guez}, {Kurtz},
  {Tafoya}, \& {Loinard}}]{zapatava2015}
{Zapata}, L.~A., {Galv{\'a}n-Madrid}, R., {Carrasco-Gonz{\'a}lez}, C., {et~al.}
  2015, \apjl, 811, L4

\bibitem[{{Zhang} {et~al.}(2011){Zhang}, {Yang}, {Xu}, {Pandian}, {Menten}, \&
  {Henkel}}]{zhang2011}
{Zhang}, S.~B., {Yang}, J., {Xu}, Y., {et~al.} 2011, \apjs, 193, 10

\bibitem[{{Zurlo} {et~al.}(2017){Zurlo}, {Cieza}, {Williams}, {Canovas},
  {Perez}, {Hales}, {Mu{\v{z}}i{\'c}}, {Principe}, {Ru{\'\i}z-Rodr{\'\i}guez},
  {Tobin}, {Zhang}, {Zhu}, {Casassus}, \& {Prieto}}]{zurlo2017}
{Zurlo}, A., {Cieza}, L.~A., {Williams}, J.~P., {et~al.} 2017, \mnras, 465, 834

\end{thebibliography}

\begin{appendix} 

\section{Sources with non-detections} \label{sec:negative_detections}
In Table~\ref{tab:non-detection}, we list 3$\sigma$ upper limits for sources without ammonia detections. Gaia alerts were chosen based on their light curves and luminosities at the time of our proposal submission. These objects were chosen because their light curves resembled those of FUors/EXors. Interestingly, no ammonia was detected towards any of the Gaia alert sources.

\begin{table*}[htbp]
    \small
    \centering
    \caption{Sources without ammonia detections in our survey.}\label{tab:non-detection}
    \begin{tabular}{cccccccccc}
    \hline \hline
        Name &  R.A. (J2000)  & Dec. (J2000) & Type & 3$\rm \sigma_{(1,1)}$ & 3$\rm \sigma_{(2,2)}$ & 3$\rm \sigma_{(3,3)}$ & $N_{\rm H_2}$ & $T_{\rm dust}$ & \multirow{2}{*}{Reference} \\
               & ($^{\rm h}$ : $^{\rm m}$ : $^{\rm s}$)   & ($^{\circ}$ : $\arcmin$ : $\arcsec$) & FUor/EXor & (K) & (K) & (K) & (cm$^{-2}$) & (K) & \\
        \hline
        V1180~Cas       & 02:33:01.53 & +72:43:26.8 & EXor     & $0.32$ & $0.32$ & $0.33$ & $-$                   & $-$ & $-$ \\
        XZ~Tau          & 04:31:40.08 & +18:13:56.6 & EXor     & $0.34$ & $0.33$ & $0.32$ & $-$                   & $-$ & $-$ \\
        UZ~Tau~E        & 04:32:43.02 & +25:52:30.9 & EXor     & $0.34$ & $0.33$ & $0.31$ & $-$                   & $-$ & $-$ \\
        VY~Tau          & 04:39:17.42 & +22:47:53.3	  & EXor   & $0.24$ & $0.25$ & $0.26$ & $-$                   & $-$ & $-$ \\
        DR~Tau          & 04:47:06.21 & +16:58:42.8 & EXor     & $0.33$ & $0.33$ & $0.35$ & $-$                   & $-$ & $-$ \\
        V582~Aur        & 05:25:51.97 & +34:52:30.0 & FUor     & $0.34$ & $0.33$ & $0.33$ & $-$                   & $-$ & $-$ \\ 
        V1118~Ori       & 05:34:44.98 & $-$05:33:41.3 & EXor   & $0.38$ & $0.45$ & $0.41$ & $3.9\times10^{21}$ & $21.9$  & 1, 2 \\
        NY~Ori          & 05:35:36.0 & $-$05:12:25.2  & EXor   & $0.41$ & $0.42$ & $0.39$ & $6.6\times10^{21}$ & $28.8$  & 1, 2 \\
        Gaia21arx       & 05:36:24.80 & $-$06:17:30.52 & unknown & $0.32$ & $0.33$ & $0.31$ & $-$                   & $-$ & $-$ \\
        V1143~Ori       & 05:38:03.89 & $-$04:16:42.8 & EXor   & $0.39$ & $0.44$ & $0.48$ & $5.4\times10^{20}$ & $19.7$  & 1, 2 \\
        V883~Ori        & 05:38:18.09 &	$-$07:02:25.9 & FUor   & $0.02$ & $0.02$ & $0.03$ & $1.5\times10^{22}$ & $18.8$  & 1, 2 \\
        HBC~494         & 05:40:27.45 & $-$07:27:30.0 & FUor   & $0.38$ & $0.42$ & $0.42$ & $-$                   & $-$ & $-$ \\
        FU~Ori          & 05:45:22.37 & +09:04:12.3 & FUor     & $0.35$ & $0.39$ & $0.37$ & $-$                   & $-$ & $-$ \\
        V1647~Ori       & 05:46:13.13 & $-$00:06:04.8 & FUor   & $0.02$ & $0.02$ & $0.02$ & $1.3\times10^{22}$ & $17.4$  & 1, 3 \\    
        V900~Mon        & 06:57:22.22 & $-$08:23:17.6 & FUor    & $0.57$ & $0.55$ & $0.57$ & $-$                   & $-$ & $-$ \\
        Gaia20bdk       & 07:10:14.92 & $-$18:27:01.04 & unknown & $0.62$ & $0.72$ & $0.67$ & $-$                 & $-$ & $-$ \\
        Gaia21aul       & 18:30:06.18 & 00:42:33.30 & unknown  & $0.34$ & $0.37$ & $0.36$ & $-$                   & $-$ & $-$ \\
        Gaia21aru      	& 19:00:56.41 & 18:48:29.20	& unknown  & $0.33$ & $0.31$ & $0.31$ & $-$                   & $-$ & $-$ \\
        Parsamian~21    & 19:29:00.84 & +09:38:43.4	& FUor     & $0.32$ & $0.31$ & $0.33$ & $-$                   & $-$ & $-$ \\
        Gaia18dvy       & 20:05:06.02 & +36:29:13.5	  & FUor   & $0.26$ & $0.26$ & $0.28$ & $-$                   & $-$ & $-$ \\
        V1515~Cyg       & 20:23:48.01 & +42:12:25.7	& FUor     & $0.32$ & $0.33$ & $0.33$ & $1.1\times10^{22}$ & $17.4$ & 1, 4  \\
        PV~Cep          & 20:45:53.9  & +67:57:38.6 & EXor     & $0.33$ & $0.30$ & $0.31$ & $1.5\times$10$^{22}$ & $16.3$ & 1, 5  \\
        Gaia19bpg       & 21:41:50.43 & 51:55:45.48	  & unknown & $0.27$ & $0.27$ & $0.29$ & $-$                   & $-$ & $-$ \\
        \hline
    \end{tabular}
      \begin{tablenotes}
      \small 
      \item \textbf{Notes.} 1 -- \citet{andre2010}, 2 -- \citet{pezzuto2021}, 3 -- \citet{konyves2020}, 4 -- \citet{cao2019}, 5 -- \citet{di-francesco2020}
      \end{tablenotes}
\end{table*}

\section{Classification and $\varv_{\rm LSR}$}
In Table~\ref{tab:appendix-long} we list all FUors/EXors from our sample including both ammonia detections and non-detections. We tabulate whether each source is an FUor or EXor and previously determined $v_{\rm LSR}$ velocities, with the line(s) used to determine these velocities noted in brackets (a dash indicates no available data). We also list the $v_{\rm LSR}$ results from our ammonia observations, where a dash indicates a non-detection. We list classifications if available in the literature, and give the references. Finally in the last column we give the distances if available, which except for RNO~1B/1C, V512~Per, Z~CMa, and HH~354~IRS are adopted from the study of \citet{audard2014}. For these four sources we use updated distances, because of water maser detections associated with these sources in our Paper\,{\footnotesize II}. In the case of V512~Per (more commonly known as SVS~13), the source is a resolved binary, consisting of VLA~4A and 4B \citep[e.g.,][]{diaz-rodriguez2022}. We found CO line data for both sources, from which we find an average value of 8.35\,km\,s$^{-1}$, similar to our unresolved single dish result.

\begin{table*}[h]
\caption{Reference classification and $\varv_{\rm LSR}$ for the FUors/EXors in our sample, including NH$_3$ detections and non-detections.}    \label{tab:appendix-long}
    \centering
    \begin{tabular}{ccccccc}
    \hline
    Name        &   Type    &   $\varv_{\rm LSR}$ &  $v_{\rm LSR}$ (NH$_3$)  &                  Classification                         & References & Distance$^{*}$  \\
                &           &    (km\,s$^{-1}$)                  &  (km\,s$^{-1}$) & (Class\,{\footnotesize 0 -- II}) & & (pc) \\
    \hline \hline
    RNO~1B/1C$^{**}$   & FUor      &   $-17.83$ ($^{13}$CO)            & $-17.83$ $(0.02)$                 & 1B: Class\,{\footnotesize 0/II},1C: Class\,{\footnotesize II}    & 1, 2 & 965 \\
    V1180~Cas   & EXor      & $-$ & $-$ & $-$ & $-$  & 600 \\
    V512~Per (SVS~13)       & EXor      & $8.35$ ($^{12}$CO)    & $8.45$ $(0.01)$                 & Class\,{\footnotesize I} & 3  & 275 \\
    PP~13S      & FUor      & $-3.5$ ($^{12}$CO)                              & $-3.62$ $(0.01)$                   & Class\,{\footnotesize I}   & 4  & 350 \\
    L1551~IRS~5 & FUor      & $6.46$ ($^{13}$CO)                              & $6.35$ $(0.01)$                   & Class\,{\footnotesize I}   & 5, 6  & $-$ \\
    XZ~Tau      & EXor      & $6.8$ ($^{12}$CO)                              & $-$                   & Class\,{\footnotesize II}  & 7, 8  & 140 \\
    UZ~Tau~E    & EXor      & $-$                               & $-$                   & Class\,{\footnotesize II}  & 9 & 140  \\
    VY Tau      & EXor      & $+18$ or $+19$ ($^{12}$CO)                    & $-$                   & Class\,{\footnotesize II}  & 10  & 140 \\
    LDN~1415~IRS& EXor      & $-5.2$ ($^{12}$CO)                             & $-5.77$ $(0.02)$                   & Class\,{\footnotesize I}   & 11  & 170 \\
    DR~Tau      & EXor      & $-$                               & $-$                       & Class\,{\footnotesize II}  & 12  & $-$ \\
    V582~Aur    & FUor      & $-10.85$ ($^{13}$CO)      & $-$                   & Class\,{\footnotesize II} & 5, 13  & $-$ \\
    V1118~Ori   & EXor      & $-$                               & $-$                   & Class\,{\footnotesize II}  & 14  & 414 \\
    Haro~5a~IRS & FUor      & $10.90$ ($^{13}$CO)                           & $10.7$ $(0.01)$                   & Class\,{\footnotesize 0/I} & 5, 16  & 450 \\
    NY~Ori      & EXor      & $-$ & $-$ & $-$ & $-$  & 414 \\
    V1143~Ori   & EXor      & $-$                               & $-$                   & Class\,{\footnotesize II} & 17  & 500 \\
    V883~Ori    & FUor      & $4.10$ ($^{13}$CO)                               & $-$                   & Class\,{\footnotesize I} & 5, 18  & 460 \\
    HBC~494     & FUor      & $\sim$4.6 ($^{12}$CO)                          & $-$                   & Class\,{\footnotesize I} & 19  & $-$ \\
    V2775~Ori   & FUor      & $3.08$ ($^{13}$CO)                             & $3.05$ $(0.01)$                   & late Class\,{\footnotesize I} & 5, 20  & 420 \\
    FU~Ori  & FUor          & $11.96$ ($^{13}$CO)                           & $-$                    & Class\,{\footnotesize II} & 5, 21  & 450 \\
    V1647~Ori   & FUor      & $10.06$ ($^{13}$CO)                             & $-$                    & Class\,{\footnotesize I/II} & 5, 22, 23 & 400 \\
    NGC~2071    & FUor      & $9.2$ ($^{13}$CO)                            & $10.4$ $(0.01)$                    & $-$ & 24 & $-$ \\
    V899~Mon    & FUor      & $9.57$ ($^{13}$CO)                               & $9.63$ $(0.01)$                   & Class\,{\footnotesize II} & 5, 25  & $-$ \\
    IRAS~06393+0913 & FUor  & $4.3$ $(0.2)$ ($^{12}$CO)                     & $7.72$ $(0.02)$                   & Class\,{\footnotesize I}  & 26, 27 & $-$ \\
    AR~6A/6B    & FUor      & $5.02$ ($^{13}$CO)                             & $5.06$ $(0.02)$                   & Class\,{\footnotesize II} & 5, 28  & 800 \\
    IRAS~06297+1021W & FUor & $5.1$ $(0.2)$ ($^{12}$CO)                     & $4.17$ $(0.01)$                   & Class\,{\footnotesize I}  & 26, 27 & $-$  \\
    V900~Mon    & FUor      & $13.77$ ($^{13}$CO)                          & $-$                   & Class\,{\footnotesize I} & 5, 29, 30 & 1100  \\
    V960~Mon    & FUor      & $23.81$ ($^{13}$CO)                              & $23.8$ $(0.02)$                   & Class\,{\footnotesize II} & 5, 31 & $-$  \\
    Z~CMa       & FUor      & $13.91$ ($^{13}$CO)                              & $13.8$ $(0.02)$                   & Class\,{\footnotesize I} & 5, 32 & 1125 \\
    iPTF~15AFQ  & FUor      & $14.04$ ($^{13}$CO)                             & $13.3$ $(0.01)$                   & Class\,{\footnotesize I}  & 5, 33 & $-$ \\
    IRAS~18270-0153W & FUor & $-$                               & $7.61$ $(0.01)$                   & Class\,{\footnotesize I}  & 34 & $-$ \\
    OO~Ser      & FUor      & $8.36$ ($^{13}$CO)                               & $8.31$ $(0.01)$                   & Class\,{\footnotesize I} & 5, 35 & 311 \\
    IRAS~18341-0113S & FUor & $-$                               & $9.27$ $(0.01)$                   & Class\,{\footnotesize I} & 34 & $-$ \\
    V371~Ser    & EXor      & $-$                               & $8.34$ $(0.01)$                   & $-$ & $-$ & 311 \\
    Parsamian~21 & FUor     & $27$ (Li I, Fe I)                 & $-$                   & Class\,{\footnotesize I}/\,{\footnotesize II}  & 36 & 400 \\
    Gaia~18dvy  & FUor      & $-$                               & $-$                   & Class\,{\footnotesize II} & 37 & $-$ \\
    V1515~Cyg   & FUor      & $5.80$ ($^{13}$CO)                & $-$                   & $-$     & 1 & 1000 \\
    PV~Cep      & EXor      & $-3$ (nearby cloud)               & $-$                   & $-$ & 38 & 325 \\
    V2492~Cyg   & EXor      & $4.97$ ($^{13}$CO)                & $4.71$ $(0.02)$                   & Class\,{\footnotesize I} & 1, 39 & 600 \\
    HBC~722     & FUor      & $4.05$ ($^{13}$CO)                & $4.93$ $(0.01)$                   & Class\,{\footnotesize II} & 1, 40 & 600 \\
    V1057~Cyg   & FUor      & $4.3$ ($^{13}$CO)                            & $4.35$ $(0.02)$                   & Class\,{\footnotesize II} & 1, 41 & 600 \\
    V2495~Cyg   & FUor      & $-$                               & $-3.83$ $(0.02)$                  & Class\,{\footnotesize I/II} & 42 & 800 \\
    RNO~127     & FUor      & $-$                               & $-2.90$ $(0.01)$                  & $-$ & $-$ & 800 \\
    CB~230      & FUor      & $2.78$ (N$_2$H$^+$)                               & $2.79$ $(0.01)$                   & Class\,{\footnotesize 0/I} & 27, 43 & $-$ \\
    V1735~Cyg   & FUor      & $4.05$ ($^{13}$CO)                & $3.80$ $(0.01)$                   & Class\,{\footnotesize II} & 1, 44, 45 & 900 \\
    HH~354~IRS  & FUor      & $-1.1$ (CS)                       & $-1.52$ $(0.01)$                 & Class\,{\footnotesize 0/I} & 46, 47 & 750 \\
    V733~Cep     & FUor     & $-17.83$ ($^{13}$CO)              & $-8.93$ $(0.01)$                  & Class\,{\footnotesize II}$^{***}$ & 1, 27 & 800 \\
    \hline      
    \end{tabular}
    \begin{tablenotes}
      \small
      \item \textbf{Notes.} The first column lists the name, the second the type of object, while third and fourth columns we list $\varv_{\rm LSR}$ from literature observations (primarily of CO) and our $\varv_{\rm LSR}$ derived from ammonia observations. The fifth column lists the classification (if available), sixth the references, and finally the distances. Errors are given in parentheses.
      \item ${^*}$ -- Adopted from \citet{audard2014}, exceptions are: RNO~1B/1C \citep{bailerjones-edr3}, V512~Per \citep{bailerjones-edr3}, Z~CMa \citep{dong2022}, HH~354~IRS \citep{reipurth1997b}. In these cases we detected water masers, and adopted updated distance values in our Paper\,{\footnotesize II}.
      \item $^{**}$ -- RNO~1B/1C is counted into the Class {\footnotesize II} statistics in Sect.~\ref{sect:discussion-class}.
      \item 1 -- \citet{feher2017}, 2 -- \citet{quanz2007}, 3 -- \citet{diaz-rodriguez2022}, 4 -- \citet{sandell1998}, 5 -- \citet{cruz-saenz2023}, 6 -- \citet{fuller1995}, 7 -- \citet{alma2015},  8 -- \citet{zapatava2015}, 9 -- \citet{mathieau1996}, 10 -- \citet{herbig1990}, 11 -- \citet{stecklum2007}, 12 -- \citet{banzatti2014}, 13 -- \citet{abraham2018}, 14 -- \citet{gianni2016}, 15 -- \citet{kospal2017b}, 16 -- \citet{kospal2021}, 17 -- \citet{parsamian2004}, 18 -- \citet{white2019}, 19 -- \citet{ruiz-rodriguez2017}, 20 -- \citet{zurlo2017}, 21 -- \citet{herbig1977}, 22 -- \citet{abraham2004b}, 23 -- \citet{principe2018}, 24 -- \citet{stojimirovic2008}, 25 -- \citet{park2021}, 26 -- \citet{wouterloot1989}, 27 -- \citet{connelley2018}, 28 -- \citet{moriarty2008}, 29 -- \citet{reipurth2012}, 30 -- \citet{takami2019}, 31 -- \citet{kospal2015}, 32 -- \citet{gramajo2014}, 33 -- \citet{miller2015}, 34 -- \citet{connelley2010}, 35 -- \citet{kospal2006}, 36 -- \citet{kospal2008}, 37 -- \citet{szegedi-elek2020}, 38 -- \citet{torrelles1986}, 39 -- \citet{hillenbrand2013}, 40 -- \citet{kospal2016}, 41 -- \citet{szabo2021}, 42 -- \citet{liu2018}, 43 -- \citet{chen2007}, 44 -- \citet{harvey2008}, 45 -- \citet{kospal2011}, 46 -- \citet{bronfman1996}, 47 -- \citet{reipurth1997a}
      \item $^{***}$ -- Based on the extreme similarities to FU~Ori from optical and NIR spectra
    \end{tablenotes}
\end{table*}

\section{H$_{2}$ column density and dust temperature maps}
Fig.~\ref{fig:sed1} shows the H$_{2}$ column density and dust temperature maps derived from the SED fitting described in Sect.~\ref{res:sed}. The H$_{2}$ column density and dust temperature values are listed in Table~\ref{tab:trot-tkin-fuors}, \ref{tab:trot-tkin-exors}.
\begin{figure*}[h]
\centering 
\vspace{0.2cm}
\includegraphics[width=0.35\textwidth]{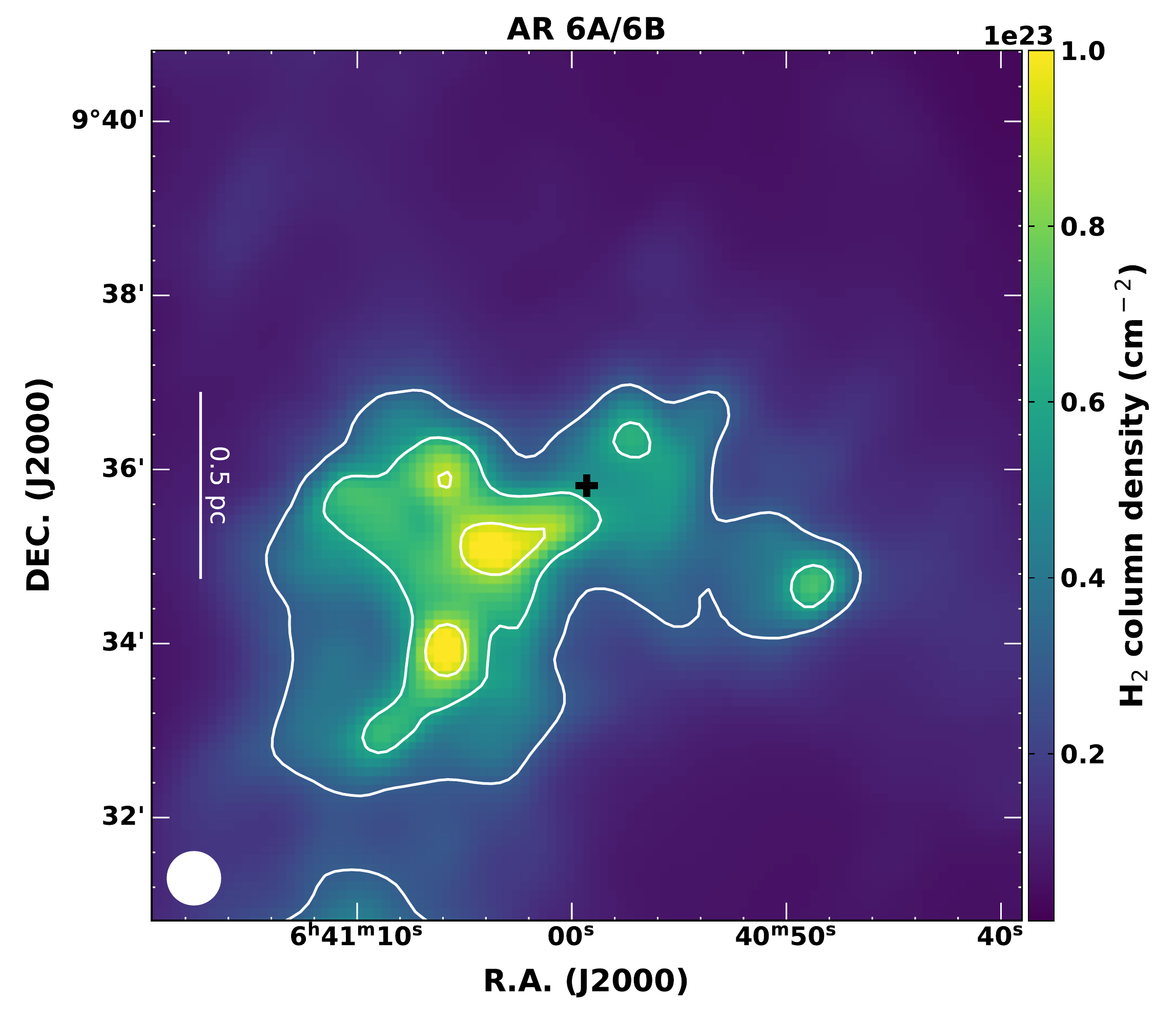}
\hspace{1cm}
\includegraphics[width=0.35\textwidth]{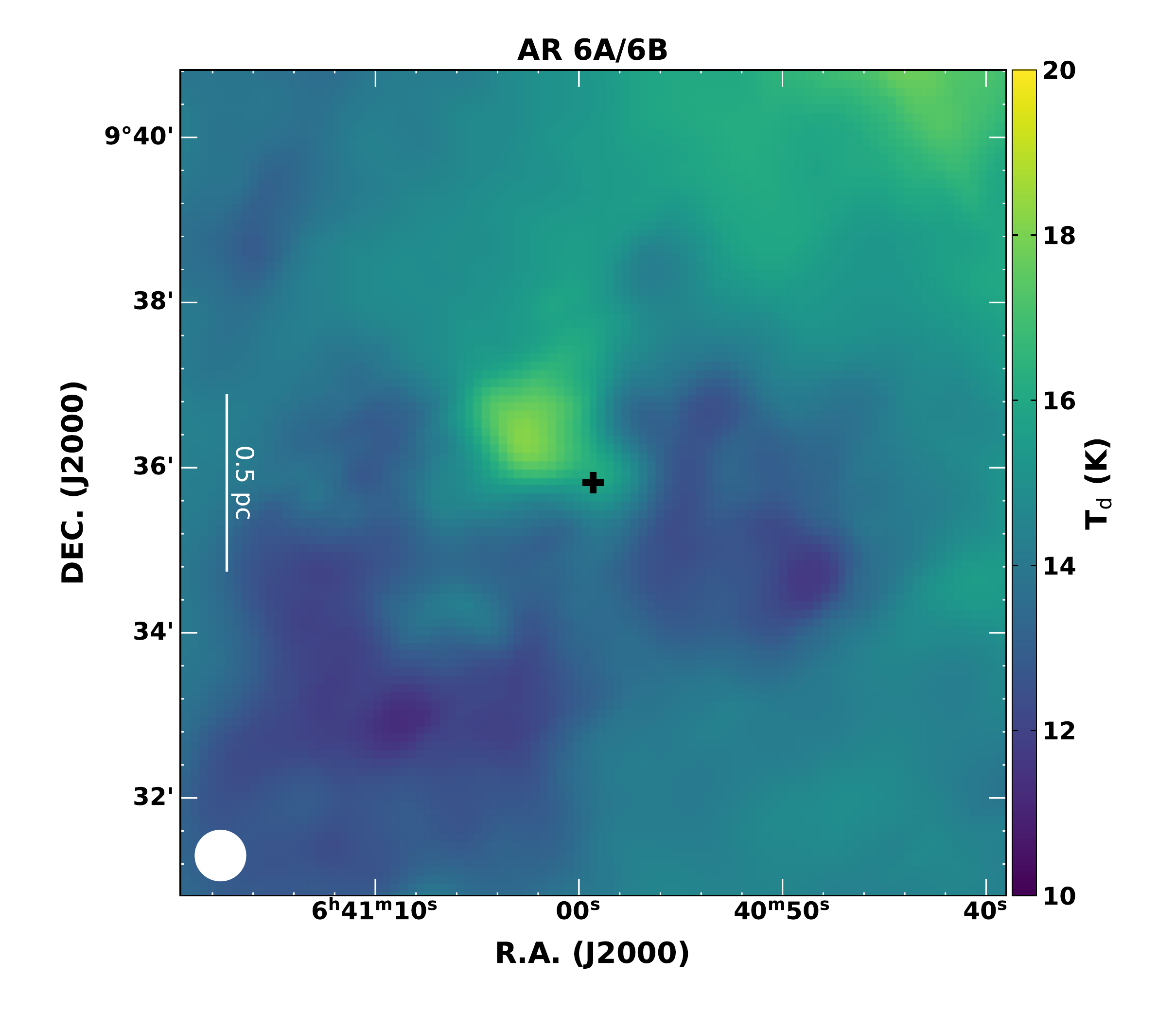} \\
\vspace{0.1cm}
\includegraphics[width=0.35\textwidth]{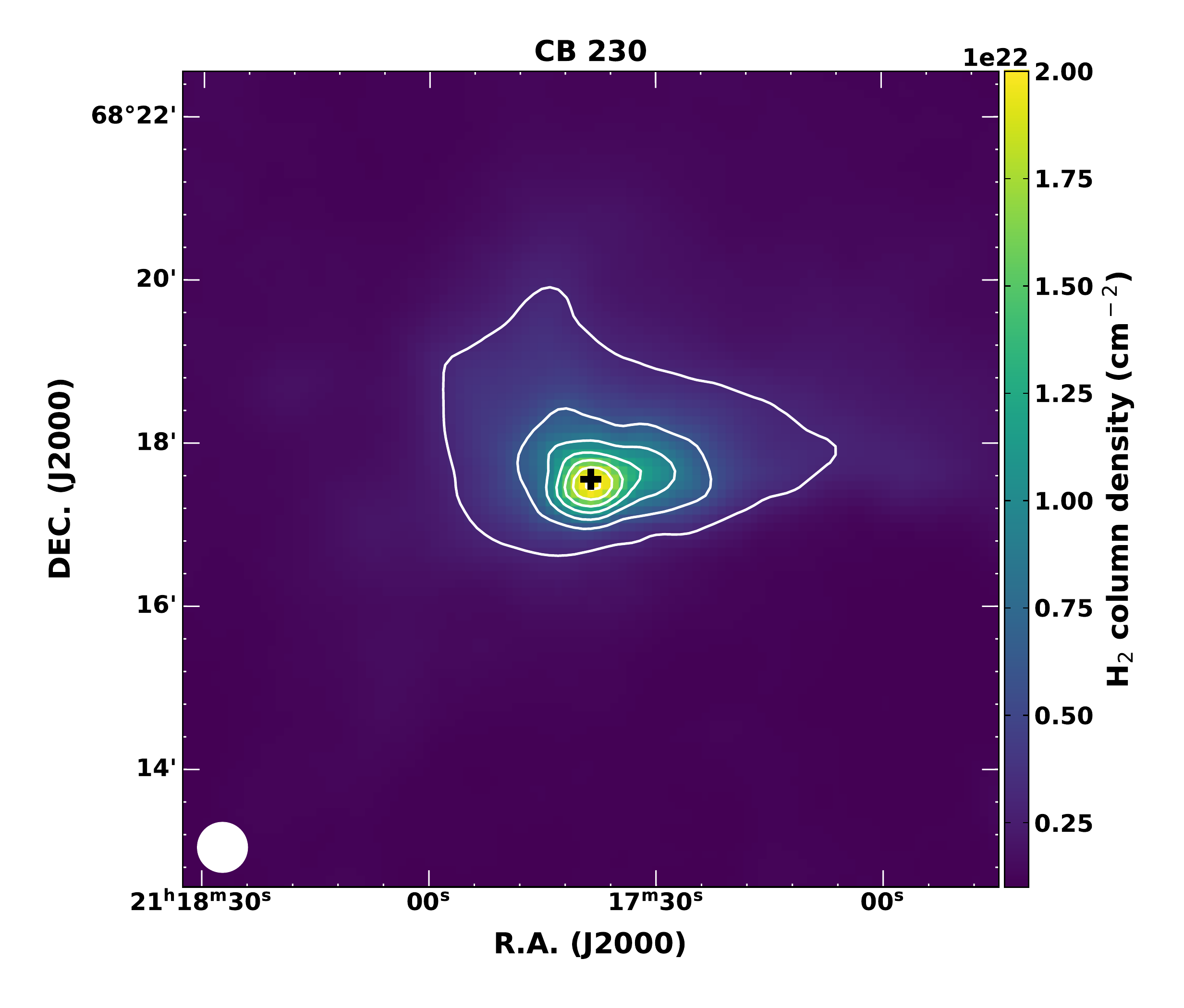} 
\hspace{1cm}
\includegraphics[width=0.35\textwidth]{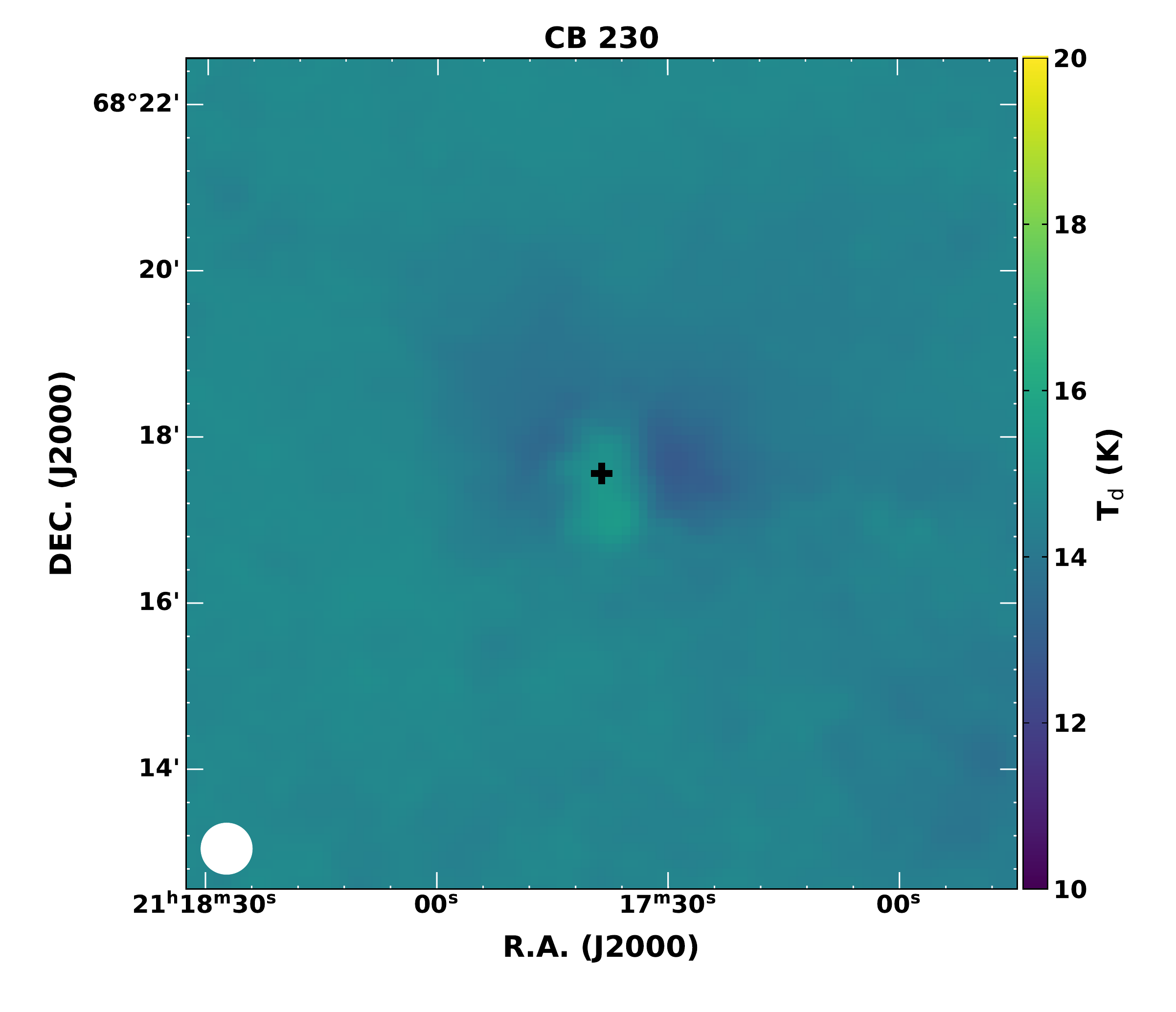} \\ 
\vspace{0.1cm}
\includegraphics[width=0.35\textwidth]{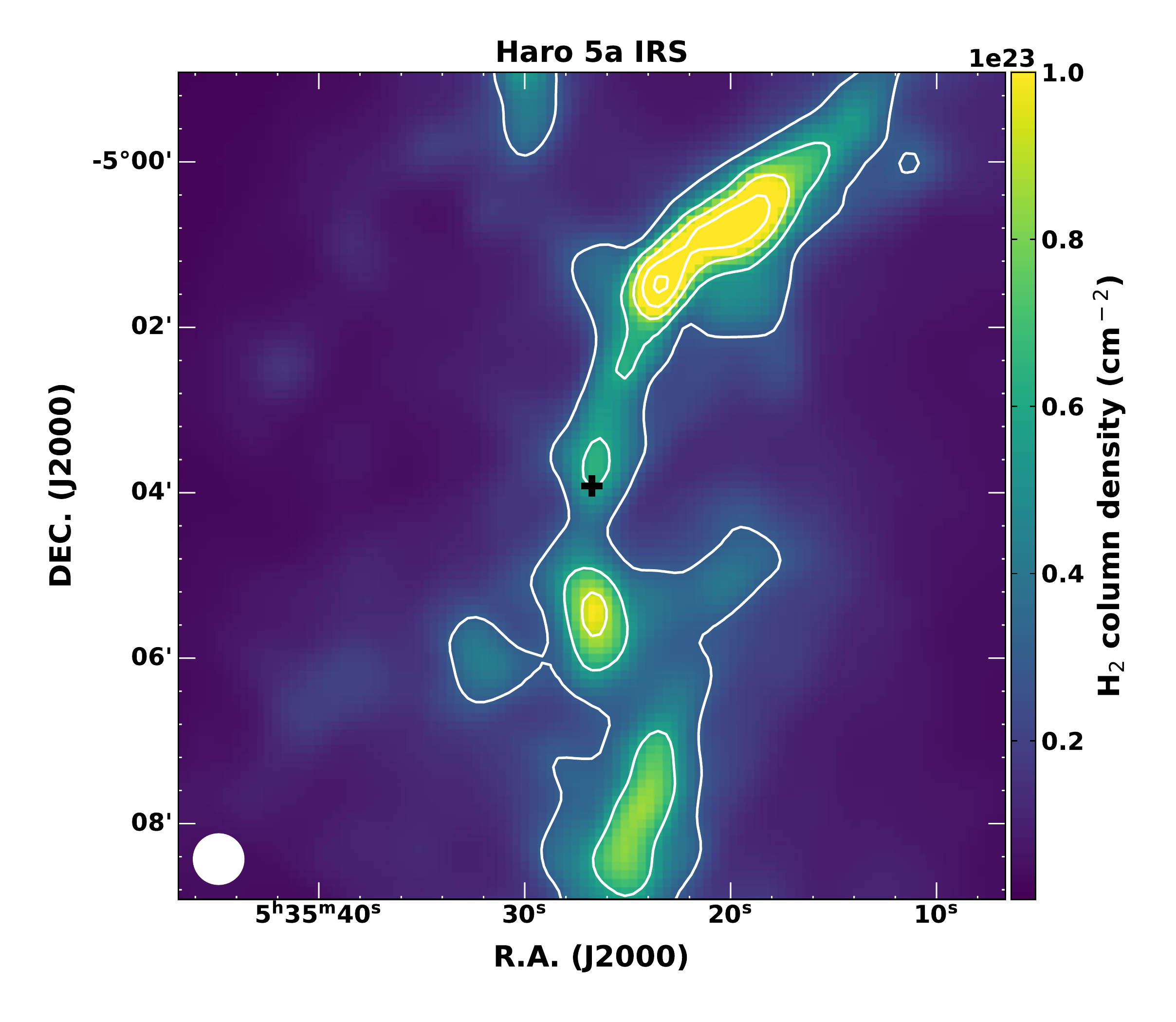}
\hspace{1cm}
\includegraphics[width=0.35\textwidth]{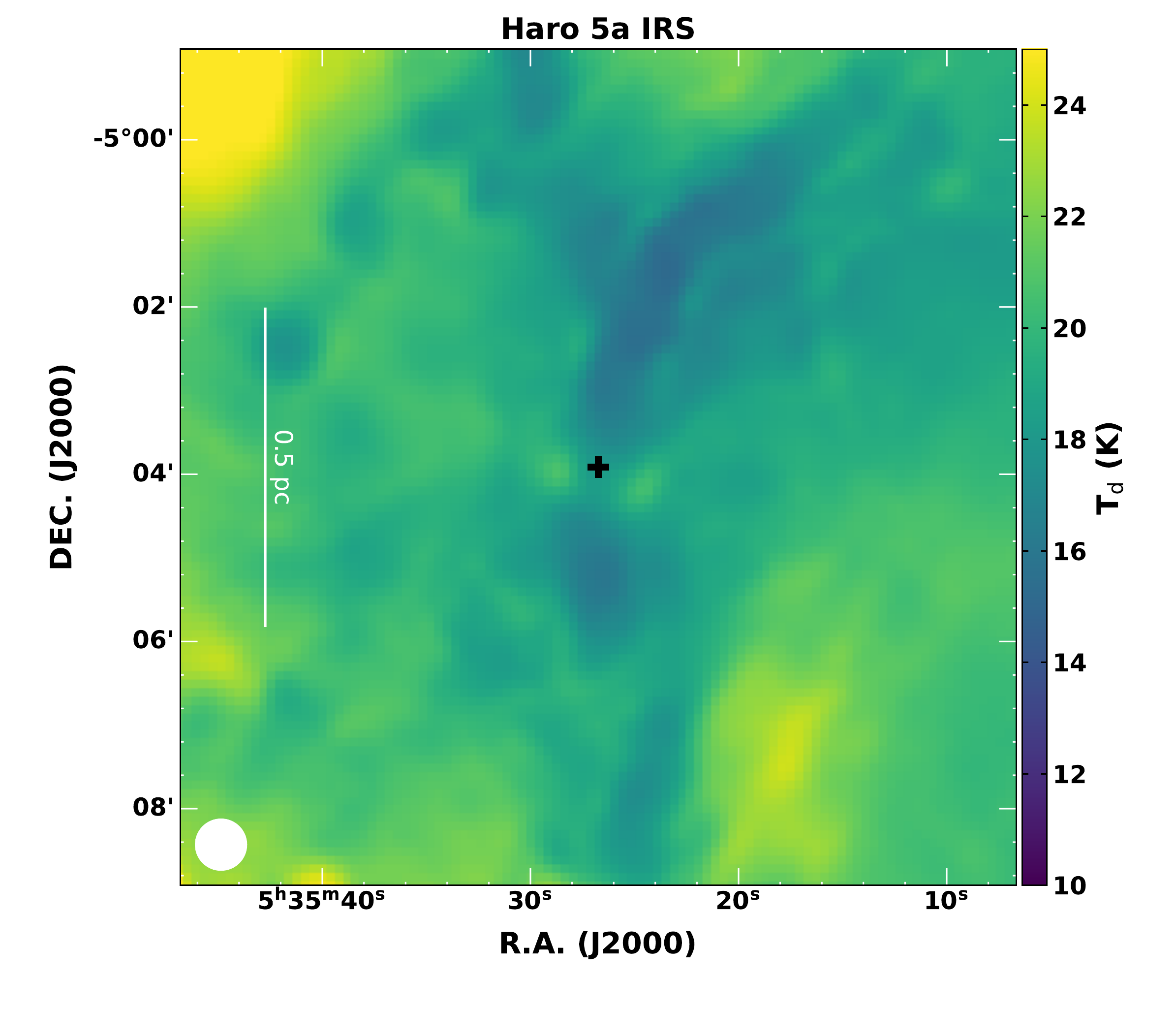} \\
\vspace{0.1cm}
\includegraphics[width=0.35\textwidth]{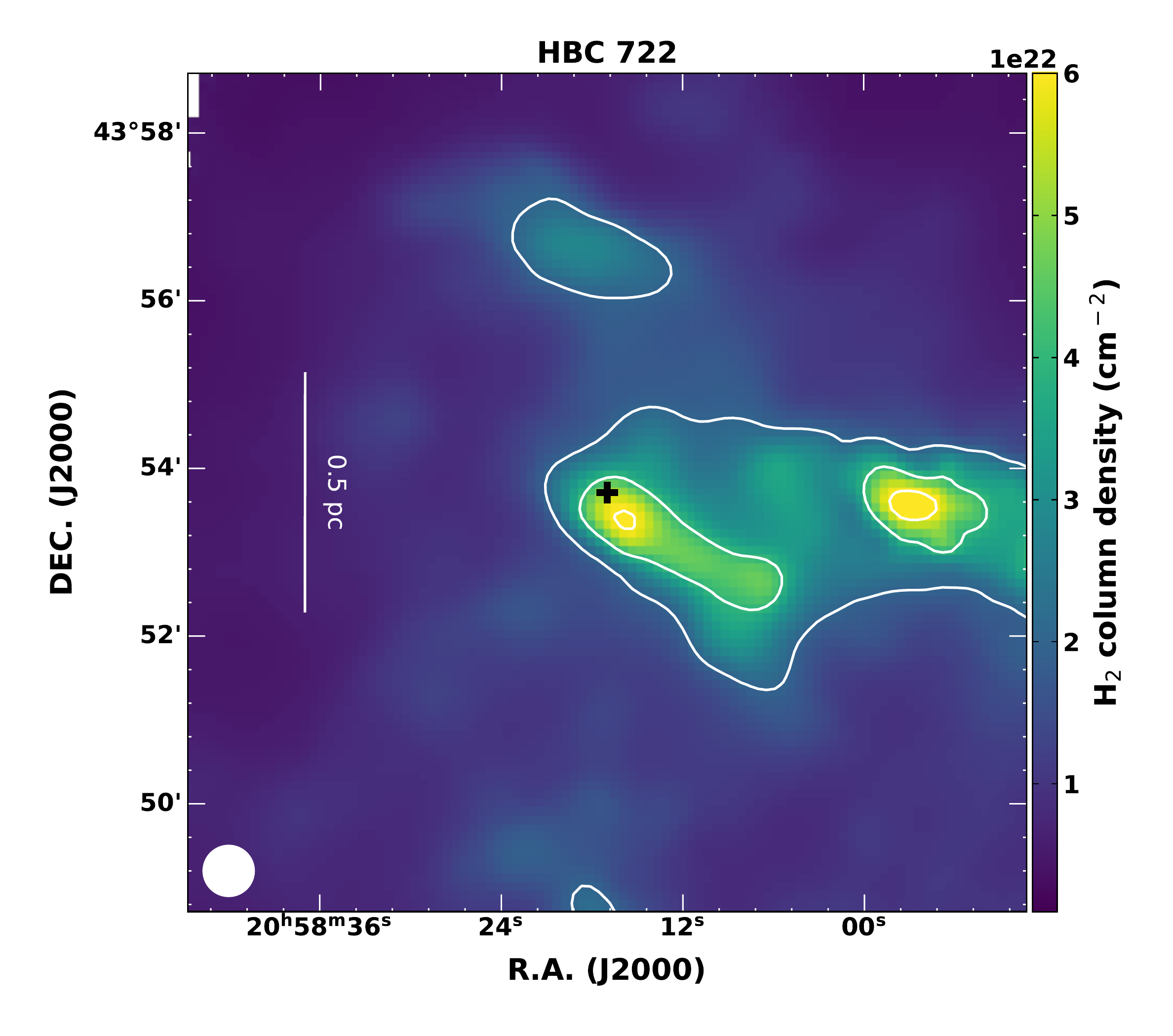}
\hspace{1cm}
\includegraphics[width=0.35\textwidth]{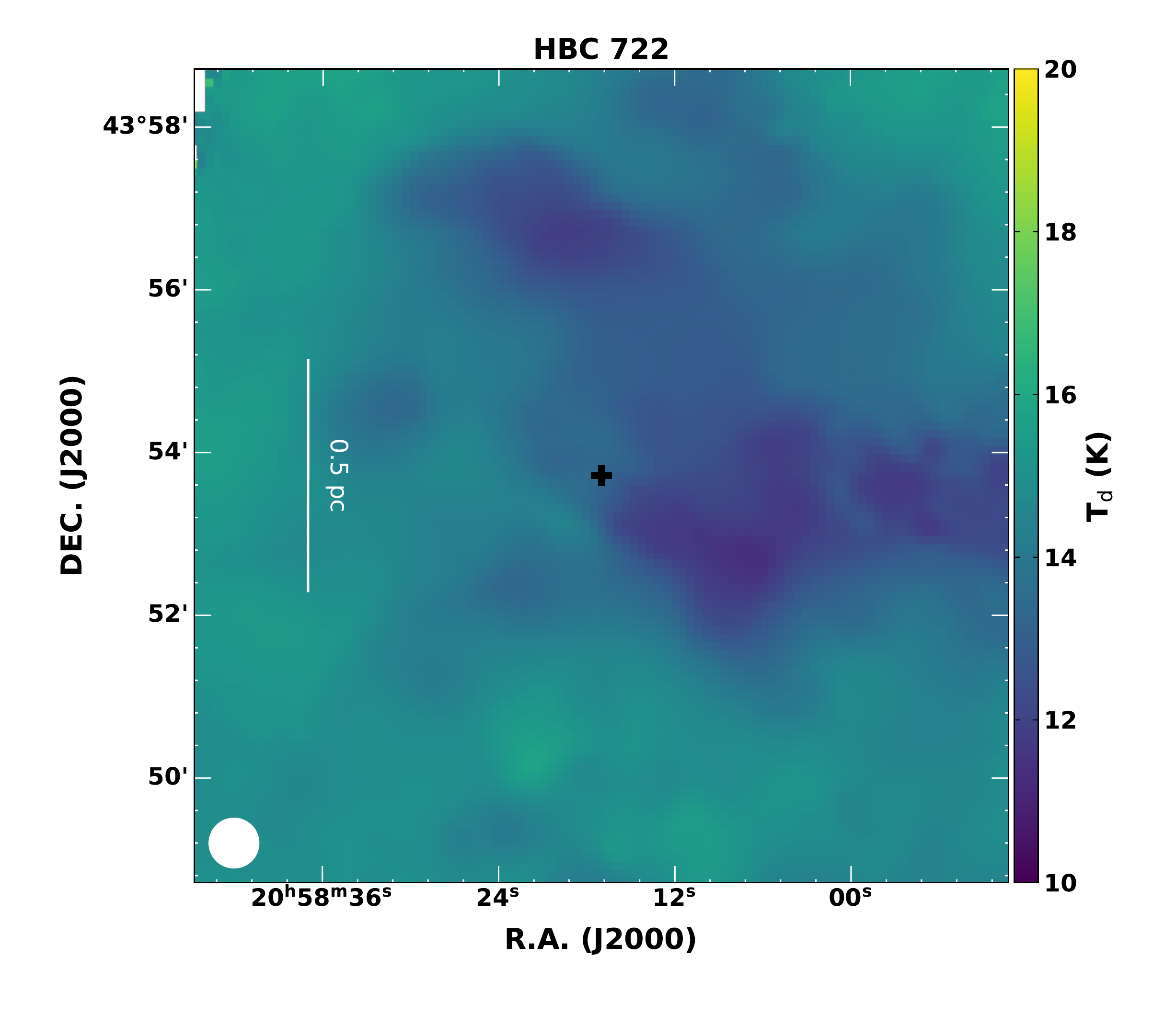} \\
\caption{H$_{2}$ column density (left) and dust temperature (right) maps derived from the pixel-by-pixel SED fitting of the \textit{Herschel} data, convolved to the Effelsberg beam (shown in the bottom left corner). The field of view is the same for all sources, corresponding to $10\arcmin\times10\arcmin$, $+$ symbols represent the pointing positions listed in Tables~\ref{tab:names-nh31,1_parameters_fuors}, \ref{tab:names-nh31,1_parameters_exors} and \ref{tab:appendix-long}, respectively. The physical scale is presented for sources with known distances, taken from the study of \citet{audard2014} or described in the notes of Table~\ref{tab:appendix-long}. The color scale is not the same for all sources.}
\label{fig:sed1}
\end{figure*}

\addtocounter{figure}{-1}
\begin{figure*}[h]
\centering 
\includegraphics[width=0.35\textwidth]{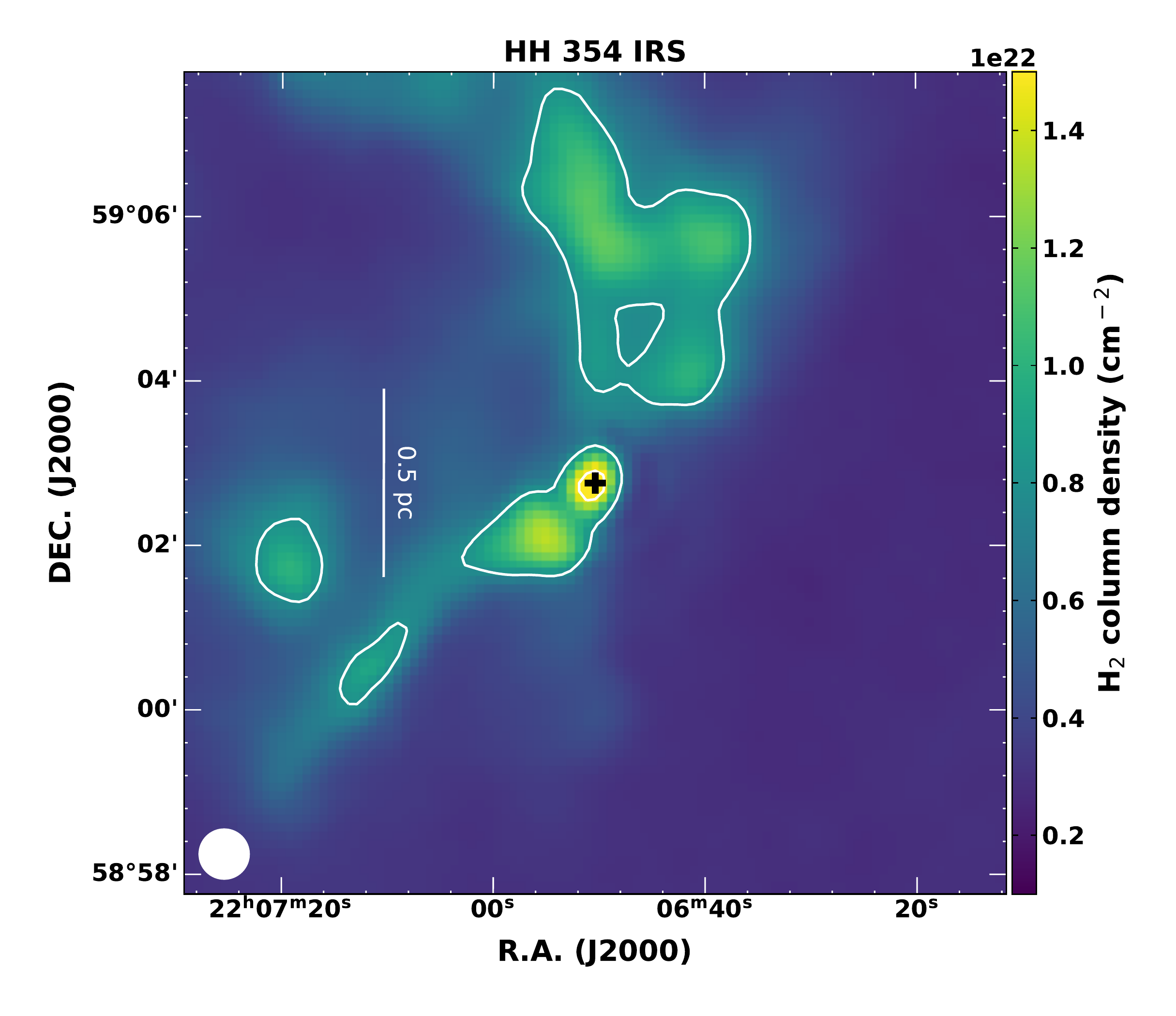}
\hspace{1cm}
\includegraphics[width=0.35\textwidth]{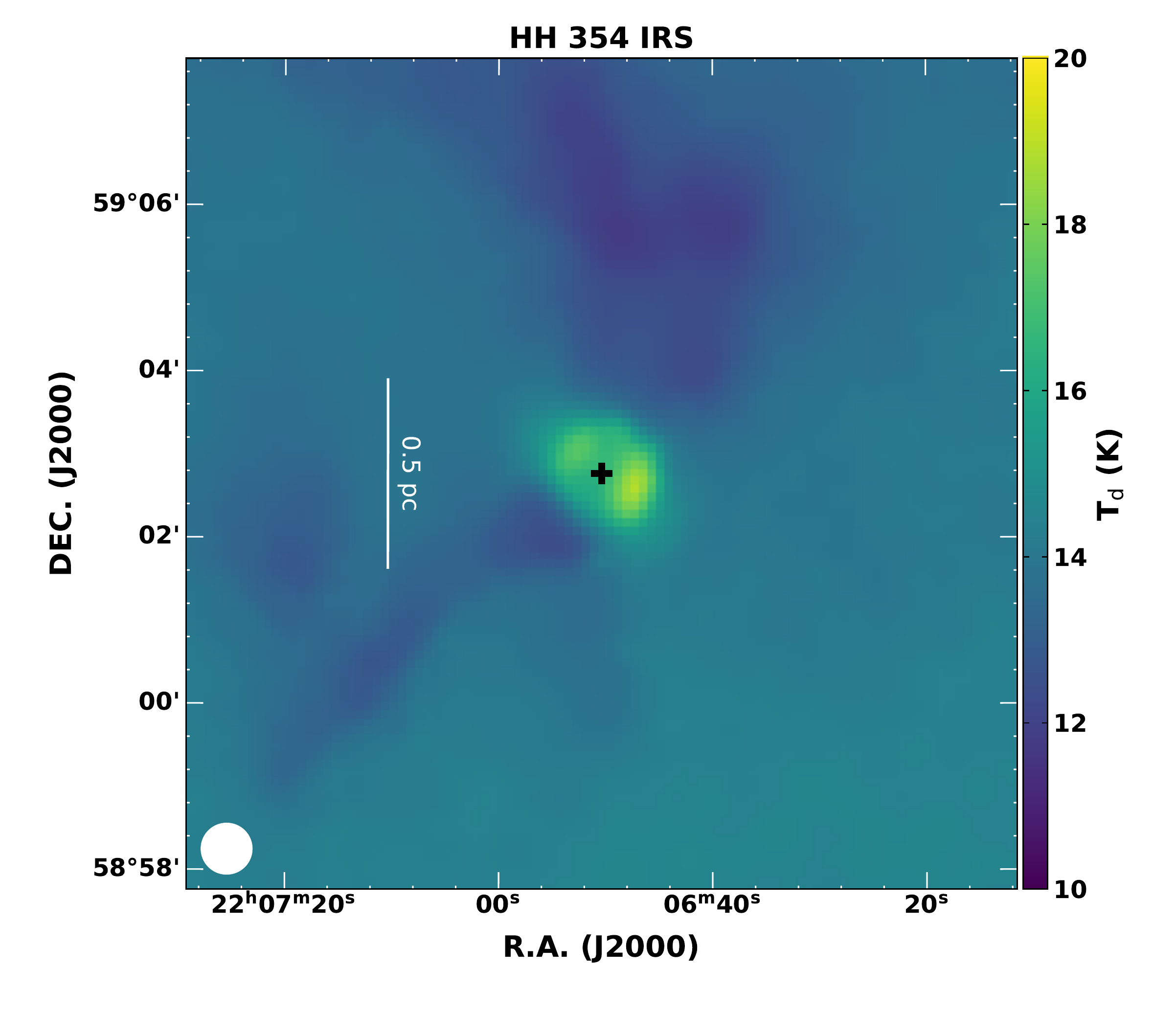} \\
\vspace{0.5cm}
\includegraphics[width=0.35\textwidth]{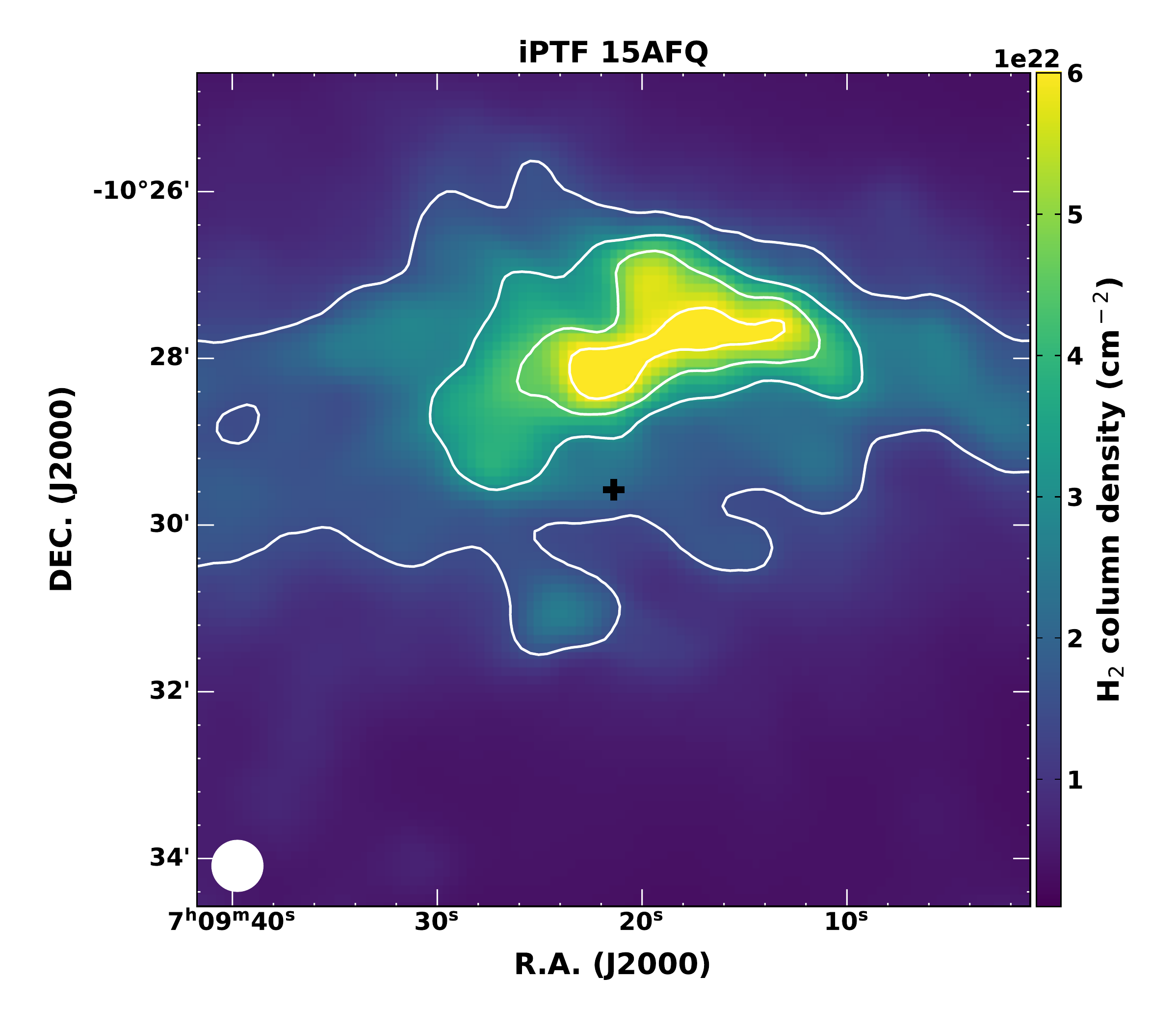} 
\hspace{1cm}
\includegraphics[width=0.35\textwidth]{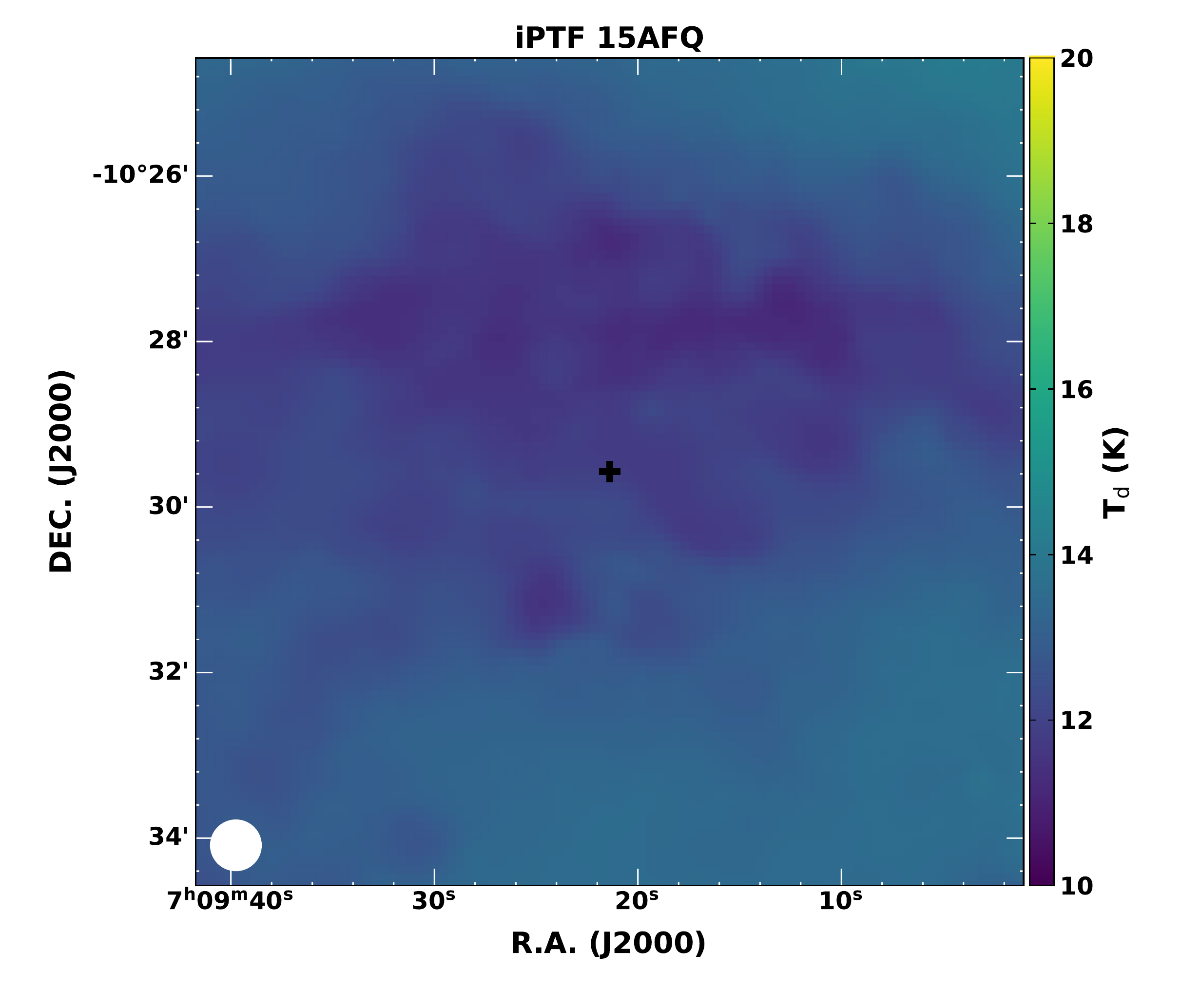} \\ 
\vspace{0.5cm}
\includegraphics[width=0.35\textwidth]{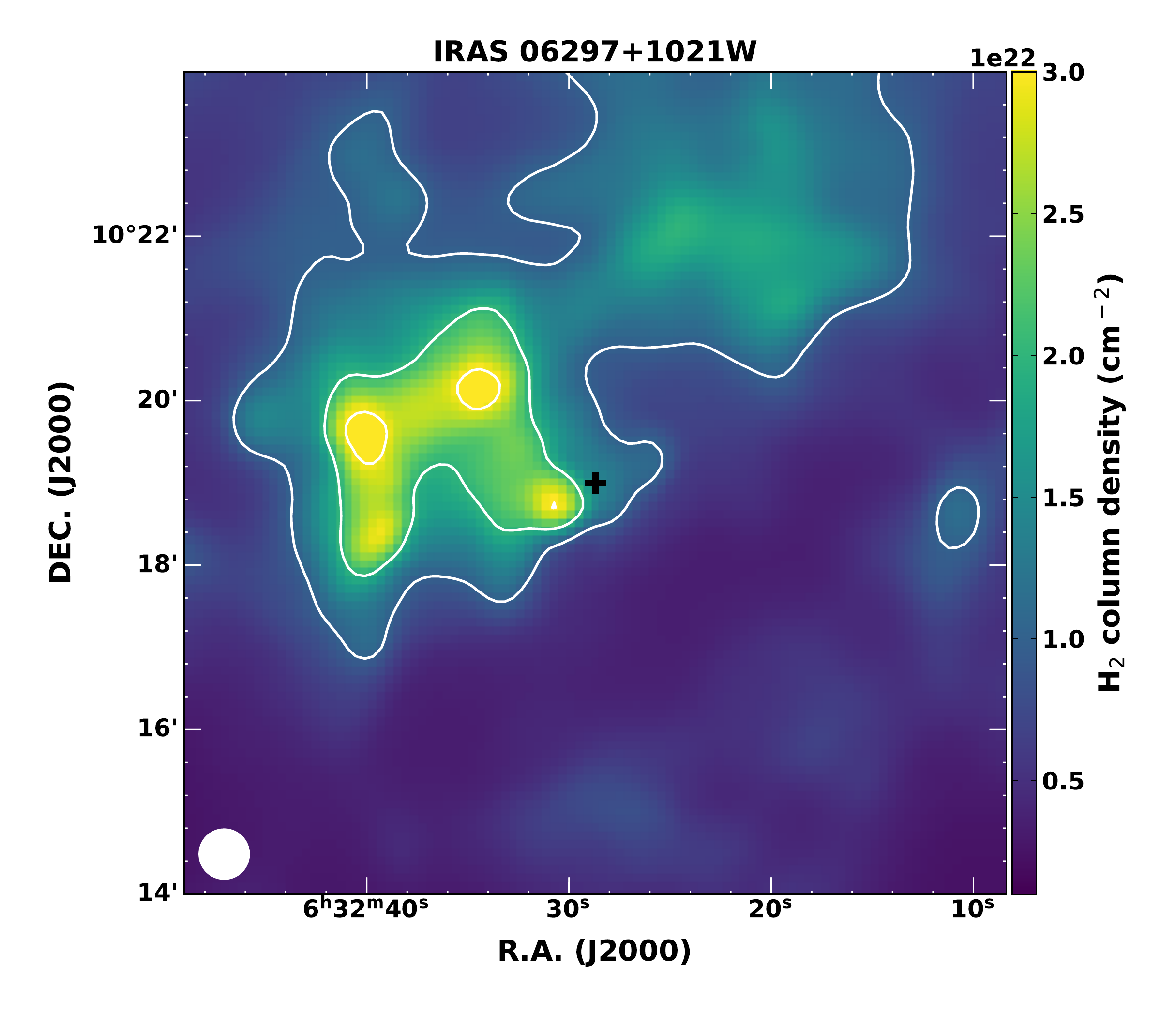}
\hspace{1cm}
\includegraphics[width=0.35\textwidth]{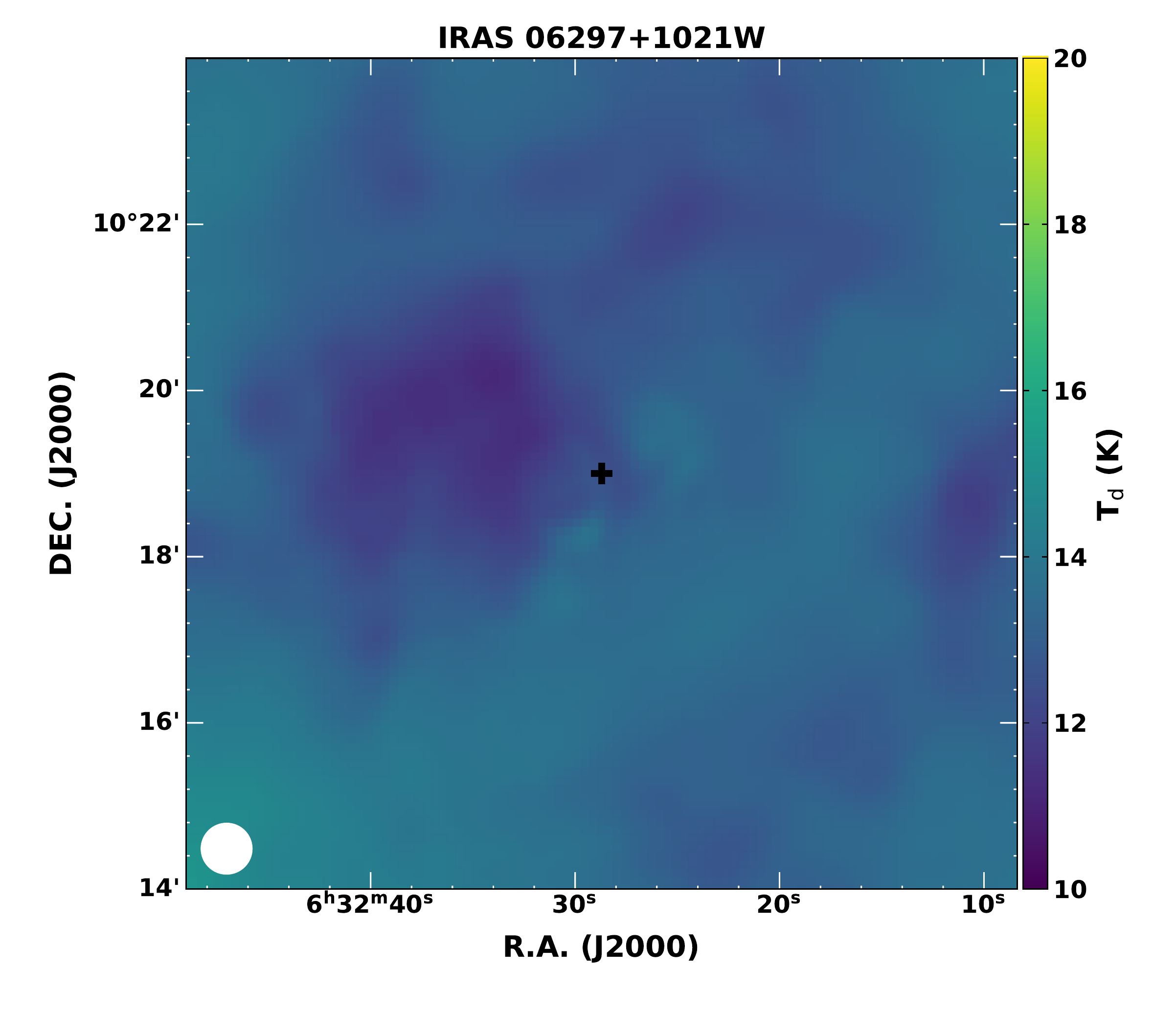} \\
\vspace{0.5cm}
\includegraphics[width=0.35\textwidth]{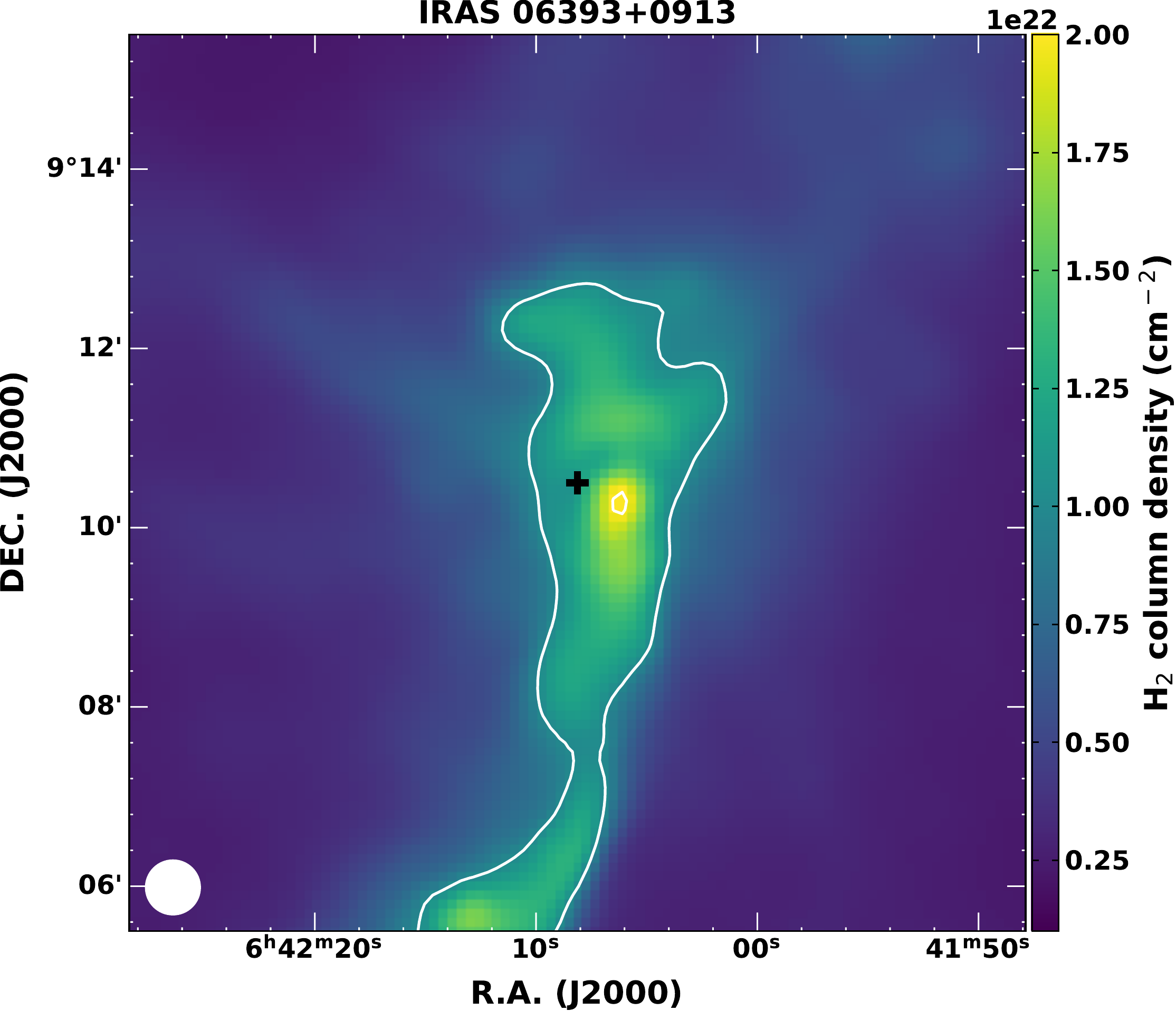}
\hspace{1cm}
\includegraphics[width=0.35\textwidth]{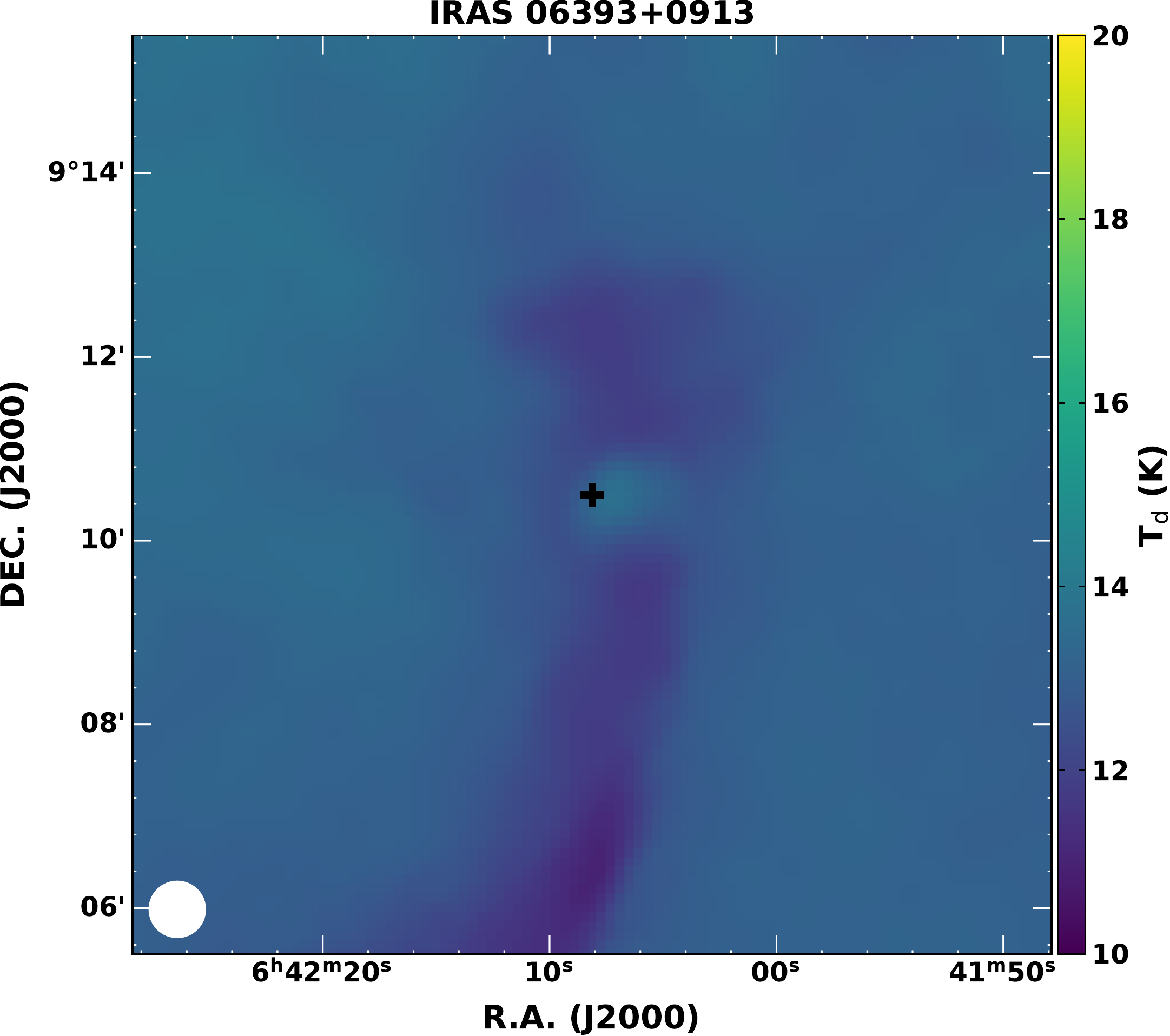} \\
\caption{(Continued.)}
\end{figure*}

\addtocounter{figure}{-1}
\begin{figure*}[h]
\centering 
\includegraphics[width=0.35\textwidth]{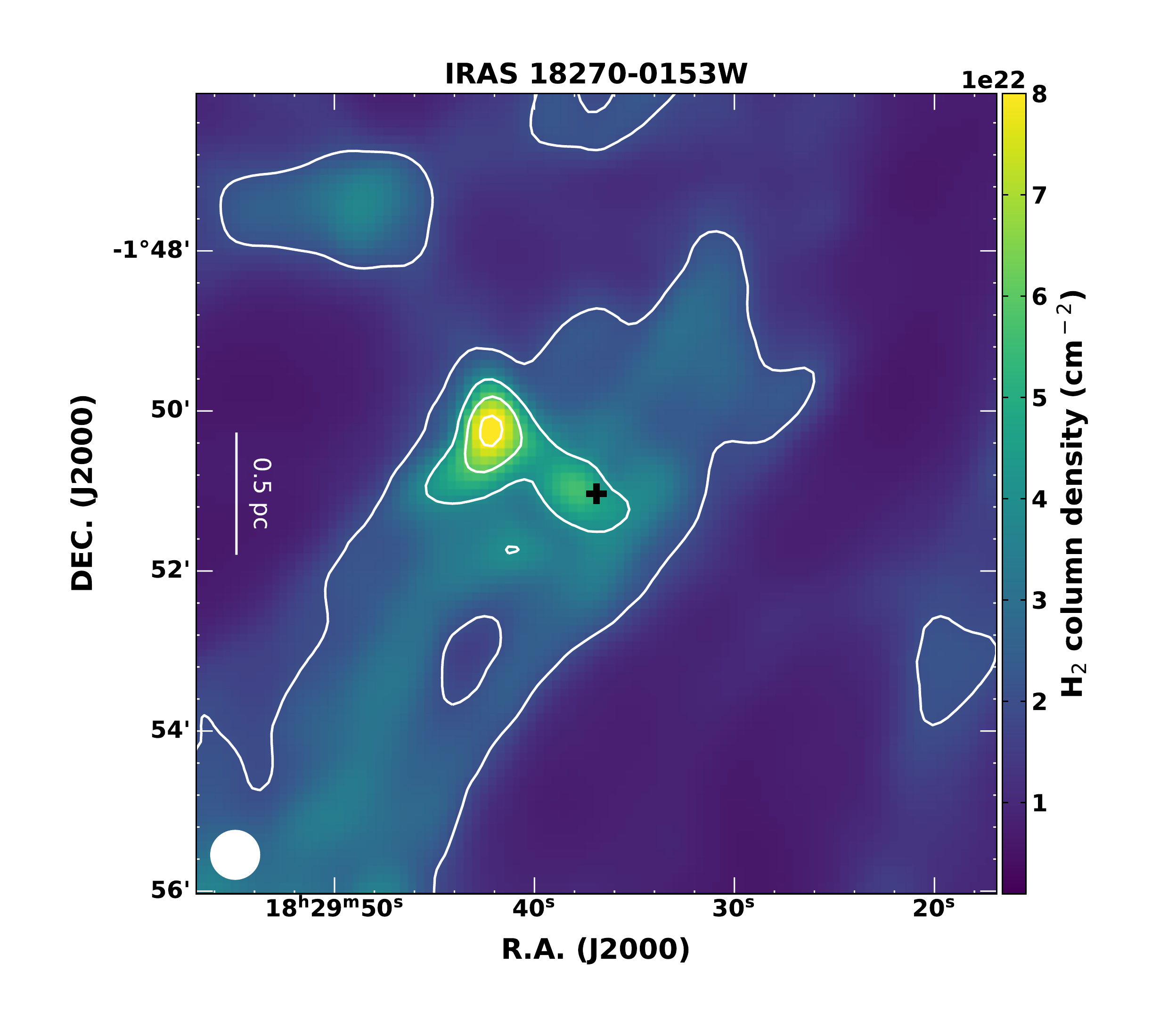}
\hspace{1cm}
\includegraphics[width=0.35\textwidth]{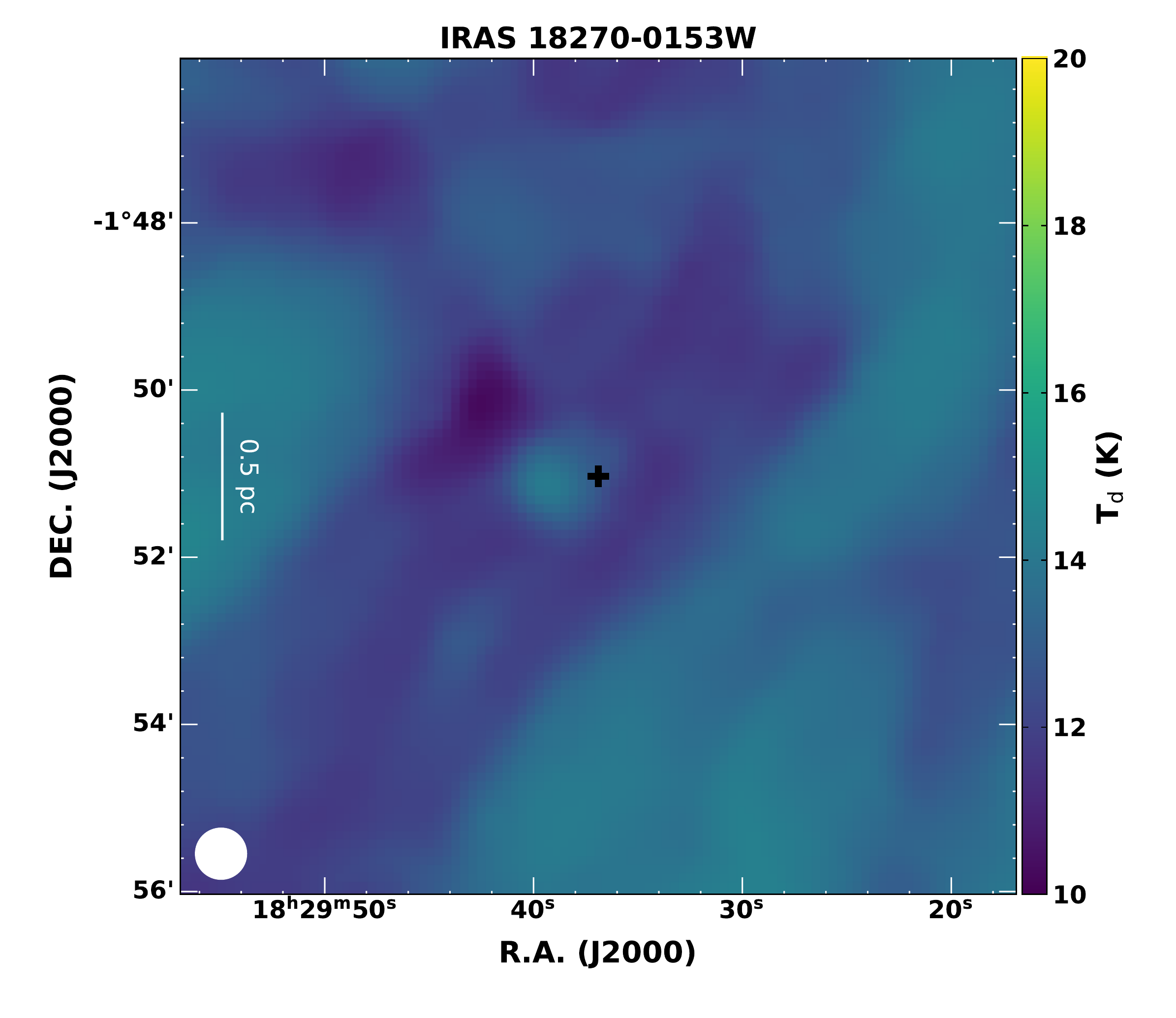} \\
\vspace{0.5cm}
\includegraphics[width=0.35\textwidth]{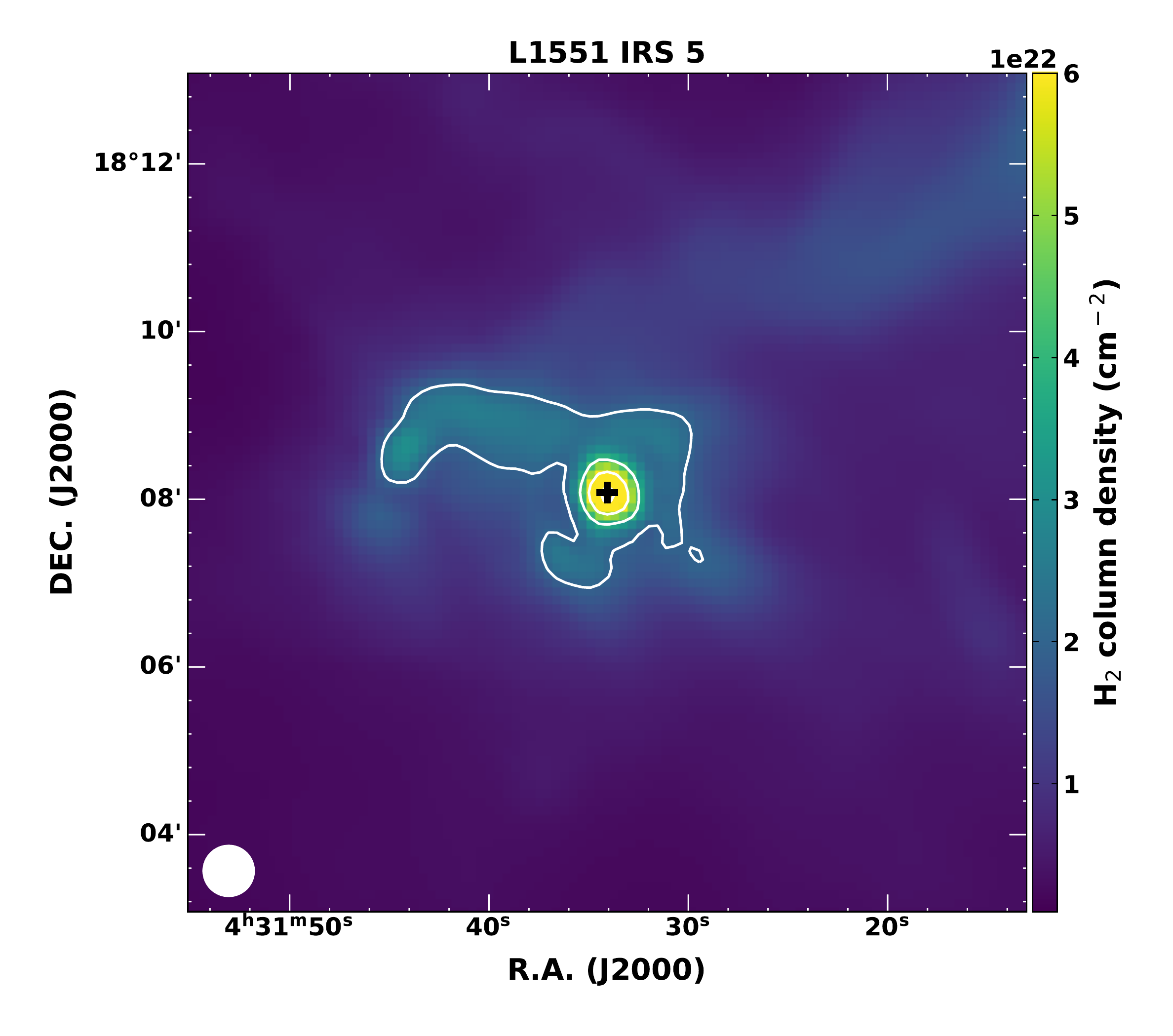} 
\hspace{1cm}
\includegraphics[width=0.35\textwidth]{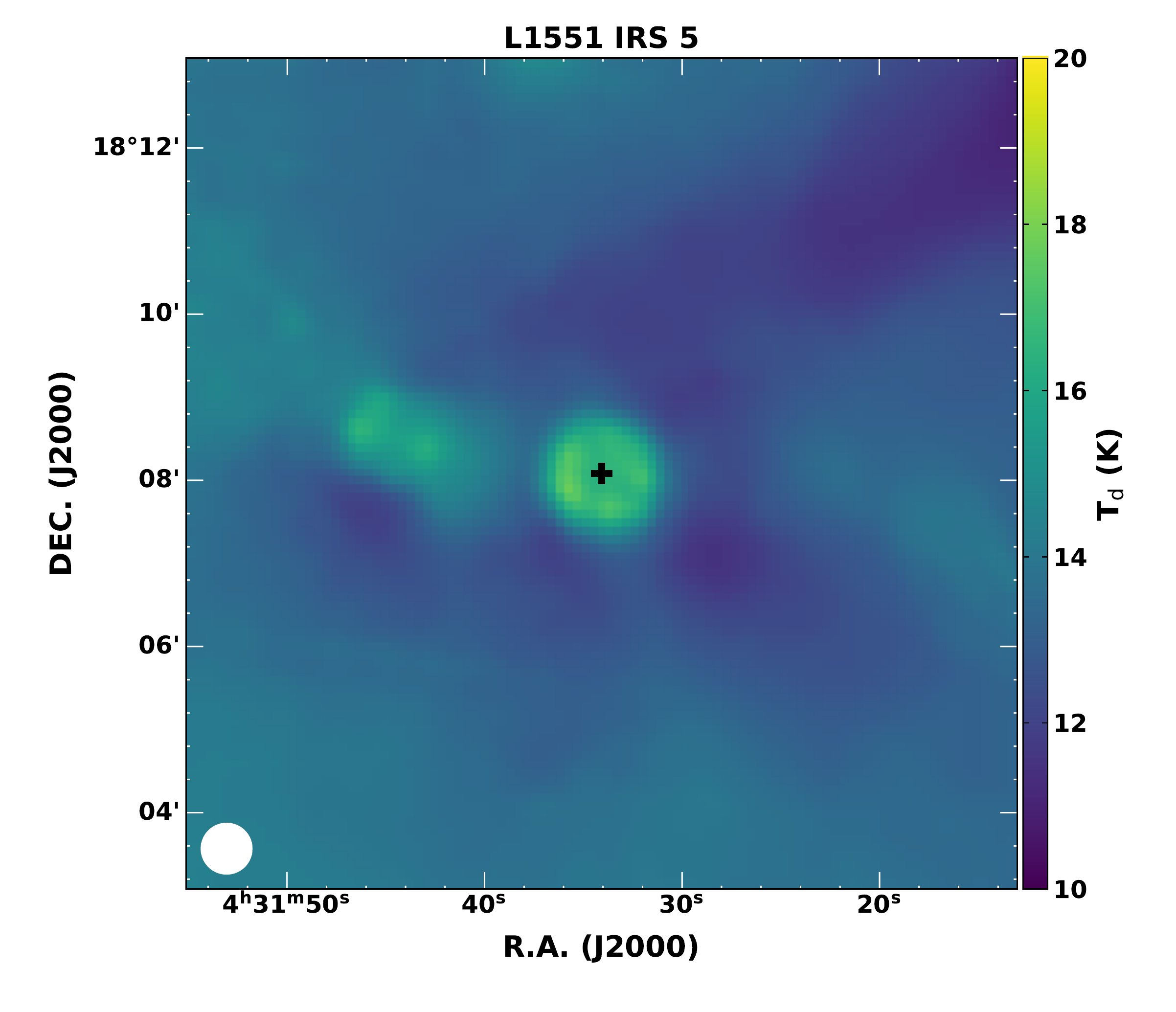} \\ 
\vspace{0.5cm}
\includegraphics[width=0.35\textwidth]{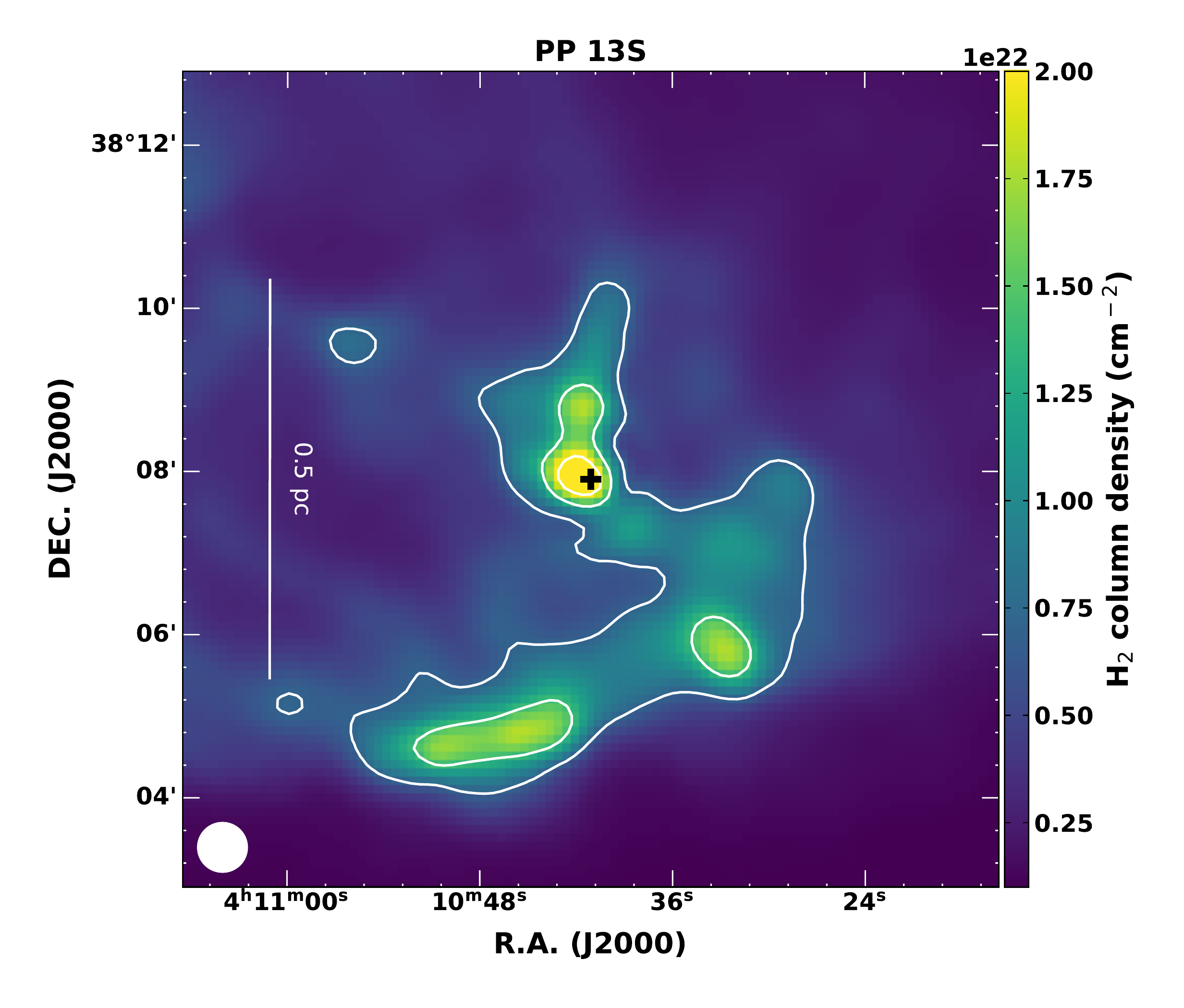}
\hspace{1cm}
\includegraphics[width=0.35\textwidth]{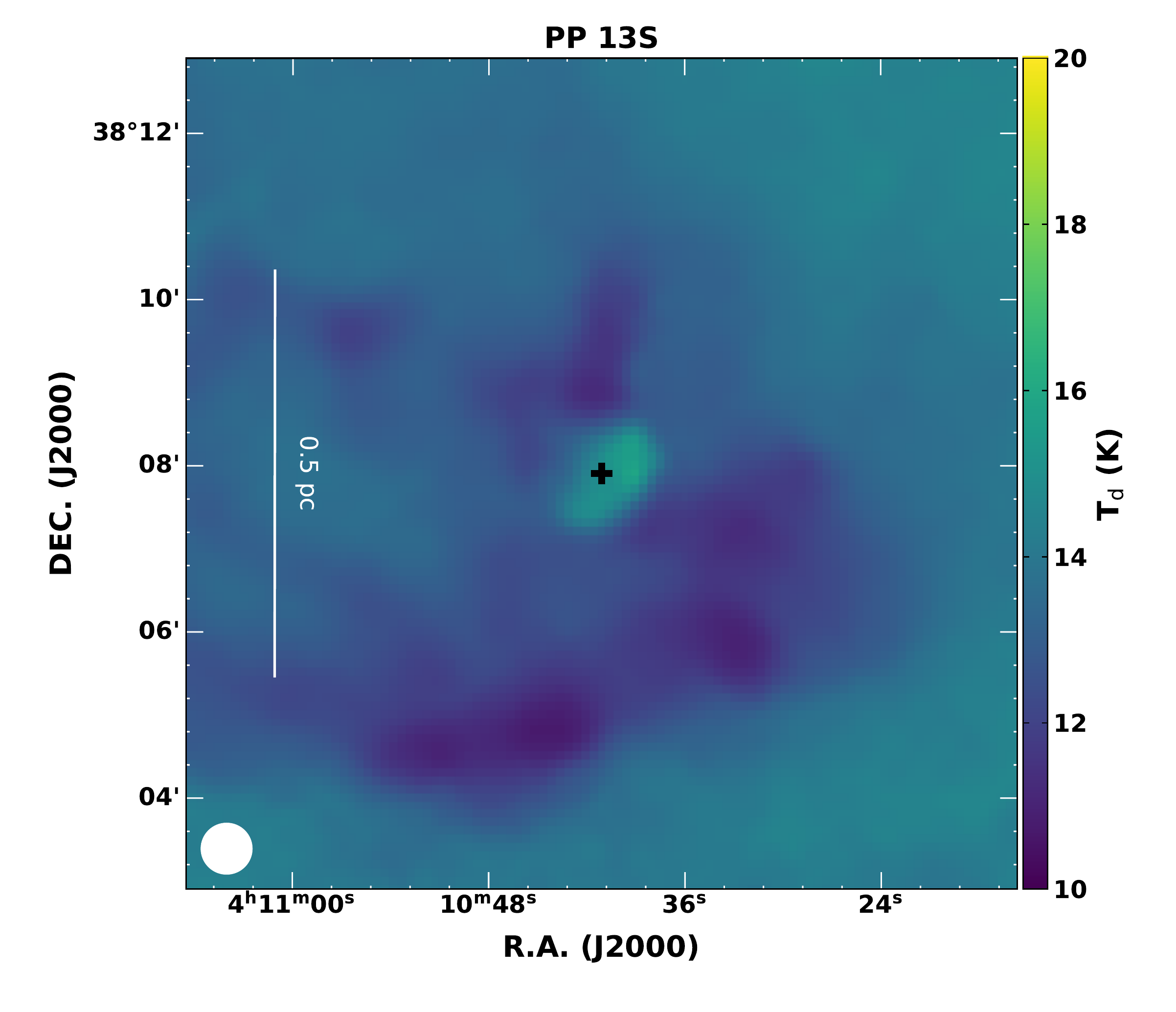} \\
\vspace{0.5cm}
\includegraphics[width=0.35\textwidth]{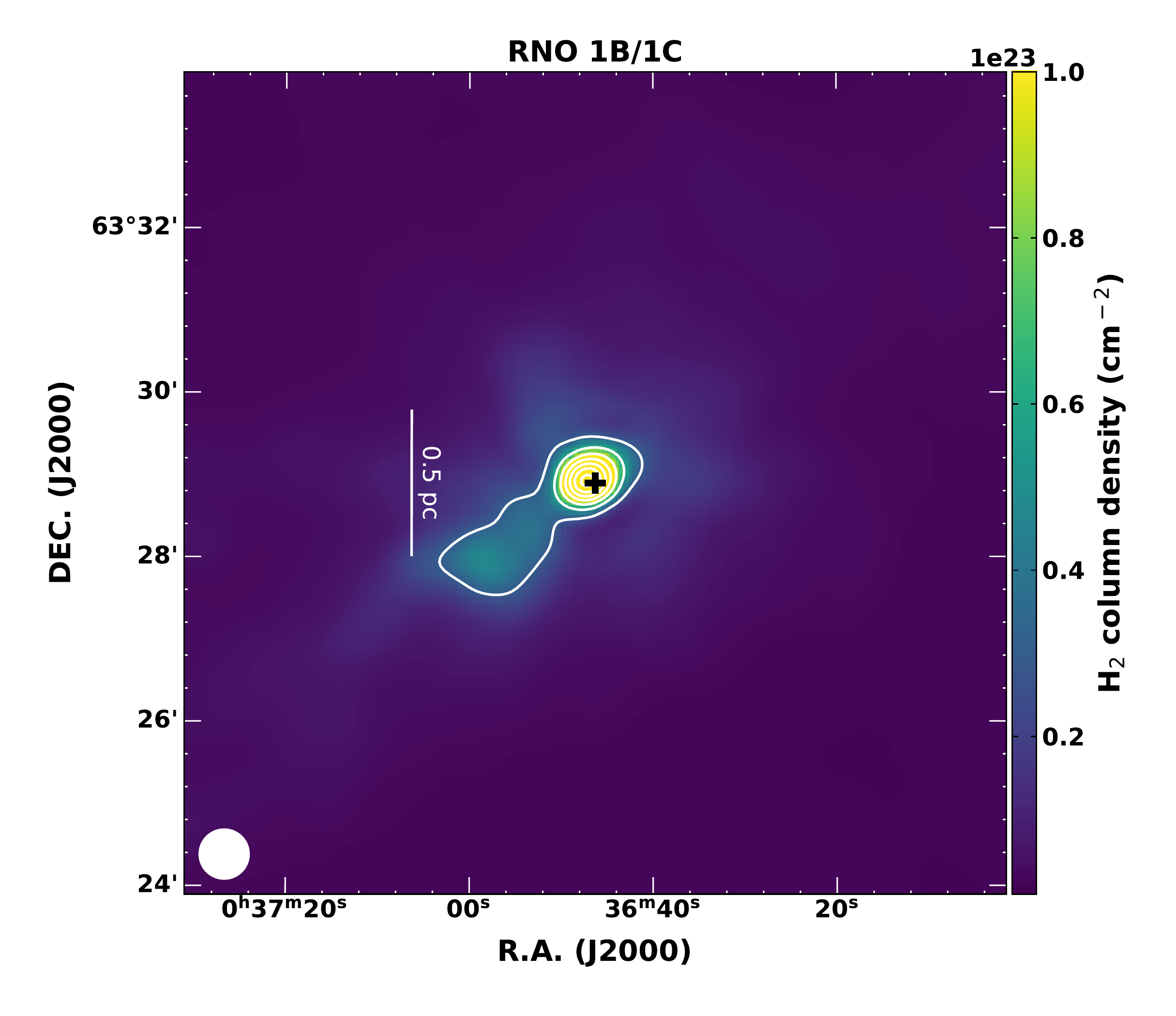}
\hspace{1cm}
\includegraphics[width=0.35\textwidth]{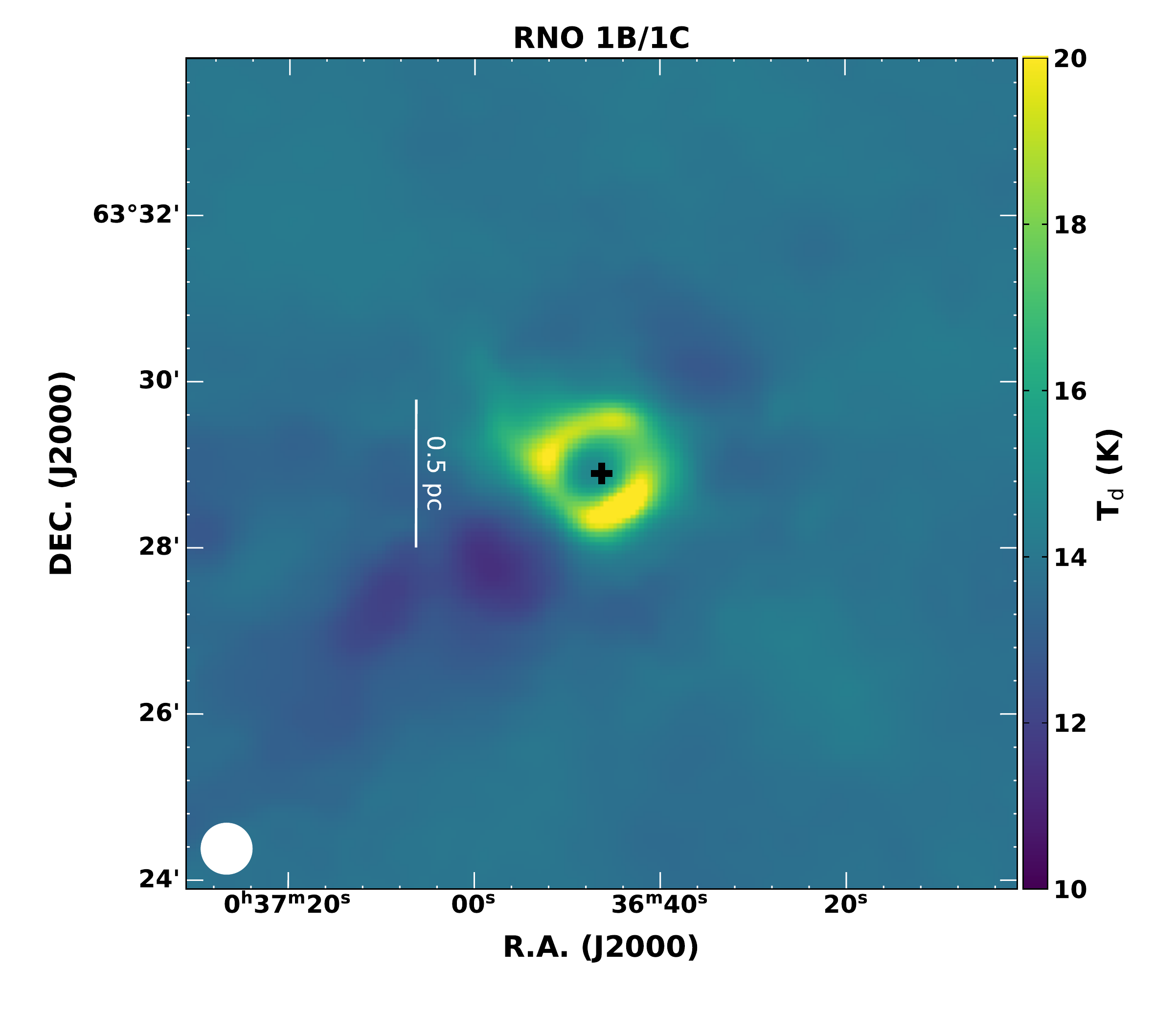} \\
\caption{(Continued.)}
\end{figure*}

\addtocounter{figure}{-1}
\begin{figure*}[h]
\centering 
\includegraphics[width=0.35\textwidth]{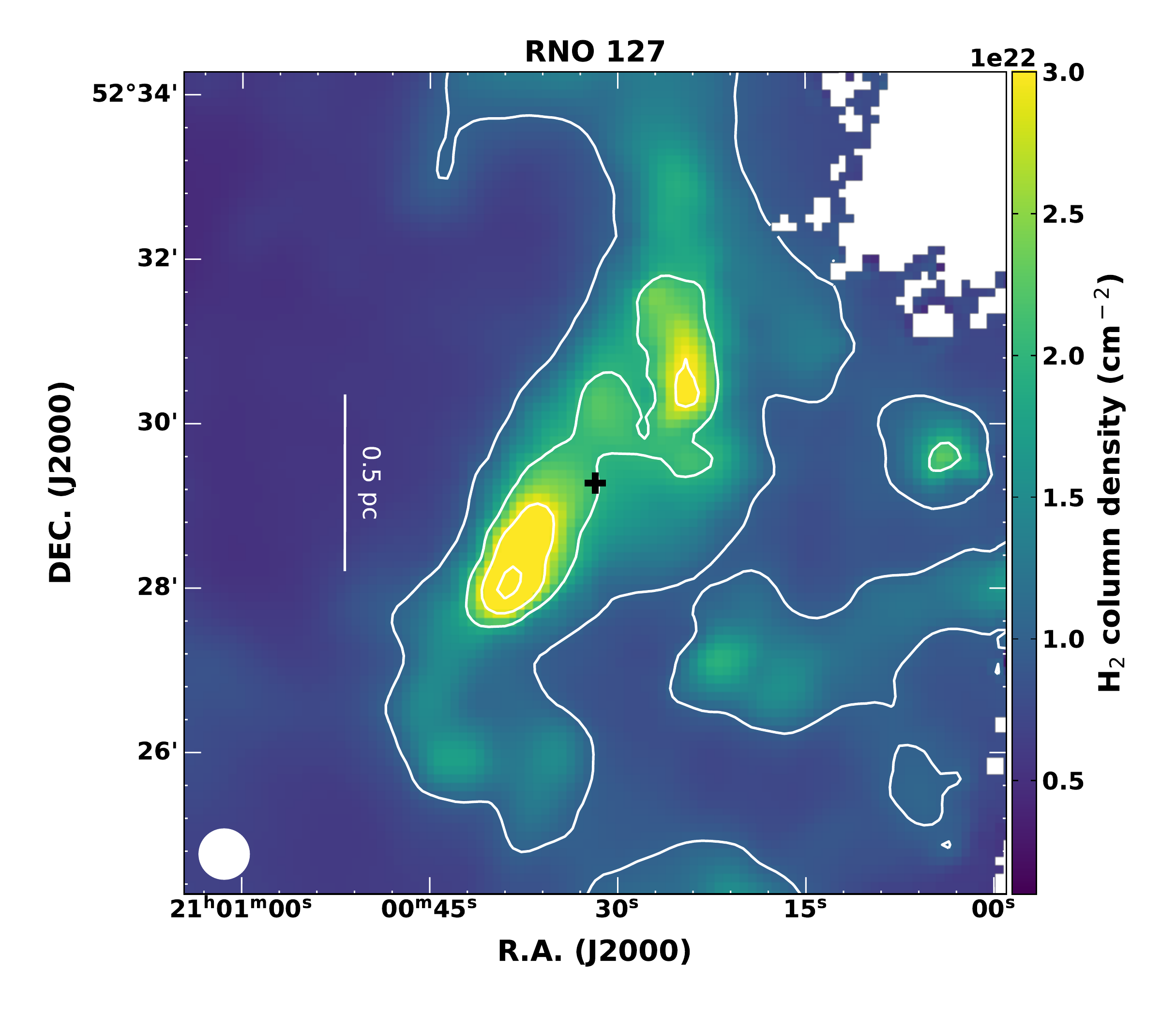}
\hspace{1cm}
\includegraphics[width=0.35\textwidth]{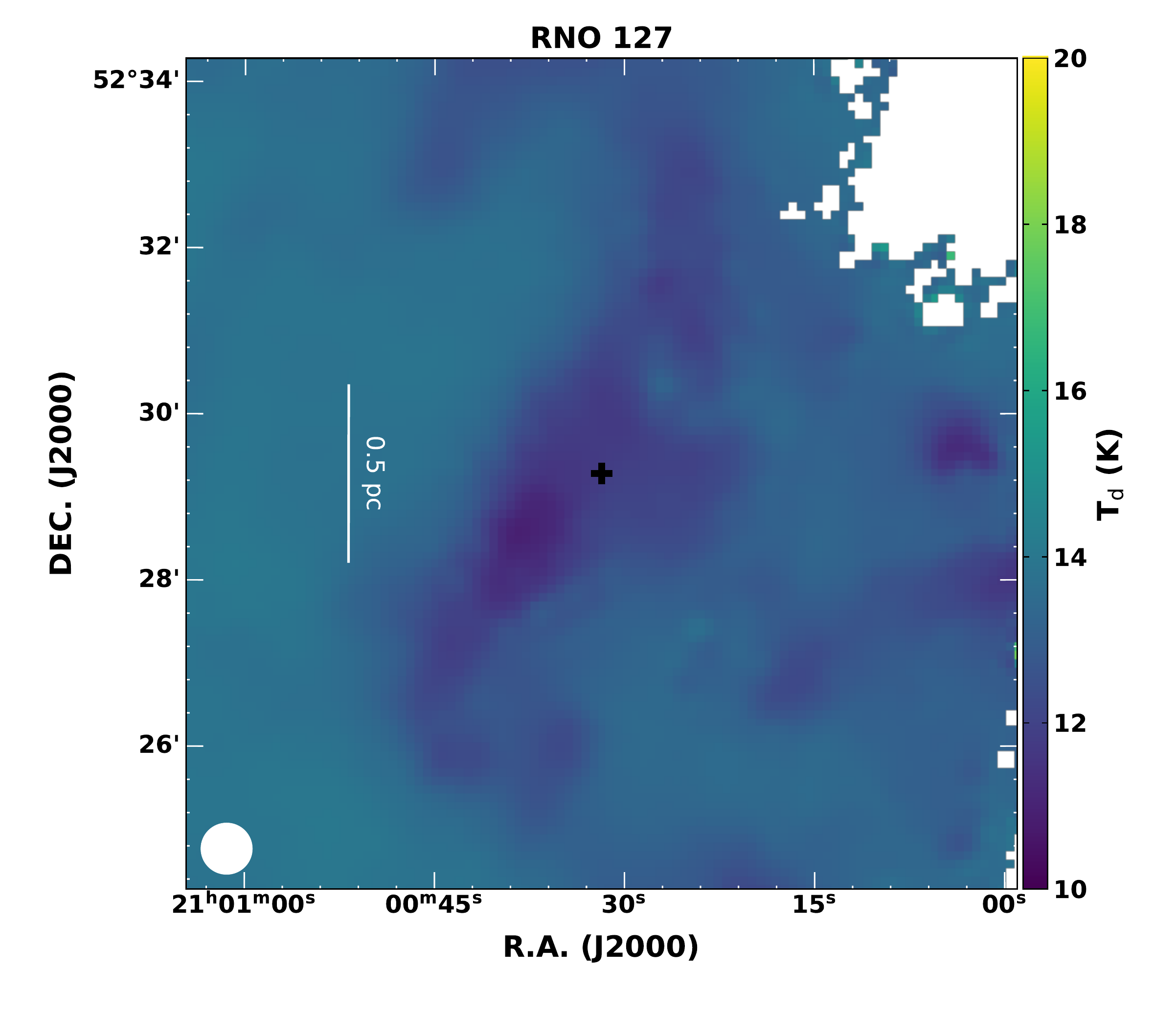} \\
\vspace{0.5cm}
\includegraphics[width=0.35\textwidth]{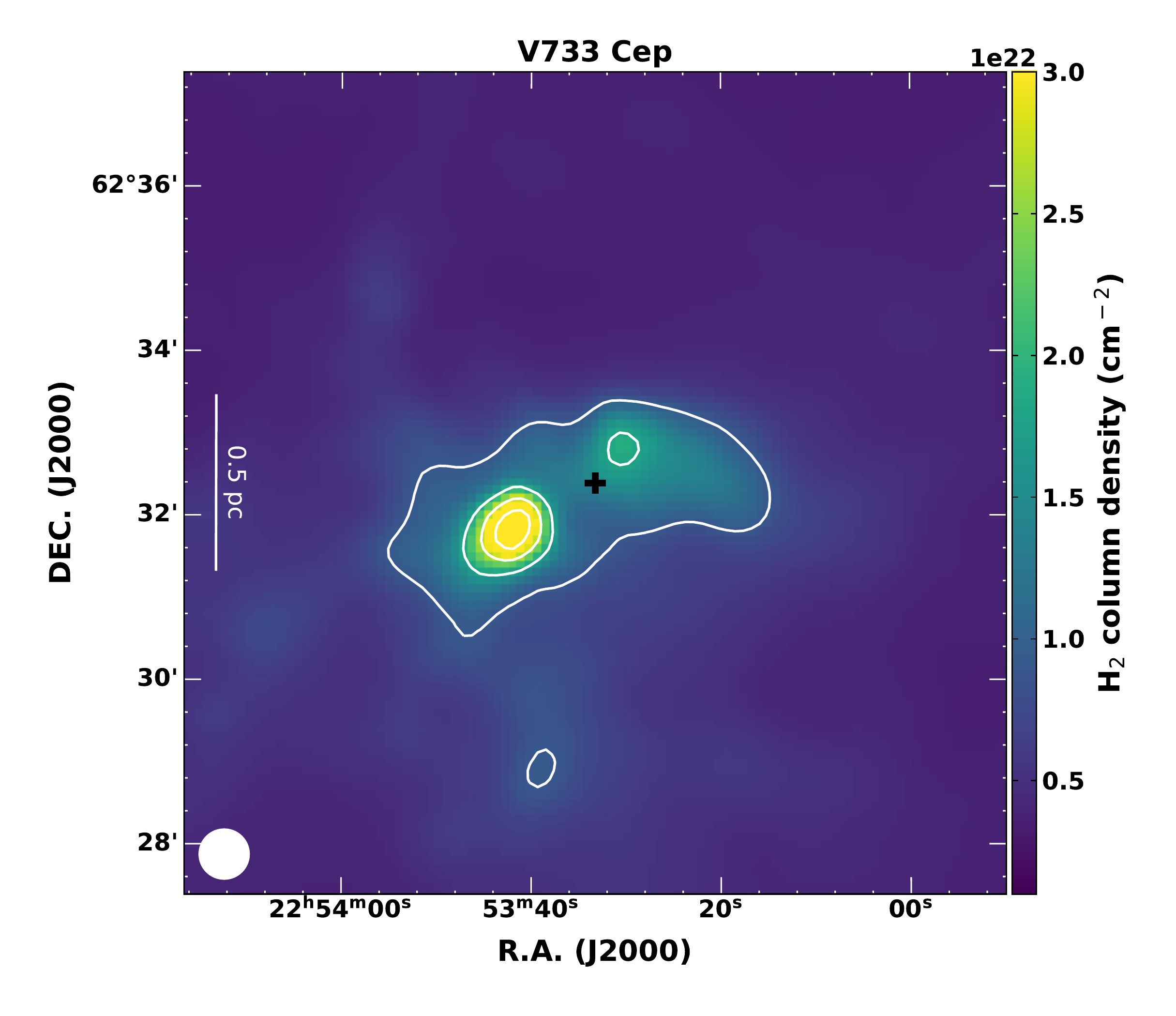} 
\hspace{1cm}
\includegraphics[width=0.35\textwidth]{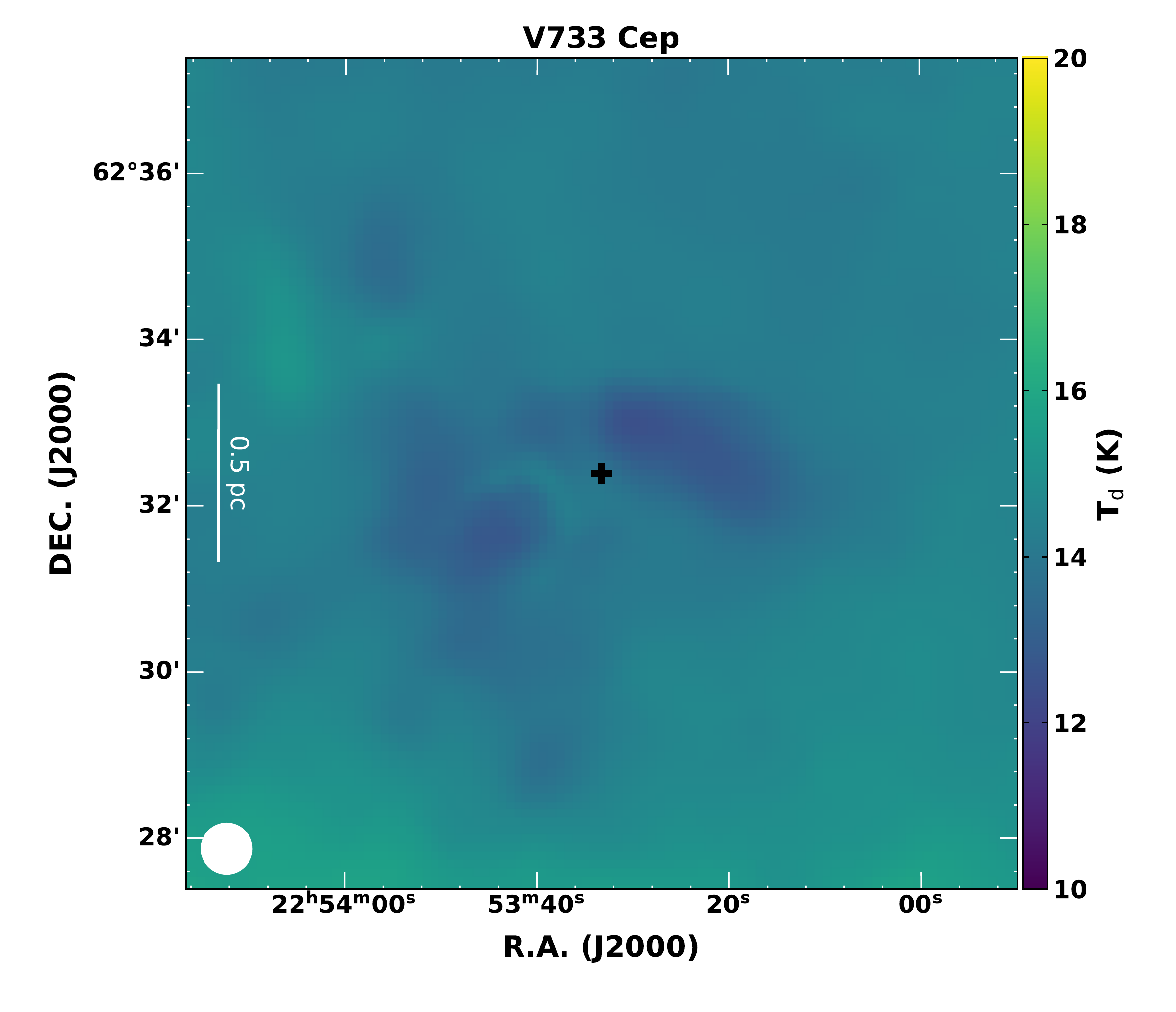} \\ 
\vspace{0.5cm}
\includegraphics[width=0.35\textwidth]{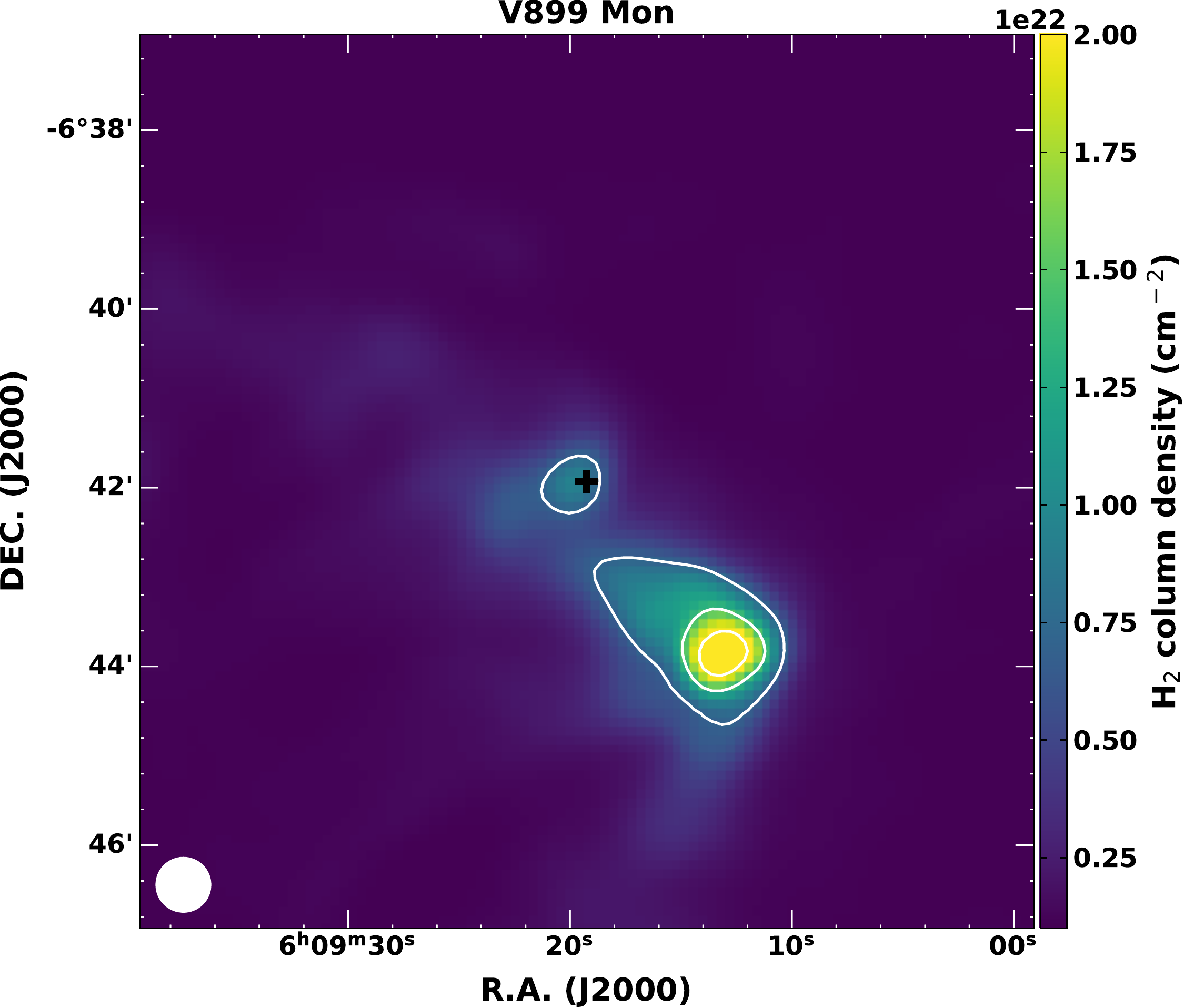}
\hspace{1cm}
\includegraphics[width=0.35\textwidth]{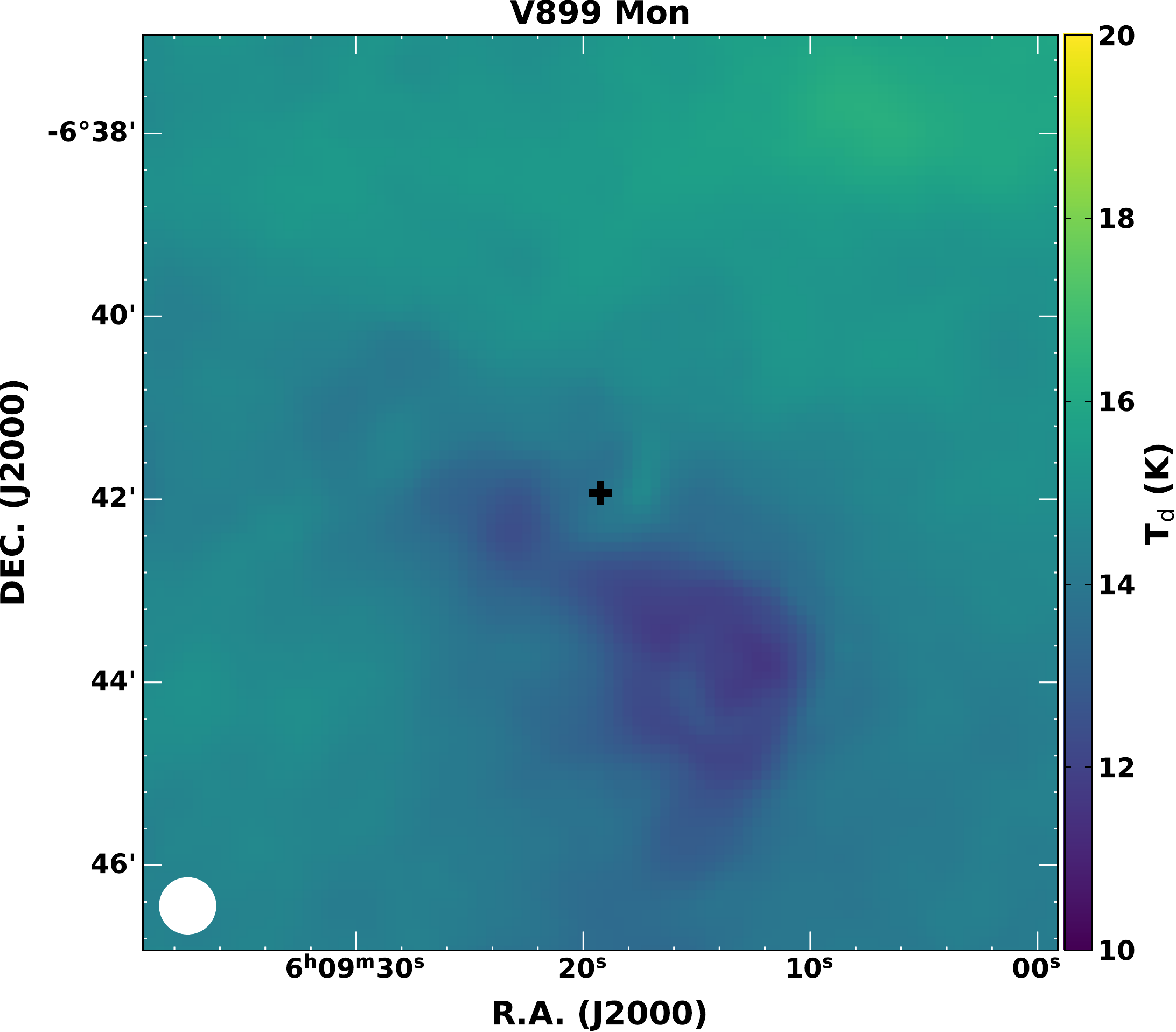} \\
\vspace{0.5cm}
\includegraphics[width=0.35\textwidth]{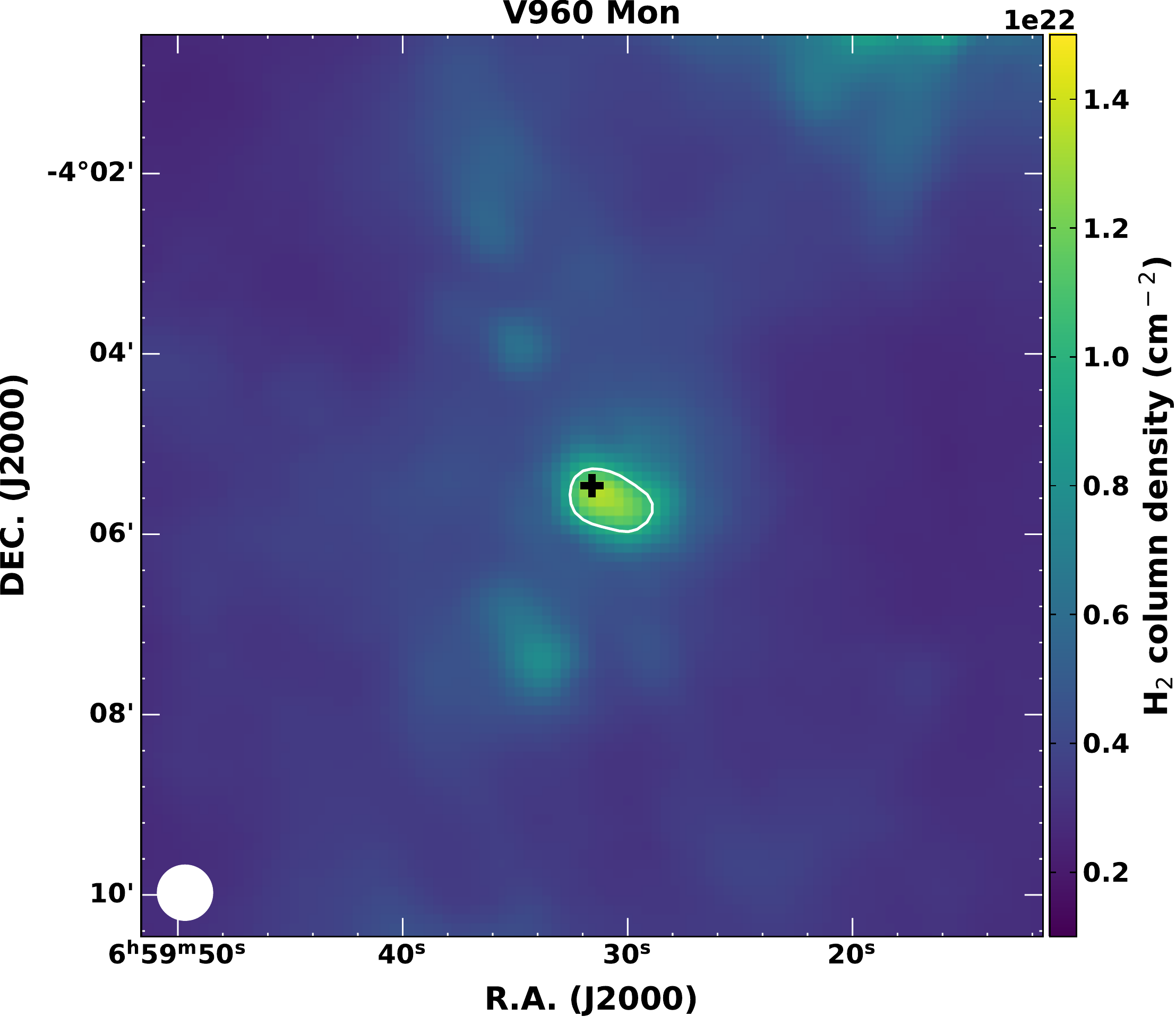}
\hspace{1cm}
\includegraphics[width=0.35\textwidth]{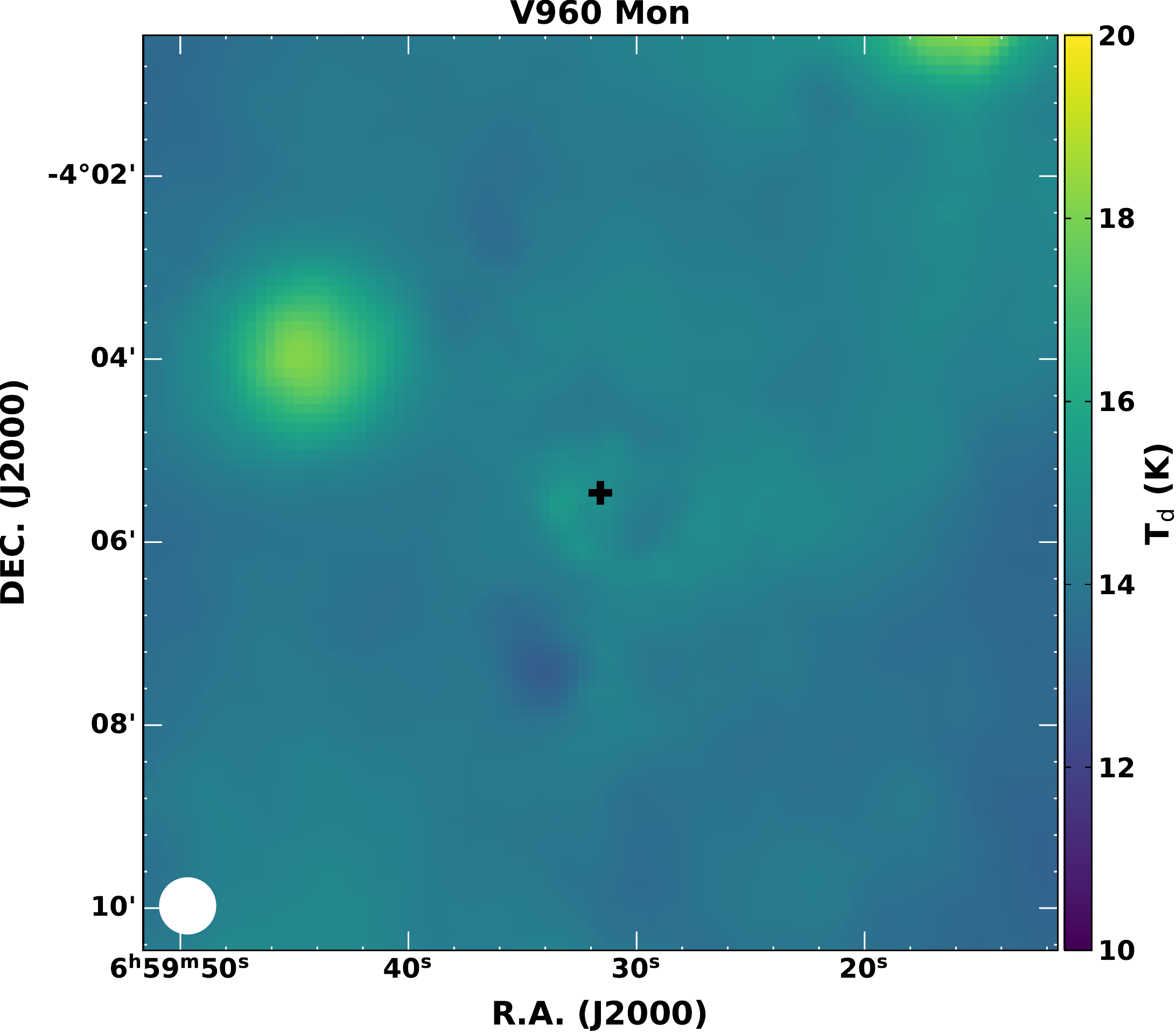} \\
\caption{(Continued.)}
\end{figure*}

\addtocounter{figure}{-1}
\begin{figure*}[h]
\centering 
\includegraphics[width=0.35\textwidth]{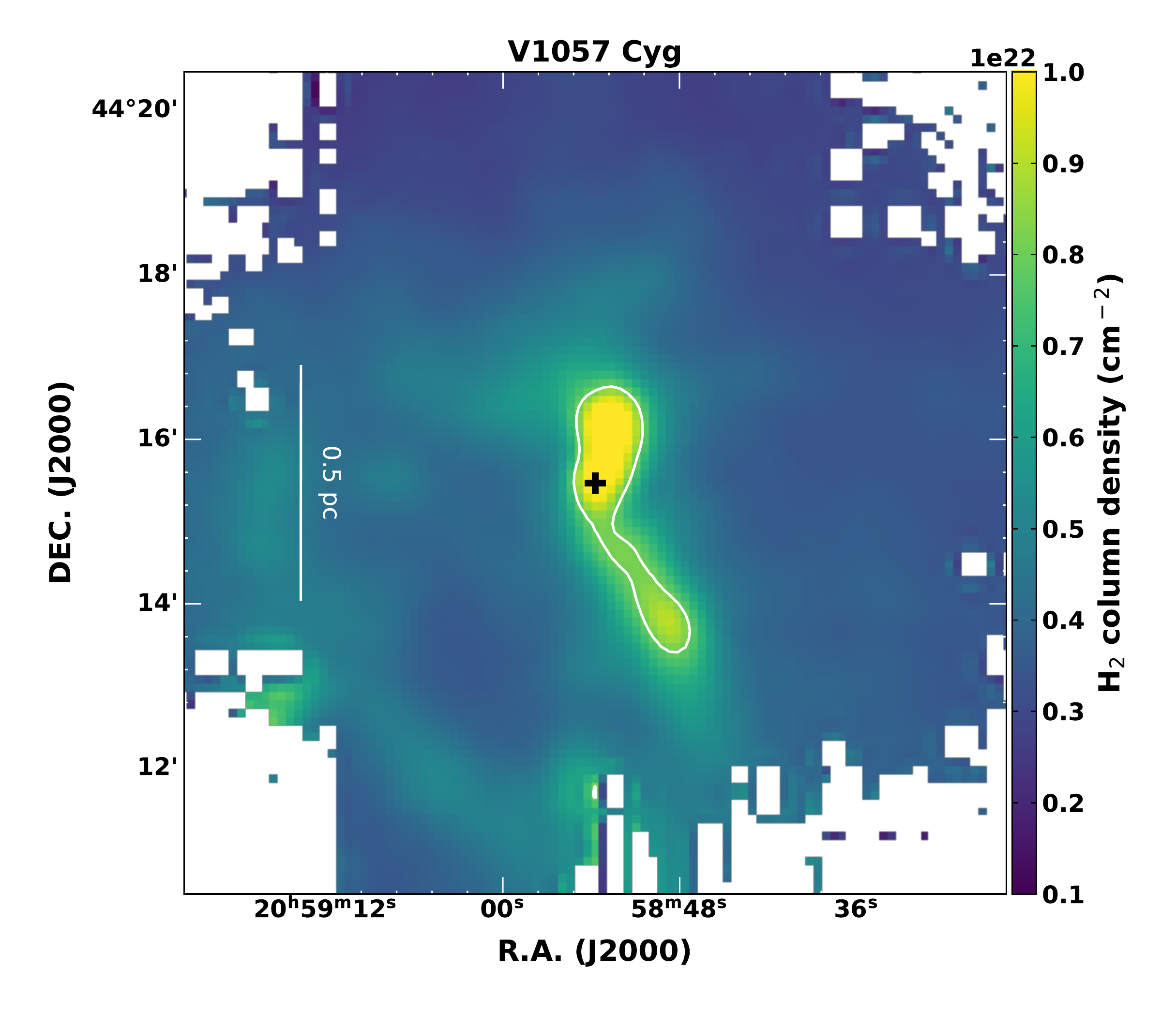}
\hspace{1cm}
\includegraphics[width=0.35\textwidth]{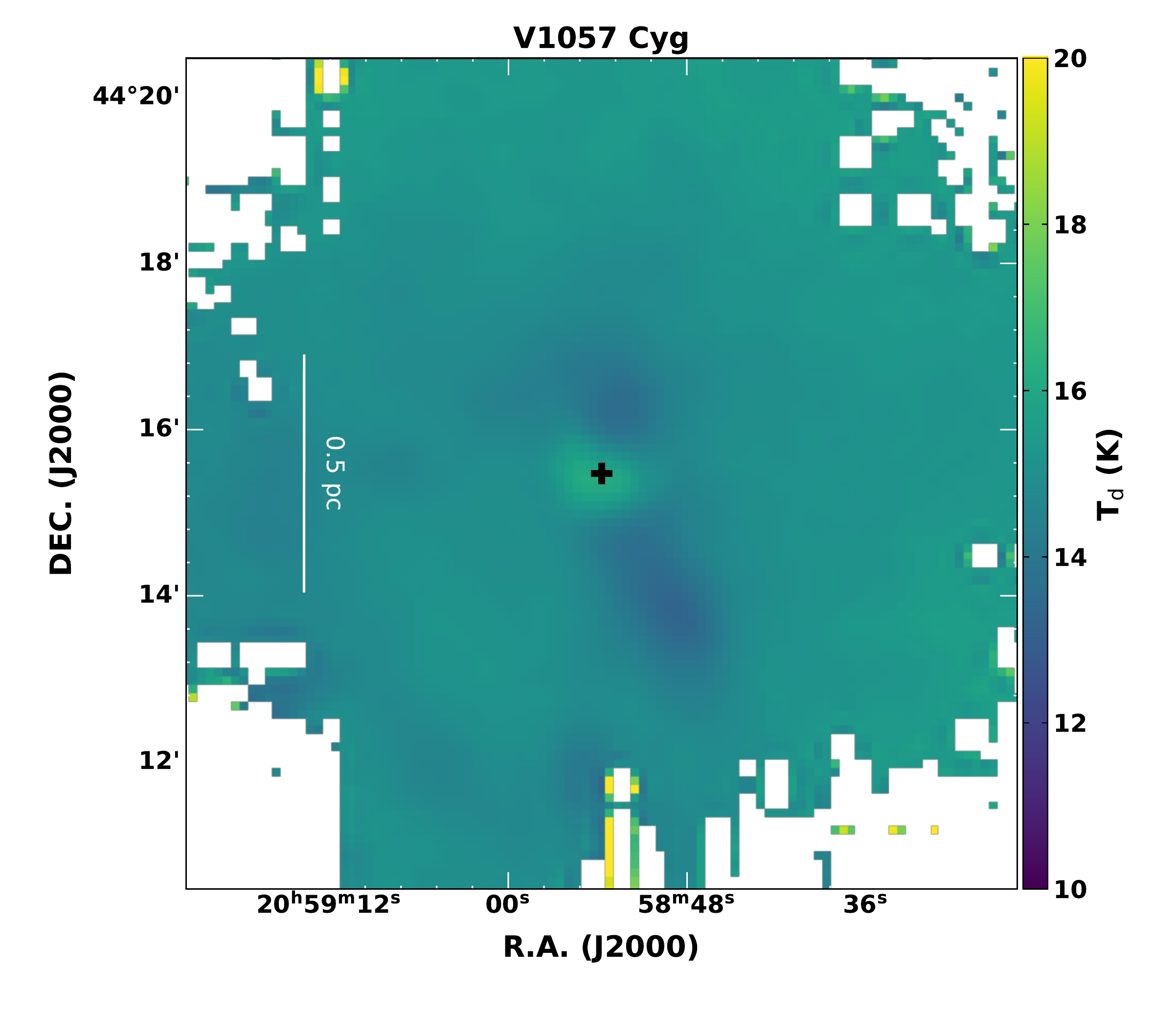} \\
\vspace{0.5cm}
\includegraphics[width=0.35\textwidth]{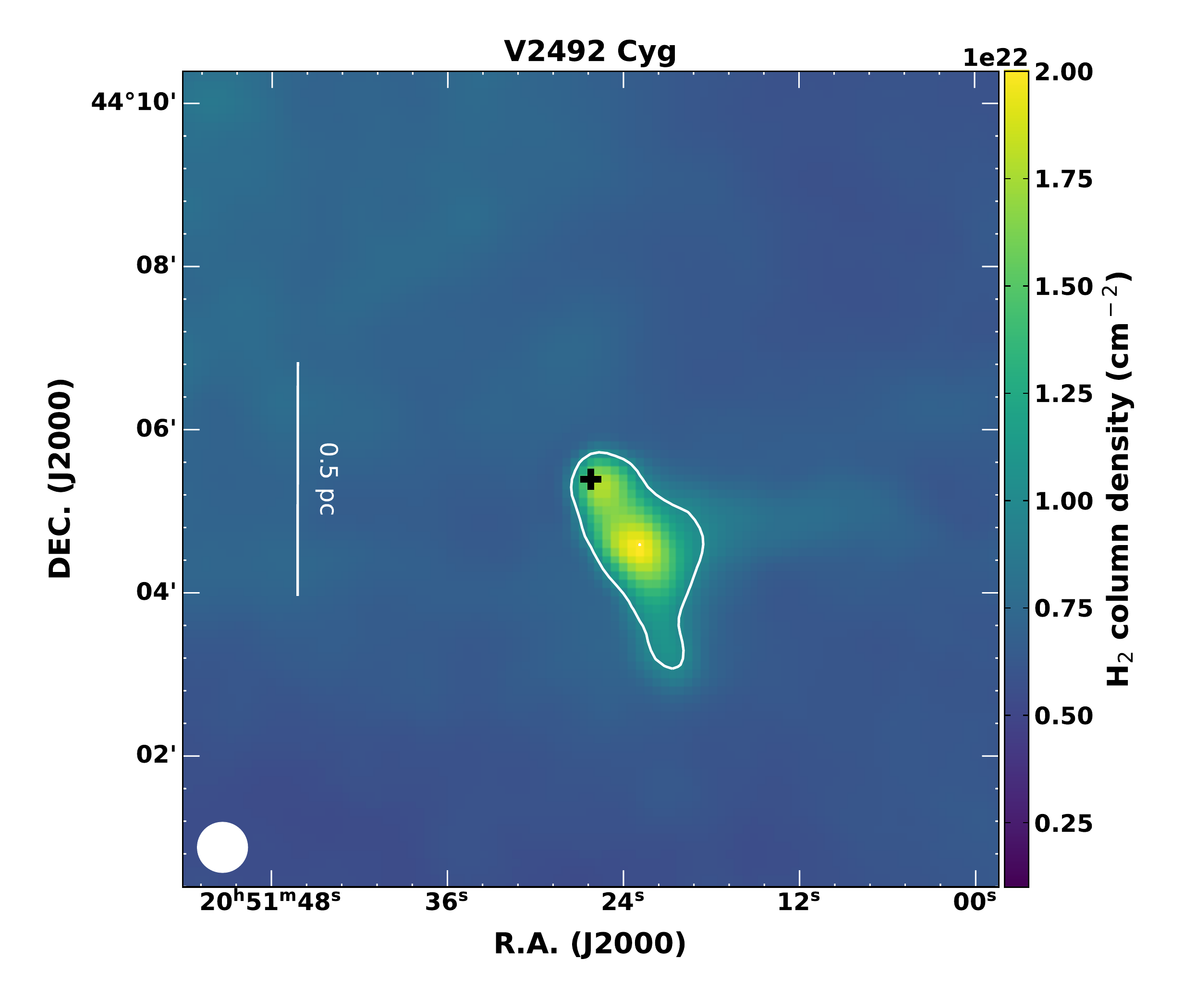} 
\hspace{1cm}
\includegraphics[width=0.35\textwidth]{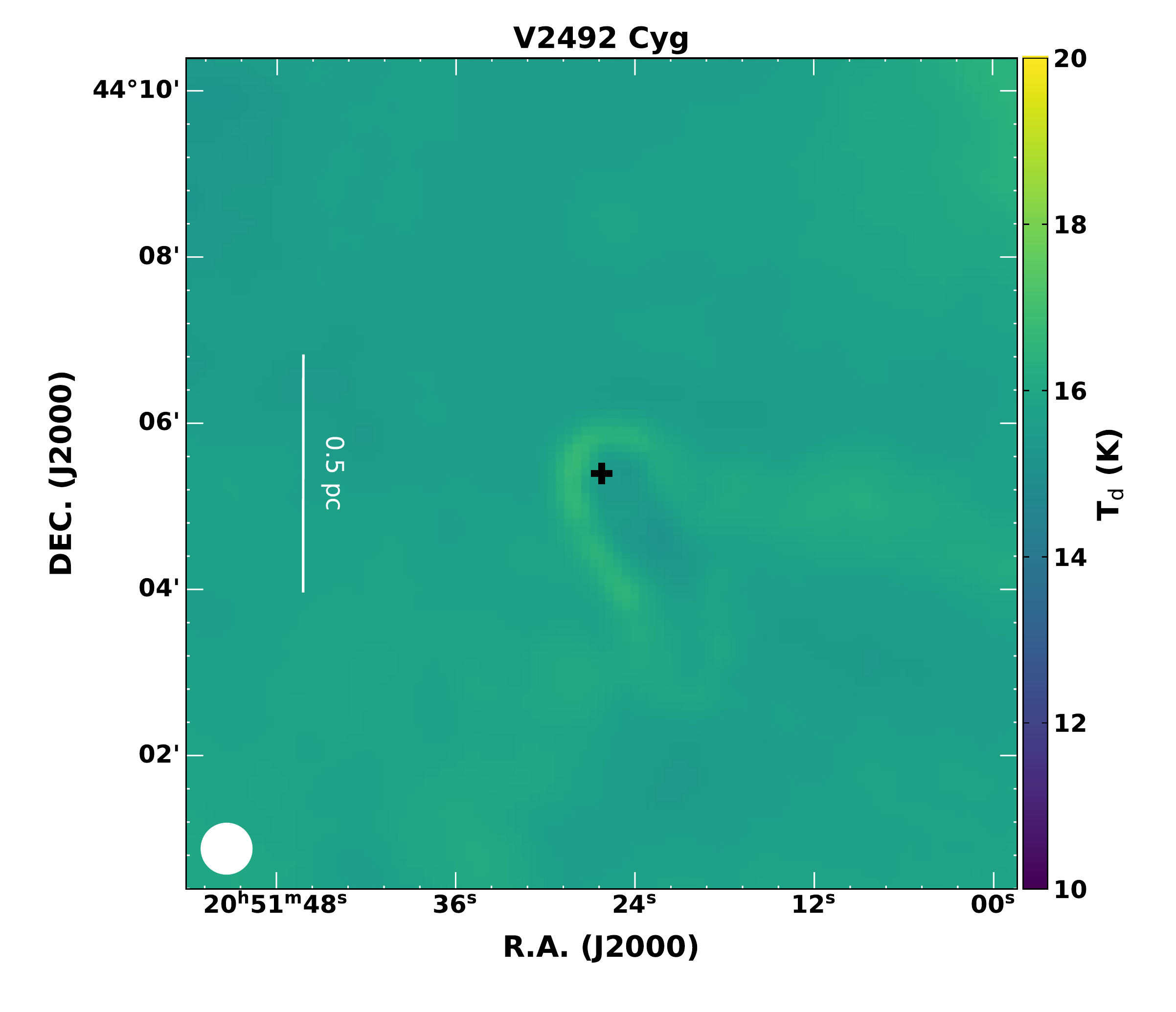} \\ 
\vspace{0.5cm}
\includegraphics[width=0.35\textwidth]{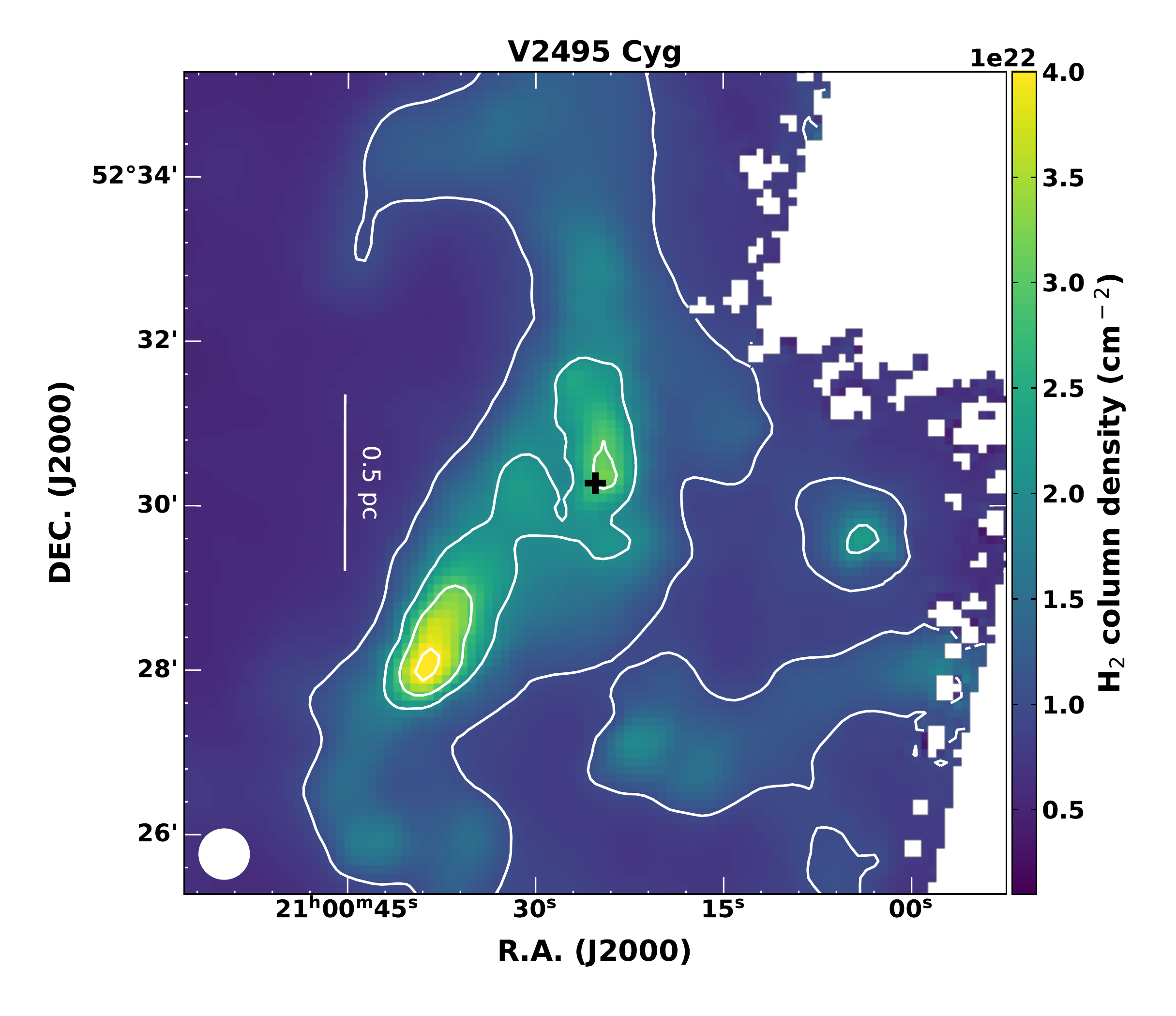}
\hspace{1cm}
\includegraphics[width=0.35\textwidth]{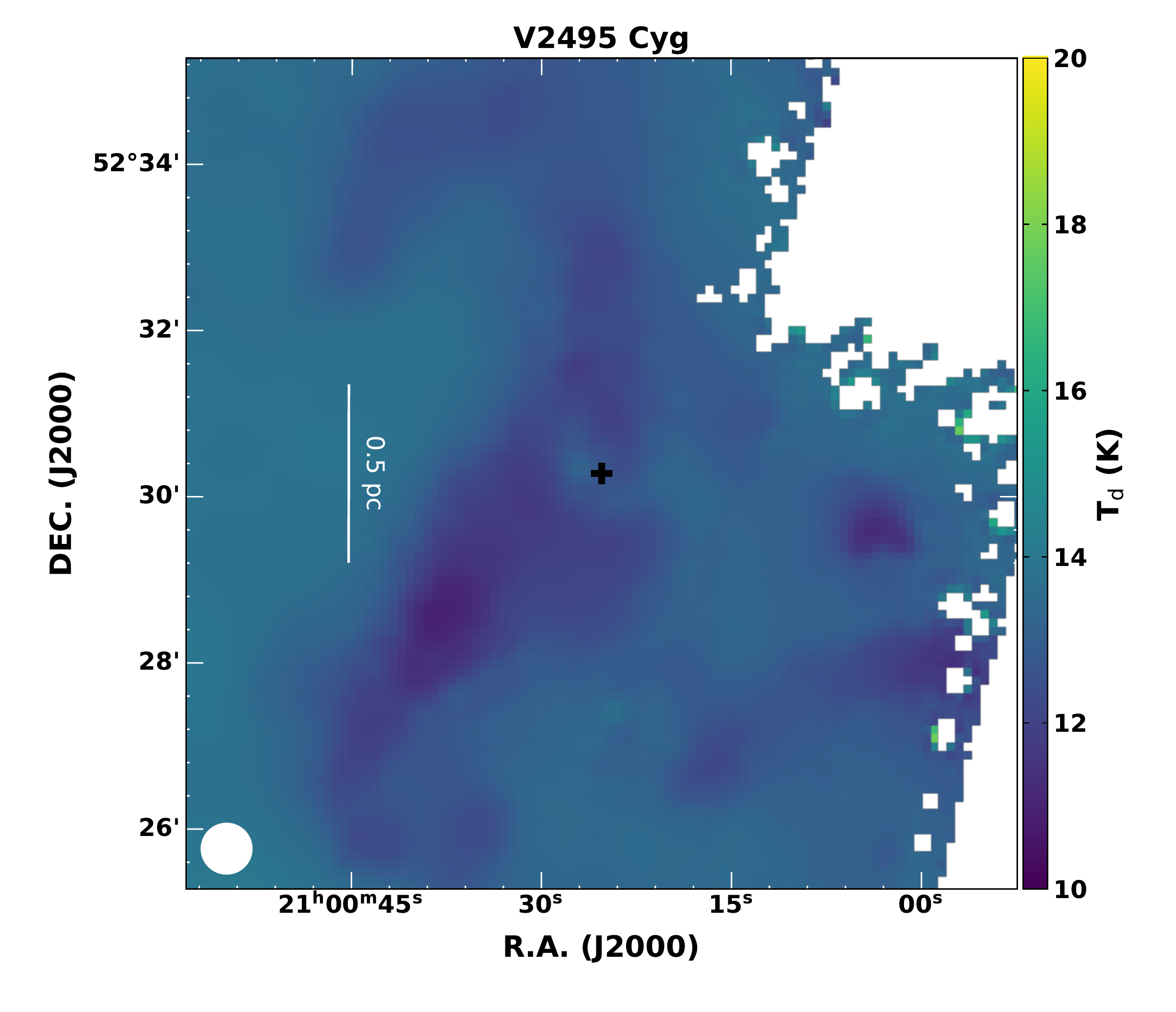} \\
\caption{(Continued.)}
\end{figure*}

\end{appendix}

\end{document}